\documentclass[final,3p,times,twocolumn,authoryear]{elsarticle}
\usepackage{xspace}
\usepackage{amssymb,latexsym,eufrak,times,amsmath}

\usepackage{ifthen}
\def\draftversion{0} % set this to 1 to display the content of the draft commands, or to 0 to hide them.

\makeatletter
\newcommand\mytoc{%
    \@starttoc{toc}%
}
\makeatother

\ifthenelse{\equal{\draftversion}{0}}{
	\usepackage{xcolor}
	\newenvironment{draft}{\begin{color}[rgb]{0,0.4,0}\begin{center}\hrule\vspace{0.5mm}DRAFT\end{center}}{\begin{center}END OF DRAFT\vspace{0.5mm}\hrule\end{center}\end{color}}

	\newcommand{\revend}[1]{\par\begin{color}[rgb]{0,0.4,0}\begin{center}\hrule\vspace{0.5mm}END OF #1's REVISIONS\vspace{0.5mm}\hrule\end{center}\end{color}\par}
	\newcommand{\todo}[1]{\begin{color}{red}\ $\bullet$ \textbf{To do: }#1 $\bullet$\ \end{color}}

	\newcommand{\del}[1]{\begin{color}[rgb]{0,0.5,0.0}\ $\bullet$ \textbf{Deleted: }#1 $\bullet$\ \end{color}}
	\newcommand{\sk}[1]{\begin{color}[rgb]{0.6,0,0.6}#1\end{color}}
	\newcommand{\toc}{\par\begin{color}[rgb]{0.6,0,0.6}\begin{center}\hrule\vspace{0.5mm}\begingroup\small\let\cleardoublepage\relax\let\clearpage\relax\mytoc\endgroup\vspace{0.5mm}\hrule\end{center}\end{color}\par}
	}{
	\newsavebox{\trashcan}

	\newcommand{\revend}[1]{}
	\newcommand{\todo}[1]{}

	\newcommand{\del}[1]{}
	\newcommand{\sk}[1]{}
	\newcommand{\toc}{}
	}

\long\def\symbolfootnote[#1]#2{\begingroup%
\def\thefootnote{\fnsymbol{footnote}}\footnote[#1]{#2}\endgroup} 

\newcommand{\aj}{AJ}% Astronomical Journal
\newcommand{\araa}{ARA\&A}% Annual Review of Astron and Astrophys
\newcommand{\apj}{ApJ}% Astrophysical Journal
\newcommand{\apjl}{ApJ}% Astrophysical Journal, Letters
\newcommand{\apjs}{ApJS}% Astrophysical Journal, Supplement
\newcommand{\apss}{Ap\&SS}% Astrophysics and Space Science
\newcommand{\aap}{A\&A}% Astronomy and Astrophysics
\newcommand{\aapr}{A\&A~Rev.}% Astronomy and Astrophysics Reviews
\newcommand{\jcap}{J. Cosmology Astropart. Phys.}% Journal of Cosmology and Astroparticle Physics
\newcommand{\mnras}{MNRAS}% Monthly Notices of the RAS
\newcommand{\na}{New A}% New Astronomy
\newcommand{\nar}{New A Rev.}% New Astronomy Review
\newcommand{\pasa}{PASA}% Publications of the Astron. Soc. of Australia
\newcommand{\pasp}{PASP}% Publications of the ASP
\newcommand{\pasj}{PASJ}% Publications of the ASJ
\newcommand{\qjras}{QJRAS}% Quarterly Journal of the RAS
\newcommand{\zap}{ZAp}% Zeitschrift fuer Astrophysik
\newcommand{\nat}{Nature}% Nature
\newcommand{\memsai}{Mem.~Soc.~Astron.~Italiana}% Mem. Societa Astronomica Italiana
\newcommand{\physrep}{Phys.~Rep.}% Physics Reports
\newcommand{\mh}{\ensuremath{\textrm{\,--\,}}}
\newcommand{\bb}[1]{\ifmmode \mbox{\boldmath $ #1$} \else  \mbox{\boldmath $#1$} \fi}

\newcommand{\dd}{\ensuremath{\,\mathrm{d}}}

\newcommand{\ave}[1]{\langle #1 \rangle}
\newcommand{\U}[1]{\ensuremath{\mathrm{~#1}}}
\newcommand{\e}[1]{\ensuremath{\times 10^{#1}}}
\newcommand{\yr}{\U{yr}}
\newcommand{\Myr}{\U{Myr}}
\newcommand{\Gyr}{\U{Gyr}}
\newcommand{\pc}{\U{pc}}
\newcommand{\kpc}{\U{kpc}}
\newcommand{\Mpc}{\U{Mpc}}
\newcommand{\msun}{\U{M}_{\odot}}
\newcommand{\Msun}{\msun}
\newcommand{\msunyr}{\Msun\yr^{-1}}
\newcommand{\Msunyr}{\Msun\yr^{-1}}
\newcommand{\cc}{\U{cm^{-3}}}

\newcommand{\hi}{H{\sc i} }
\newcommand{\hii}{H{\sc ii} }

\newcommand{\rh}{\ensuremath{r_\mathrm{h}}}

\newcommand{\code}[1]{\texttt{#1}\xspace}

\newcommand{\trh}{\tau_{\rm rh}}

\newcommand{\eqn}[2][]{Equation#1~\ref{eqn:#2}} % for plural form, use: \eqn[s]{emc2} and (\ref{eqn:emc3})     to get  Equations (1) and (2)
\newcommand{\fig}[2][]{Figure#1~\ref{fig:#2}}

\newcommand{\sect}[2][]{Section#1~\ref{sec:#2}}

\renewcommand{\eqn}[2][]{equation#1~(\ref{eqn:#2})}
\renewcommand{\fig}[2][]{Fig#1.~\ref{fig:#2}}

\journal{New Astronomy Reviews}
\begin{document}
 
%%%%%%%%%%%%%%%%%%%%%%%%%%%%%%%%%%%%%%%%%%%%%%%%%%%%%%%%%%%%%%%%%%%%%%%%%%%%%%%%%%%%%%%%%%%%%%%%%%%%%%%%%%%%%%%%%%%%%%%%%%%%%%%%%%%%%%%%%%%%%%%%%%%%%%%%%%%%%%%%%%%%%%%%%%%%%
%%%%%%%%%%%%%%%%%%%%%%%%%%%%%%%%%%%%%%%%%%%%%%%%%%%%%%%%%%%%%%%%%%%%%%%%%%%%%%%%%%%%%%%%%%%%%%%%%%%%%%%%%%%%%%%%%%%%%%%%%%%%%%%%%%%%%%%%%%%%%%%%%%%%%%%%%%%%%%%%%%%%%%%%%%%%%
%%%%%%%%%%%%%%%%%%%%%%%%%%%%%%%%%%%%%%%%%%%%%%%%%%%%%%%%%%%%%%%%%%%%%%%%%%%%%%%%%%%%%%%%%%%%%%%%%%%%%%%%%%%%%%%%%%%%%%%%%%%%%%%%%%%%%%%%%%%%%%%%%%%%%%%%%%%%%%%%%%%%%%%%%%%%%

\begin{frontmatter}
\title{Star Clusters in Evolving Galaxies}
\author{Florent Renaud}
\address{Department of Astronomy and Theoretical Physics, Lund Observatory, Box 43, SE-221 00 Lund, Sweden\\Department of Physics, University of Surrey, Guildford, GU2 7XH, UK}
\ead{florent@astro.lu.se}

\begin{abstract}
Their ubiquity and extreme densities make star clusters probes of prime importance of galaxy evolution. Old globular clusters keep imprints of the physical conditions of their assembly in the early Universe, and younger stellar objects, observationally resolved, tell us about the mechanisms at stake in their formation. Yet, we still do not understand the diversity involved: why is star cluster formation limited to $10^5 \Msun$ objects in the Milky Way, while some dwarf galaxies like NGC~1705 are able to produce clusters 10 times more massive? Why do dwarfs generally host a higher specific frequency of clusters than larger galaxies? How to connect the present-day, often resolved, stellar systems to the formation of globular clusters at high redshift? And how do these links depend on the galactic and cosmological environments of these clusters? In this review, I present recent advances on star cluster formation and evolution, in galactic and cosmological context. The emphasis is put on the theory, formation scenarios and the effects of the environment on the evolution of the global properties of clusters. A few open questions are identified.
\end{abstract}
\end{frontmatter}

\mytoc

%%%%%%%%%%%%%%%%%%%%%%%%%%%%%%%%%%%%%%%%%%%%%%%%%%%%%%%%%%%%%%%%%%%%%%%%%%%%%%%%%%%%%%%%%%%%%%%%%%%%%%%%%%%%%%%%%%%%%%%%%%%%%%%%%%%%%%%%%%%%%%%%%%%%%%%%%%%%%%%%%%%%%%%%%%%%%
%%%%%%%%%%%%%%%%%%%%%%%%%%%%%%%%%%%%%%%%%%%%%%%%%%%%%%%%%%%%%%%%%%%%%%%%%%%%%%%%%%%%%%%%%%%%%%%%%%%%%%%%%%%%%%%%%%%%%%%%%%%%%%%%%%%%%%%%%%%%%%%%%%%%%%%%%%%%%%%%%%%%%%%%%%%%%
%%%%%%%%%%%%%%%%%%%%%%%%%%%%%%%%%%%%%%%%%%%%%%%%%%%%%%%%%%%%%%%%%%%%%%%%%%%%%%%%%%%%%%%%%%%%%%%%%%%%%%%%%%%%%%%%%%%%%%%%%%%%%%%%%%%%%%%%%%%%%%%%%%%%%%%%%%%%%%%%%%%%%%%%%%%%%
\part{Introduction}

%%%%%%%%%%%%%%%%%%%%%%%%%%%%%%%%%%%%%%%%%%%%%%%%%%%%%%%%%%%%%%%%%
\section{Outline, framework and objectives}

Over the last years, the boundary between the fields of galaxy formation and stellar populations started to fade. There is now a strong need to merge the two and build a more holistic picture of star clusters within their galactic and cosmological environments. However, bridging the gap between previously distinct fields of research is a demanding task. With this review, I aim to draw a picture of our current knowledge and understanding of this topic, by focussing on the multi-scale and multi-physics aspects of the question and the coupling between the two sides of this coin. This fields being vast, technical aspects and details on the results presented here are often skipped, and I strongly encourage the reader to consult the studies mentioned and the references therein. This contribution could not be more than a mere starting point to explore a field too rich and diverse to make an exhaustive review of it.

The focus being on the link between the clusters and their galactic and cosmological context, it does not address the topics of internal cluster physics and their stellar populations in details, nor the state of the art on the observational side of the subject. I encourage the reader to explore complementary reviews on these topics, in particular in
\begin{itemize}
\item \citet{Naab2017}: galaxy formation
\item \citet{Hennebelle2012}: turbulence of the interstellar medium
\item \citet{Krumholz2014}: star formation
\item \citet{Brodie2006}: formation of extragalactic clusters, mainly from an observational perspective
\item \citet{Charbonnel2016}: multiple stellar populations
\item \citet{Portegies2010} and \citet{Adamo2015}: young massive clusters
\item \citet{Heggie2003} and \citet{Vesperini2010}: internal dynamics of clusters
\item \citet{Aarseth2003}: $N$-body simulations of clusters.
\end{itemize}

The interested reader could also compare our current understanding of this field with the challenges identified more than a decade ago by \citet{Davies2006}, and even confront the timeline they proposed for progress in each domain to what actually happened...

In this introduction, I set the stage for this review, list a few possible definitions of star clusters, and highlight connections with dwarf galaxies. The second part addresses our theoretical understanding of star cluster formation in different galactic environments, in the Milky Way, in interactions and mergers, in the local Universe, at high redshift etc. The main focus of this part is the physical phenomena responsible for the accumulation of star forming dense gas, and the first Myr of the resulting stars clusters, until they become gas-free. I also propose short incursions in the neighbour topics of nuclear clusters and ultra-compact dwarf galaxies. The third part concerns the evolution of gas-free clusters, focussing on the effects of the galactic and cosmological environment but leaving aside most of the cluster internal physics. I address here the processes modifying the orbits of clusters, like dynamical friction, and the tidal effects (adiabatic and impulsive) that alter their properties. Finally, I identify in the last part a few open questions on these topics, the state of the art and why they still keep us puzzled.

%%%%%%%%%%%%%%%%%%%%%%%%%%%%%%%%%%%%%%%%%%%%%%%%%%%%%%%%%%%%%%%%%
\section{A multi-scale and multi-physics topic}

Many introductions of papers and proposals about star clusters would tell you they are the building blocks of galaxies, that most if not all stars form in clusters and that globular clusters have witnessed all steps in the evolution of their host. Yet, the formation and evolution of star clusters remains a poorly known topic. The reason for this situation probably lies in the multi-scale and multi-physics nature of the field. Star clusters form in dense gas clouds, of which assembly is triggered, set and regulated by galactic-scale hydrodynamics, itself influenced by the inter-galactic and cosmological environment. Once formed, stars in clusters alter the morphology, energy and chemistry of their nurseries through stellar feedback. Details on the injection of feedback are set by stellar evolution and depends on the clustering of star formation, the presence of binary and multiple stars, the metallicity etc. These effects propagate to larger scales, up to e.g. driving galactic outflows and enriching the inter-galactic medium. The details depend on the porosity of the interstellar medium (ISM), the cooling rate, the turbulence and other processes which are themselves largely set at galactic scales. In parallel, the internal physics of clusters (regulated by stellar evolution and star-star interactions) rules the evolution of these stellar systems and plays an important role in the evolution of their mass and size, on top of external factors like tides.

In the end, the formation and evolution of star clusters is the results of a complex interplay of dynamical and hydrodynamical (and probably magnetic) processes from intergalactic scales ($\sim 10 \Mpc$, $\sim 10 \Gyr$), down to stellar scales ($\sim 10^{-3} \pc$, $\sim 1 \U{day}$), all this happening for most of the life of the Universe. This wide diversity of scales and physical processes makes it difficult to propose holistic theories. Yet, significant progresses on individual concepts have been made in the last few decades, and coupling several aspects together is now routinely achieved in theories, simulations and interpretation of observations. By focussing on the interactions between clusters and their galactic context, I will highlight here some of these progresses and pin down a few possible directions for future research.

%%%%%%%%%%%%%%%%%%%%%%%%%%%%%%%%%%%%%%%%%%%%%%%%%%%%%%%%%%%%%%%%%
\section{What is a star cluster?}

This section gives the general tone of this review: asking a simple question and showing that the complexity of the problem still causes a (frustrating but exciting) lack of definite answer.

Until about a decade ago, star clusters in general and globulars in particular were considered as simple (and sometimes bland) objects, made of gravitationally bound stars sharing a common origin. However, the improvements in observational resources and techniques allowing us to deeper probe these objects and study the composition of their stars, and this in an always widening range of environments, have revealed a much more complex picture. Up to the point that, today, there is no clear definition of what a star cluster is. More than just semantics (or our urge to sort the Universe), the underlying question is to establish whether star clusters and galaxies follow comparable physical formation processes, and to which extend parallels can be drawn between theses classes of objects.

%%%%%%%%%%%
\subsection{Bound and part of a galaxy?}

As it is often the case in astrophysics, the community attempted to define this concept in contrast with other objects. For instance, star clusters differ from stellar associations and moving groups by being gravitationally bound \citep[see][and references therein]{Zuckerman2004}. \citet{Gieles2011} further proposed that objects with crossing times longer than the age of their stars would likely be unbound and should thus be classified as associations. One can also attempt to discriminate clusters and galaxies using a hierarchical argument: one contains the other. In that case, some dwarf galaxies like Aquarius and Tucana, which do not host any cluster \citep{Forbes2005}, would not qualify as galaxies. Furthermore, some of the so-called ultra-faint dwarfs count fewer stars than massive globular clusters and can even be fainter than a single star \citep{Belokurov2007}.

%%%%%%%%%%%
\subsection{Small and compact?}

\begin{figure*}
\includegraphics{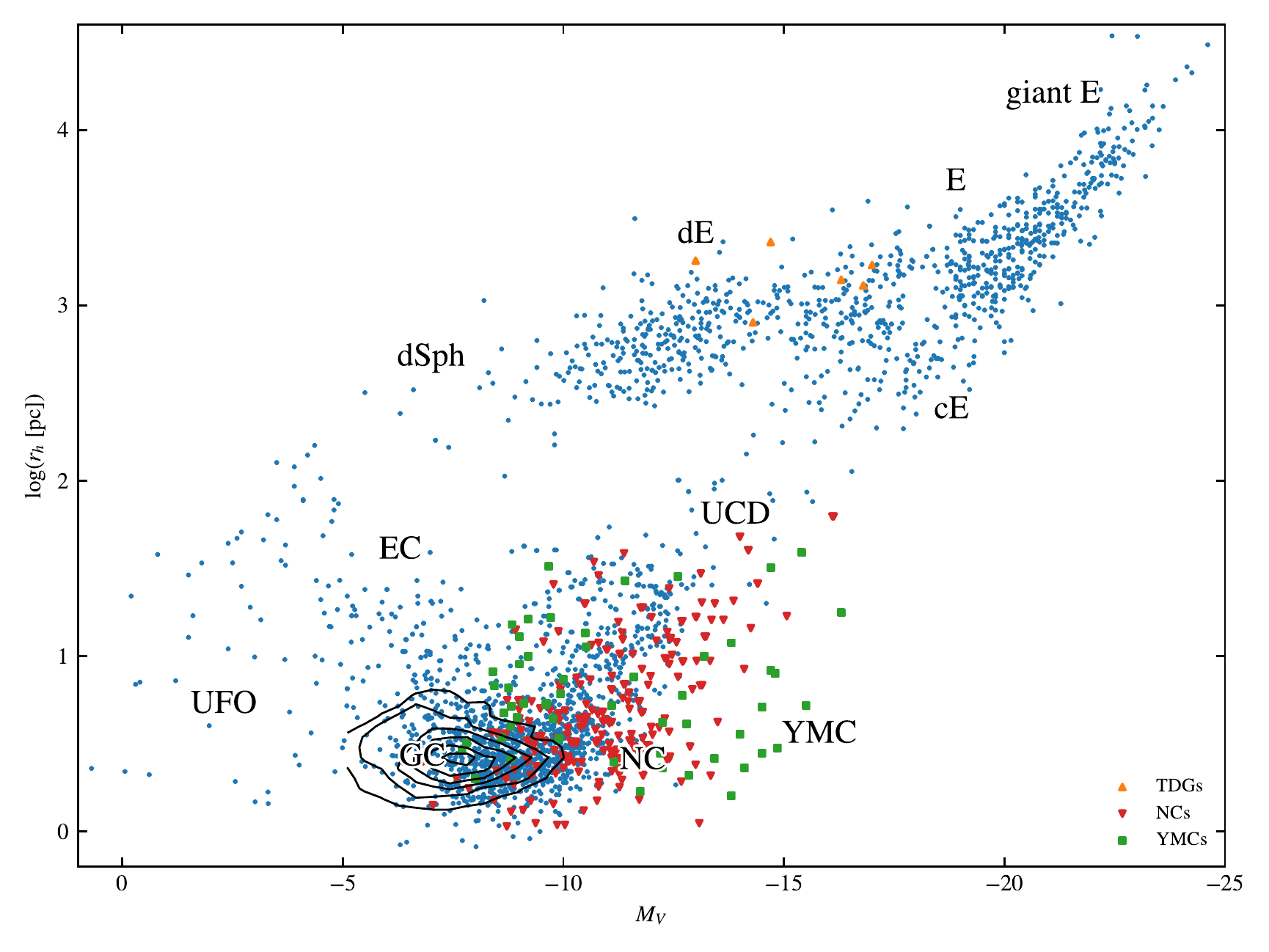}
\caption{V-band absolute magnitude versus half-light radius of dynamically hot stellar objects. Contours indicate the number density distribution of the $\sim 13000$ globular candidates in galaxies of the Virgo cluster from \citet{Jordan2009}, that would overload the figure if plotted individually. We give rough indications of the loci of the different classes of objects: globular clusters (GCs), extended clusters (ECs, also known as faint fuzzy clusters), ultra-faint objects (UFOs), dwarf spheroidal galaxies (dSphs), dwarf elliptical galaxies (dEs), tidal dwarf galaxies (TDGs) nuclear clusters (NCs), ultra-compact dwarf galaxies (UCDs), Young massive clusters (YMCs), compact ellipticals (cEs), ellipticals (and lenticulars) and giant elliptical galaxies (Es), as well as a few examples that illustrate the difficulties of drawing boundaries between these classes. We distinguish some objects that are undoubtedly classified, because of their position in their host galaxy (TDGs, NCs) or the age of their stars (YMCs). 
Data points from \citet{Portegies2010}, the SAGES database (\citealt{Brodie2011}), \citet{McConnachie2012}, \citet{Duc2014} \citet{Norris2014}, \citet{Georgiev2014}, \citet{Homma2016}, \citet{Contenta2017} and from Karina Voggel (private communication, as an update of \citealt{Voggel2016}) and references therein.}
\label{fig:masssize}
\end{figure*}

After the hierarchical proposition above, it is natural to try to discriminate star clusters and galaxies in term of size and luminosity. \fig{masssize} compiles a large variety of stellar systems in the magnitude - size plane. A few years ago, the dense clusters seemed to be separated from the galaxies, both on the bright side (GCs and NCs vs. cEs) and the faint one (GCs vs. dSphs). These gaps between the classes disappeared, thanks to deeper observations in a variety of environments. The ``chronological evolution'' of this figure can be seen by looking at previous versions (each from their own perspective) in e.g. \citet{Mackey2005}, \citet{Belokurov2007}, \citet{Misgeld2011}, \citet{Brodie2011}, \citet{Hwang2011}, \citet{Forbes2013}, \citet{Norris2014}, \citet{Voggel2016} and \citet{Contenta2017}. However, there is still a clear minimum in the size distribution of objects around $100 \pc$. \citet{Gilmore2007} proposed this to be a transient zone, populated by objects out of equilibrium and being close to dissolution, but recently found UCDs challenge this interpretation at the bright end of the distribution \citep[e.g.][]{Voggel2016}.

Filling these gaps has made challenging (and even impractical) to define boundaries between these classes in term of luminosity (or mass) and size (or density). For instance, in \fig{masssize} the loci of nuclear clusters (which can otherwise be unambiguously defined based on their central position in their host galaxies) overlap with that of ultra-compact dwarf galaxies, and the brightest end of the globulars. However, different formation scenarios have been proposed for these classes of objects (see Part II).

%%%%%%%%%%%
\subsection{Dark matter-free?}
\label{sec:dm}

Recently discovered extended clusters and faint dSphs around nearby galaxies appear to make a continuous connection between the bulks of their respective families. In this regime, the difference is often made on the mass-to-light ratio. When high, it is assumed to depict the presence of dark matter that would only be found in galaxies. Following such a definition, tidal dwarf galaxies, that originate from galactic disc material being tidally stripped during an interaction \citep{Duc1999} and that hence are free of dark matter, should be classified as clusters. Furthermore, \citet{Baumgardt2008b} argued that globular clusters, if formed in dark matter-dominated environments and evolving in weak tides, would retain a dark halo around them (but not in their central parts, see also \citealt{Bekki2012}). This could be the case of massive globular clusters like Omega Cen which is suspected to be the core remnant of a tidally stripped dwarf \citep{Freeman1993, Bekki2003}. \citet{Ibata2013b} estimated that NGC~2419 could have a dark halo twice more massive than its stellar component. This would then explain the flattening of the velocity dispersion profiles observed in the outskirts of some clusters \citep[see e.g.][and \sect{pe}]{Scarpa2007}. However, such clusters would have a different formation scenario than less massive clusters, but possibly more similar to that of UCDs \citep[see e.g.][and the opposite conclusion in \citealt{Caso2014}]{Mieske2012, Pfeffer2013, Renaud2015}. Here again, the classification is not obvious.

Furthermore, a high mass-to-light ratio can be interpreted differently. For instance, the high ratio observed in UCDs \citep{Hasegan2005, Dabringhausen2008, Mieske2008} can be explained by invoking an initial stellar mass function (IMF) with an excess of massive stars. In old UCDs \citep{Evstigneeva2007}, the massive stars have transformed into the remnants of their evolution, like black holes and neutron stars, thus increasing the mass-to-light ratio \citep{Dabringhausen2009, Dabringhausen2012}. This interpretation would imply that the star formation in UCDs differs from that of solar neighbourhood, where the IMF appears to be invariant (\citealt{Bastian2010}, see also \citealt{Leigh2012} on a dynamical reconstruction of the IMF in globular clusters). It is still not clear whether the same could apply to massive globular clusters.

%%%%%%%%%%%
\subsection{Relaxed?}
\label{sec:relax}

In their attempt to define galaxies, \citet{Forbes2011b} proposed to compare the relaxation time of objects to the age of the Universe. The relaxation time quantifies how long a stellar system takes to erase signs of a perturbation, through star-star encounters and exchanges of energy toward equipartition (see \sect{collisional} for more details). For old systems (i.e. with an average stellar mass of $0.5 \Msun$) in the mass range of $10^{5 \mh 8} \Msun$, a relaxation time shorter than a Hubble time corresponds to a half-light radii of $1 \mh 6 \pc$. Most of the globular and nuclear clusters do indeed yield relaxation times shorter than a Hubble time (which qualifies them as collisional systems), while it is not the case for the more diffuse, collision-less, UCDs and cEs. The overlap of these classes in \fig{masssize} implies that adopting such a definition would require to update the classification of some of these objects, but the majority would keep their historical type. If convenient from a semantic point of view, this definition however sets a somewhat arbitrary boundary, and hence could not be used to distinguish (or not) different physical processes at stake during the formation and/or evolution of these objects. This unsatisfactory situation pushed the community to propose alternative definitions, based on a common composition or a common origin of stars.

%%%%%%%%%%%
\subsection{Hosting a simple population?}

For long, globular clusters have been thought to consist of stars with a common origin, i.e. that formed at the same place, same time and thus from the same material \citep[e.g.][]{Ashman1998}. On the contrary galaxies yield more complex star formation histories and thus encompass a variety of stellar populations. The complexity of the stellar populations would then set a clear boundary between the two classes. The situation is however more complicated.

Anti-correlations between the abundances of several chemical elements (Na-O, C-N, but also Mg-Al to some extent, \citealt{Pancino2017}) seem to exist in all globular clusters \citep[see e.g.][and references therein]{Cohen1978, Kraft1997, Gratton2001, Carretta2009}. Such variations, from stars to stars, results from precise reaction rates depending on the abundance of the catalysts species during the CNO and NeNa cycles of hydrogen burning at high temperature \citep{Denisenkov1990, Kraft1994}. Therefore, theses abundances and the anti-correlations have been set at the earliest stages of the clusters, by the now long-gone massive (and thus hot) stars, and are therefore expected to leave a unique fingerprint of each cluster. Anti-correlations are indeed used to estimate cluster memberships in the always deeper surveys where contamination of background stars must be corrected for \citep{Meszaros2015}. However, these variations imply a sequential star formation, with a first generation of stars producing the chemical elements (with the above mentioned spreads and others) used to form a second population, and even more. This translates into multiple main sequences and subgiant branches visible in colour-magnitude diagrams \citep{Piotto2007, Milone2008}.

To date, the field of multiple stellar populations has raised more questions than it has provided answers. While several formation scenarios have been proposed, our lack of detailed understanding of stellar evolution, star cluster formation and feedback has prevented us to reach a comprehensive theory that would account for all the strong observational constraints (abundance ratios and spreads, relative importance of the populations etc.). Discussing these scenarios and their respective issues is out of the scope of the present contribution and the reader is referred to extensive reviews by \citet{Gratton2012} and \citet{Charbonnel2016}.

All clusters more massive than $10^4 \Msun$ seem to yield a Na-O anti-correlation \citep{Carretta2010}, while it is absent in less massive objects like open clusters \citep[e.g.][]{Bragaglia2014}, yet with the exception of NGC 6791 \citep{Geisler2012}. Therefore, the presence of multiple stellar populations has been proposed to define the class of globular clusters. This would exclude a few objects like Palomar 3, 12, 14, Terzan 7, 8 or Rup 106 \citep{Cohen2004, Sbordone2004, Koch2009, Caliskan2012, Villanova2013, Carretta2014}, notably smaller than the average globular in the Milky Way. 

Star clusters would then be almost coeval collections of stars, with comparable but not exactly identical chemistries and ages. However, massive objects like NCs and UCDs (\sect{nc}) potentially assembled by merging several clusters, either shortly after the formation in a common gaseous structure, or later elsewhere in their galaxy, and could thus fulfill the above-mentioned criteria to define clusters.

%%%%%%%%%%%
\subsection{Is there a definition?}

The situation depicted above leads to the absence of a clear, unambiguous and practical definition of star clusters (and thus also of galaxies, see \citealt{Forbes2011b}). Even defining star cluster members as being gravitationally bound can be a dangerous shortcut. Stars with sufficient energy to potentially escape their cluster can be trapped for a significant amount of time and appear as cluster members (see \sect{pe}). In this review, I cowardly adopt flexible definitions by focusing on the most commonly agreed star clusters and mentioning, here and there, a few more debated cases.

%%%%%%%%%%%%%%%%%%%%%%%%%%%%%%%%%%%%%%%%%%%%%%%%%%%%%%%%%%%%%%%%%
\section{Cluster mass function}
\label{sec:cmf}

Probing the exact nature of the stellar systems formed in a diversity of galactic and cosmological environments is made difficult by limitations on both observational and numerical resolutions. For instance, depending on the physical conditions in their gas nurseries, star clusters could either form monotonically, or through the merger of sub-systems, parts of larger molecular complexes. The latter was suggested by \citet{Bonnell2003} and \citet{Fellhauer2005} as a formation channel of the most massive objects (see also \citealt{Fujii2015}, \citealt{Smilgys2017}, and an observational confirmation in 30 Doradus by \citealt{Sabbi2012}). Simulations resolving these sub-structures account for both formation channels being active simultaneously in different locations of interacting galaxies \citep[e.g.][]{Renaud2015}. The young age of the YMCs in the Antennae galaxies and comparable mergers imposes that the merging process should happen rapidly ($\lesssim 10 \Myr$). 

At the heart of these questions lies the initial cluster mass function (ICMF) and its evolution. The nature (shape and bounds) of the ICMF can be used as a hint on the formation mechanism of clusters, but an important step must be taken to infer this from observational data, since observations can only probe the present-day cluster mass function (CMF), i.e. an evolved version of the ICMF. Even young clusters experience physical mechanisms altering their masses. It is thus far from obvious to infer similarities and/or differences on the formation process(es) of e.g. old globular clusters and YMCs by simply comparing their observed present-day mass functions, as described in the rest of this review.

The present-day mass function of globular clusters in the Milky Way is described by a log-normal distribution, peaking at $\approx 2\times 10^5 \Msun$ (\citealt{Whitmore1999}, see also \citealt{Gnedin2014}, i.e. at the magnitude $M_V \approx -7.5$ in the corresponding cluster luminosity function, \citealt{Harris1991}). Uncertainties arise when converting the observed luminosity function into the mass function, i.e. when adopting a mass-to-light ratio derived from single-age and mono-abundance models. These models assume a stellar initial mass function and often a simple stellar population, i.e. coeval stars in each cluster, despite a growing body of evidence for abundance spreads and multiple stellar populations, and a still open debate on the single-age assumption \citep[see e.g.][]{Mackey2008b, Dib2013, Mucciarelli2014, Piatti2017}. Additionally, for lower mass systems, the stochastic presence of massive stars may cause systematic errors and biases in estimating the properties of the clusters, and care must be used \citep[see a discussion and method in][]{Fouesneau2010}. Other effects, like those originating from the sample size, are discussed in \citet{Adamo2015}.

\citet{Mandushev1991} showed that the mass-to-light ratio of clusters increases with mass. As a consequence, and as predicted by models and simulations \citep[e.g.][]{Ostriker1972, Baumgardt2003, Gieles2009} and detailed in Part~\ref{part:evolution}, it is very likely the result of several evolutionary processes most efficient at damaging the low-mass clusters. As a consequence, the low-mass end of the CMF is very different from that of the ICMF, while the high-mass end (made of more robust clusters) has only be mildly affected \citep{Boutloukos2003, Gieles2009}.

When considering young systems however (i.e. getting closer to birth epoch and thus to the ICMF), the mass function follows a Schechter function, i.e. a power-law with an exponential cut-off at the high mass end:
\begin{equation}
\label{eqn:schechter}
\frac{\dd N}{\dd M} \propto M^{-\beta} \exp{\left(-\frac{M}{M_c}\right)}.
\end{equation}
This is comparable to the shape of the mass function of galaxies, derived from the \citet{Press1974} formalism. This function is parametrised by the slope of the power-law $\beta \approx 2 \pm 0.3$ \citep{Zhang1999, Bik2003, Hunter2003, Lada2003, Kennicutt2012}, likely set by the organisation of the ISM regulated by turbulence \citep[e.g.][]{Fujii2015} and the efficiency of cluster formation, and the characteristic mass $M_c$ of the exponential cut-off of which physical origin is still debated (see \sect{mc}). However, due to low number statistics, the very presence of a truncation has not been unambiguously demonstrated (\citealt{Adamo2017}, see also several examples in \citealt{Portegies2010}, their figure 10, \citealt{Chandar2017}). Demonstrating the presence of a physical truncation at high mass will possibly allow to tell apart the formation mechanisms of clusters from those of dwarf galaxies.

The lower limit in this function (in nearby galaxies) goes down to $\sim 10^2 \Msun$, where the definition (and possibly formation mechanisms) of clusters conflicts with that of associations \citep[e.g.][]{Piskunov2006, deGrijs2006, Baumgardt2013, Fouesneau2014}. Such a functional form provides a good description of young clusters across a number of galaxies, including mergers \citep[see][]{Portegies2010}, with $M_c$ (or the maximum mass detected, in case of non-truncated mass functions) being of the order of $\sim 10^5 \Msun$ for spiral galaxies \citep[e.g.][]{Larsen2009, Chandar2010}, but significantly larger for starbursting galaxies \citep[$\gtrsim 10^6 \Msun$, e.g.][]{Bastian2008, Whitmore2010, Linden2017}.

Simulations showed that stellar systems as massive and extended as UCDs can form during galaxy interactions \citep{Renaud2015}. These could, for instance, match the properties of W3 ($8\e{7}\Msun$, $17 \pc$, \citealt{Fellhauer2005, Maraston2004}) detected in the merger NGC~7252. In this case, these objects merely lie at the end of the mass and size distributions of star clusters, as noted observationally (recall \fig{masssize}, and see also \citealt{Mieske2012}), and share formation mechanisms with young massive clusters. The galactic interactions would then represent the necessary trigger to reach the required extreme physical conditions, but other formation channels of UCDs have been proposed (see \sect{ucd}). Here again, the absence of a clear, physically motivated definition of star cluster introduces potential biases, in particular in the high-mass end of the CMF. The same question arises about the (dark matter free) tidal dwarf galaxies and lower mass clusters detected as beads on a string along tidal tails of interacting systems \citep[e.g.][]{deGrijs2003, Mullan2011, Knierman2003}. Among other scenarios, the formation of such objects would be favoured by the accumulation of gas near the tip of the tails, in which gravitational collapse would then proceed in a similar manner as for massive, ordinary clusters \citep{Duc2004}.

The situation is more complex in dwarf galaxies, where (keeping in mind low-number statistics effects) only a couple of massive clusters dominate the (I)CMF \citep[e.g.][]{Anders2004, Larsen2012, Pasquali2011}. This unexpected specific frequency (i.e. the number of massive cluster per galactic stellar mass, \citealt{Georgiev2010, Harris2013, Larsen2012, Larsen2014}) is yet to be explained, possibly through formation mechanisms leading to a non-Schechter, top-heavy ICMF in such environments.

Making a connection between the Schechter-like ICMF and the peaked present-day CMF requires a description of the mass-loss and dissolution processes experienced by clusters along their entire evolutions (i.e. possibly over $\sim 10\mh 12 \Gyr$), and the dependence of these mechanism on mass. \citet{Jordan2007} proposed an evolved version of the Schechter ICMF (\eqn{schechter}), replacing the initial mass $M$ with a $M +\Delta M$ term, but keeping the same functional form and parameters. This thus assumes that the mass-loss $\Delta M$ is independent of the mass, contrarily to predictions by simulations of a number of disruption mechanisms and series of their coupling (see Part~\ref{part:evolution}).

The mere exploration of the properties of the ICMF and its evolution reveals the need for an holistic description of the physics governing both the formation and the evolution of star clusters, and equality importantly, how these mechanism depends on the environment. Our current understanding of these questions is presented in the rest of this contribution.

%%%%%%%%%%%%%%%%%%%%%%%%%%%%%%%%%%%%%%%%%%%%%%%%%%%%%%%%%%%%%%%%%%%%%%%%%%%%%%%%%%%%%%%%%%%%%%%%%%%%%%%%%%%%%%%%%%%%%%%%%%%%%%%%%%%%%%%%%%%%%%%%%%%%%%%%%%%%%%%%%%%%%%%%%%%%%
%%%%%%%%%%%%%%%%%%%%%%%%%%%%%%%%%%%%%%%%%%%%%%%%%%%%%%%%%%%%%%%%%%%%%%%%%%%%%%%%%%%%%%%%%%%%%%%%%%%%%%%%%%%%%%%%%%%%%%%%%%%%%%%%%%%%%%%%%%%%%%%%%%%%%%%%%%%%%%%%%%%%%%%%%%%%%
%%%%%%%%%%%%%%%%%%%%%%%%%%%%%%%%%%%%%%%%%%%%%%%%%%%%%%%%%%%%%%%%%%%%%%%%%%%%%%%%%%%%%%%%%%%%%%%%%%%%%%%%%%%%%%%%%%%%%%%%%%%%%%%%%%%%%%%%%%%%%%%%%%%%%%%%%%%%%%%%%%%%%%%%%%%%%
\part{Formation}

The first effect of the cosmological and galactic environment on the history of star clusters is the assembly of their formation sites. When dense enough, gas ignites thermo-nuclear reactions that actually make stars radiate energy and produce chemical elements. In this part, I focus on this first phase of cluster evolution, from the stage of increasing the gas density, to that when gas and stars becomes decoupled.

One key question in this field is the formation of the globular clusters in the early Universe. However, the lack of direct observational constraints makes this a difficult topic. Our current approach is thus to study it indirectly, by exploring cluster formation in the local Universe, where is can be resolved, and extrapolating (and often speculating) to higher redshift.

%%%%%%%%%%%%%%%%%%%%%%%%%%%%%%%%%%%%%%%%%%%%%%%%%%%%%%%%%%%%%%%%%
\section{Resolved star formation, hints from local galaxies}

%%%%%%%%%%%%%%%%%%%%%%%%%%%%%%%%
\subsection{Molecular clouds, star formation and feedback}

Each of these topics easily deserves its own review. A lot of progresses have been made over the last years, thanks to increasingly powerful resources like the Hershel Space Telescope, ALMA, MUSE, NOEMA, and supercomputers capable of handling the large amount of data on these interconnected subjects and running always improving models. I present here only a tiny fraction of our current understanding of these aspects and their interplay.

The first step of star formation consists in gathering gas at high enough densities to start the collapse process that would eventually lead to the formation of pre-stellar cores, and then proto-stars. Leaving aside the questions of chemical composition and internal properties of clusters, I skip here considerations on the initial mass function, binary stars and many others, to focus on the assembly of the star forming sites themselves.

Observational surveys in the Milky Way show that the dense, cold gas (e.g. traced by CO) is organised in clumps with typical masses and radii of $\sim 10^6 \Msun$ and $\sim 10-30 \pc$ \citep{Solomon1987, Dame2001}. These giant molecular clouds (GMCs), embedded in envelops of atomic gas (with comparable mass, see e.g. \citealt{Blitz1990}), host dense enough gas to potentially form star clusters ($\gtrsim 100\mh 1000 \cc$). Measuring  observationally the inner properties of these clouds (and a fortiori the sites of individual star formation) is limited to the Milky Way and nearby galaxies (LMC, SMC, M~31, M~33, etc., see e.g. \citealt{Rosolowsky2007, Rosolowsky2007b}, \citealt{MivilleDeschenes2017} with estimates of cloud properties over the entire Milky Way disc, and \citealt{Hughes2013} in the more distant M~51), all representing relative quiescent environments, with low star formation rates ($\sim 1 \msunyr$). Comparable studies in more active galaxies, including mergers and starbursts like M~82, will only become possible in a few years.

All stars seem to form in clusters \citep{Lada2003}, and even though all these clusters are not necessarily bound and could be dynamically destroyed within a few Myr \citep{Bressert2010}, the important point here is that star nurseries host the formation of more than one star, in turbulent molecular clouds \citep{Hennebelle2012}. It must be noted that no star formation mechanism actually requires the gas to be molecular to form stars. Yet, after the cooling from free-free emission in charged plasma ($\gtrsim 10^6 \U{K}$, i.e. important to cool gas in galactic halos), the collisional recombination (i.e. the inverse of the ionization process) and the atomic de-excitation ($\sim 10^{4\mh 5} \U{K}$, \citealt{Baugh2006}), the scarcity of free particles in atomic gas to exchange energy with (and thus radiate energy away and cool down) makes atomic cooling rather inefficient below $\sim 10^4 \U{K}$.\footnote{Note however the non-negligible cooling effect provided by the fine-structure energy levels of e.g. carbon and oxygen.} Molecular gas (and dust) however yields more numerous and lower energy levels that allow efficient cooling below $10^4 \U{K}$ (e.g. the CO rotational line emission), thus reducing the internal pressure of clouds and allowing self-gravity to increase the density enough to trigger fragmentation and collapse into cores \citep[see e.g.][]{Glover2016}. 

Star formation then only proceeds in the densest cores \citep{McKee2007, Andre2014}, meaning that a large fraction of the gaseous surrounding medium is not converted into stars. This is quantified by the star formation efficiency (SFE), i.e. the mass ratio of stars formed and the initial gas mass. This ratio obviously depends on the scale over which it is measured but for typical molecular clouds ($\sim 10\mh 100 \pc$) up to galactic scales, values ranging from 1 to 30 percent are commonly accepted \citep{Krumholz2007, FaucherGiguere2013}.

Metallicity plays an important role in shaping the resulting star clusters, including the speeds of the stellar winds \citep[e.g.][]{Goldman2017} and the escape velocity for ejecta \citep{Georgiev2009b}, which possibly influences self-enrichment and the properties (or the existence) of multiple stellar populations.

The internal kinematics of the cloud may imprint rotation on the cluster, by conservation of the angular momentum during collapse \citep{Lee2016}, and/or through large scale tidal torques when massive clusters assemble hierarchically \citep{Mapelli2017}. Such a rotation has been detected in young \citep{HenaultBrunet2012}, intermediate \citep{Mackey2013} and old clusters \citep[e.g.][]{Davies2011, Bianchini2013, Lardo2015}. However, \citet{Vesperini2014} noted that rotation, specially in the outer parts, of old globulars could results from the interplay of internal dynamics (relaxation) and external effects (tides, see Part~\ref{part:evolution}), whereas \citet{Gavagnin2016} invoke mergers of clusters to induce rotation. Rotation of clusters, potentially set at birth, is thought to accelerate the pace of their evolution \citep{Einsel1999, Hong2013} and modify their velocity dispersion profiles, thus potentially biasing the interpretation of observational data \citep{Varri2012, Bianchini2013}.

Once formed, stars alter their surrounding through proto-stellar outflows, photo-ionisation, radiation, winds, the emission of cosmic rays, supernovae blasts and chemical enrichment (see, among many others, \citealt{Kessel-Deynet2003, Joung2006, Nakamura2007, Krumholz2012b, Dale2012, Offner2009, Grisdale2017} and a review in \citealt{Dale2015}). Because of the wealth of mechanisms and the non-linearity of their coupling \citep[e.g.][]{Hopkins2012, Agertz2015, Gavagnin2017}, the exact role of stellar feedback is not fully understood yet. We know it can include stopping nearby star formation, destroying clouds, ionising the ISM, injecting turbulence, launching galactic outflows and even quenching star formation activity at galactic scale \citep{Murray2010, Krumholz2014, Hopkins2014, Dale2015, Grisdale2017, Semenov2017}, but can also have a positive effect by compressing the gas (e.g. in shock fronts) and triggering sub-sequent star formation \citep{Koenig2012, Shima2017}. Furthermore, in dense environments (e.g. gas-rich galaxies, mergers), feedback might not be powerful enough to clear all the gaseous left-overs from the star forming sites \citep[Guillard et al., in preparation]{Rahner2017, Rahner2018, Howard2017}, possibly allowing for the retention and/or the re-accretion of gas and, the formation of several populations of stars \citep{Pflamm-Altenburg2009}. Such mechanism could, at least qualitatively, participate in the formation of multiple stellar populations detected in globular clusters (thus formed at high redshift in gas-rich, dense regions) and also in some young clusters \citep{Vinko2009, DeMarchi2011, Beccari2017}. The other types of feedback, active before the supernovae (i.e. before $10 \Myr$ on average), can however form hot, ionized cavities around the young stars (e.g. \hii regions), in which the supernova blasts expand more efficiently, thus increasing their net impact on the ISM \citep[e.g.][]{Hopkins2012}. Finally, the accretion onto black holes also represents a channel to remove the gas left-overs \citep{Krause2012, Leigh2013}.

Star forming regions which can be observationally resolved are restricted to the quiescent environment of the Milky Way and nearby galaxies where the gas density of clouds and their surroundings is low enough to allow feedback to efficiently remove gas from the star forming sites and preventing further accretion (but see e.g. \citealt{Rahner2017}). When the gas is removed, the gravitational potential of the region rises rapidly, such that the stars left alone, and which were in a relative dynamical equilibrium before, yield an excess of kinetic energy with respect to the new potential energy. Depending on the SFE (i.e. the relative importance of gas left-overs in the net gravitational potential), the re-adjustment of the stars to the new potential can lead to the dissolution of the cluster, a phenomenon known as the infant mortality \citep{Goodwin1997, Boily2003, Boily2003b, Smith2013b}. Even if the cluster is self-bound enough to survive, it likely loses a significant fraction of its mass \citep{Bastian2006c}, such that after the average timescale for supernova feedback ($\approx 10 \Myr$), only $\sim 10\%$ of all stars remain in bound stellar structures \citep{Lada2003}. Therefore, the SFE is supposedly higher in regions producing bound clusters \citep{Parmentier2009}.

In such a picture, both the geometry and the timescale of star formation (and feedback) are critical in setting the properties of the stellar objects formed, in term of binarity \citep{Goodwin2005, Raghavan2010, Kuruwita2017} and even the IMF \citep{Chabrier2014}, but also in term of the sub-structures encompassed in a given cloud. Once the first star or stellar aggregate formed, its feedback affects the surrounding media which can alter, quench or trigger the formation of other stars. The assembly of a cluster is thus highly dependent on these aspects \citep{Smilgys2017}. Yet, the structuring of the ISM in a hierarchical manner \citep{Elmegreen1996, Burkhart2013} being first set by large-scale mechanisms ($\gtrsim 100 \pc$, like flows, shear, tides and turbulence), the details of star and star cluster (or stellar association) formation are likely strongly dependent on the galactic environment \citep[e.g.][]{Rey-Raposo2017}.

%%%%%%%%%%%%%%%%%%%%%%%%%%%%%%%%
\subsection{The role of the environment}
\label{sec:environment}

In isolated galaxies, the exact role of the kpc-scale environment on star forming clouds and thus star cluster formation remains unclear. On the one hand, number of observations reported that the average properties and the scaling relations (e.g. mass-size, size-velocity dispersion, see \citealt{Larson1981, Solomon1987}) of molecular clouds are remarkably similar in a diversity of galaxies, ranging from dwarfs to spirals \citep{Rosolowsky2007, Bolatto2008, Heyer2009, Gratier2012, DonovanMeyer2013, Hughes2013, Faesi2016, Freeman2017}.

On the other hand, although technical limitations have restricted for long the theoretical studies of star formation to isolated, low mass clouds ($\lesssim 10^3 \msun$), modern simulations now emphasize the importance of galactic (hydro-)dynamics on the properties of clouds and thus on the process of star (cluster) formation \citep{Fujimoto2014, Smilgys2017}. For instance, \citet{Rey-Raposo2015} showed that clouds experiencing shear or compression would host star formation at different rates than equivalent clouds in isolation (with the same virial parameter). 
\begin{figure*}
\begin{center}
\includegraphics[width=\textwidth]{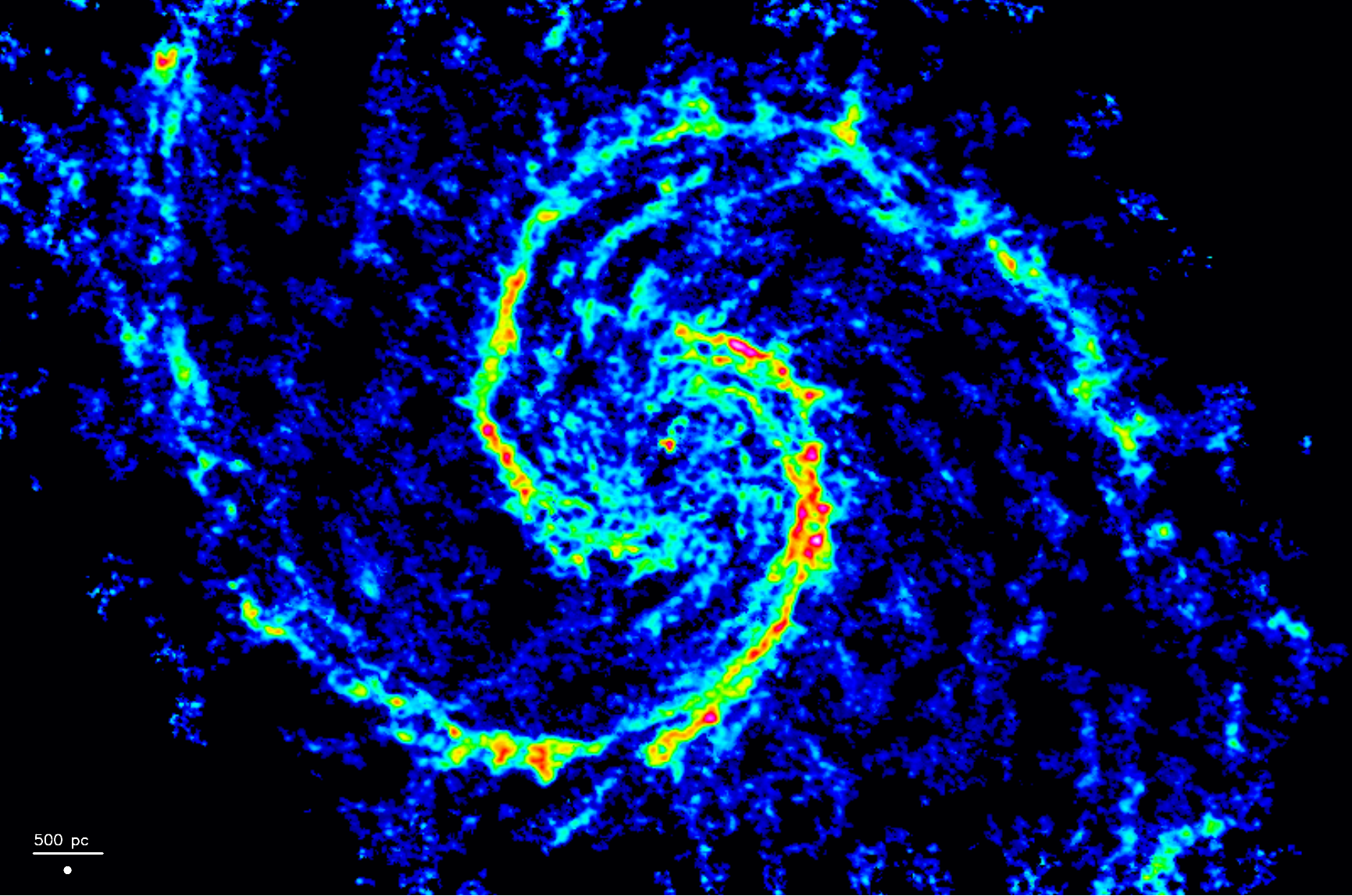}
\end{center}
\caption{CO(1-0) line emission across the disc of M~51, seen by the PAWS survey \citep[their figure 1, \copyright AAS, reproduced with permission]{Schinnerer2013}. The morphology of the dense gas clouds varies significantly in between and along the spiral arms. Note for example the elongated feathers at the top-middle zone, in contrast with the regularly spaced beads on a string in more central areas. Not only the morphologies of the GMCs vary, but also their densities and their positions with respect to the spirals, which likely translates into different dynamical evolutions.}
\label{fig:paws}
\end{figure*}

Observationally, \citet{MivilleDeschenes2017} measured a large range of masses and sizes of clouds (between 10 and $10^7\Msun$ and 0.5 and $200 \pc$) over the Milky Way disc. The outer Galaxy ($\gtrsim 15 \kpc$) also counts clouds significantly smaller and lighter than the rest of the Galaxy \citep[e.g.][]{Izumi2014, Sun2017}, indicating an influence of the galactic environment on the properties of clouds.

In this context, M~51 represents a prime target for observations, being seen face-on and close enough to resolve individual clouds, showing a grand-design spiral pattern and thus offering a strong contrast in environmental physical conditions \citep[see \fig{paws},][]{Koda2009}. Furthermore, it hosts a large number of clusters \citep[e.g.][]{Bastian2005, Scheepmaker2009, Chandar2011}. \citet[see also \citealt{Meidt2013}]{Schinnerer2013} particularly highlighted the role the galactic potential and gas streaming motions in shaping the molecular clouds of M~31. The effect of the spiral arms and the differences with the inter-arm volumes is discussed in the next section.

Simulations of isolated galaxies further emphasized the role of the kpc-scale effects, through bar and spiral patterns and the global dynamics of the galaxy \citep[e.g.][]{Dobbs2012, Benincasa2013, Renaud2013b, Fujimoto2014, Tasker2015, Ward2016, Nguyen2017}. In addition to affecting the large scales flows of gas, the gathering of cloud material \citep[e.g.][]{Dobbs2011} or their destruction by shear \citep[e.g.][but see also \citealt{Dib2012}]{Weidner2010, Emsellem2015}, galactic dynamics also influence the collision rate between clouds \citep[see e.g.][and \sect{ccc}]{Renaud2015d, Fujimoto2016}.

When ignoring the environment, the break point in the balance between internal pressure and self-gravitation can be expressed by considering a density perturbation propagating in a homogeneous, isolated medium. This leads to the \citet{Jeans1902} formalism, providing the maximum scale-length and mass stable against collapse. Such a formalism neglects the large scale galactic effects, in particular, shear, tides and the external pressure which significantly modify the stability criteria \citep{Dobbs2011, Field2011}. These points seem to (at least qualitatively) explain the observed spreads and deviations from the classical scaling relations of molecular clouds in regions where environmental effects play major roles \citep[e.g.][]{Leroy2015, Grisdale2018}. To reconcile this diversity of clouds with our search of a universal star formation law, \citet{Meidt2013} suggested that accounting for the dynamical pressure on star forming regions would unify the otherwise diverse star forming relations in different environments.

Although we still ignore how and to what extend this diversity of star forming regions translates into a diversity of star clusters, it is worth noticing that \citet[see also \citealt{Konstantopoulos2013}]{Bastian2012} detected variations a factor $\sim 3$ of the truncation mass of the ICMF (recall \sect{cmf}) with galactocentric radius within a given galaxy (M~83 and NGC~4041). Variations from galaxies to galaxies can be even stronger and are discussed in \sect{mc}.

%%%%%%%%%%%%
\subsection{Arms and inter-arm regions}
\label{sec:spiral}

Observations resolving the inner structures of galaxies (spiral arms, bars) reveal the importance of the environment in setting the star formation activity. For instance, \citet{Schinnerer2013} showed the connection between CO emission tracing dense molecular gas and the gravitational features of M~51. By concentrating gas, spirals arms yield an excess of molecular cloud formation with respect to the surroundings, as commonly observed \citep{Heyer1998, Hou2014} and modelled \citep{Dobbs2006}. This however does not imply that either the efficiency of cloud formation and that of star (cluster) are increased in arms. \citet{Foyle2010} reported indeed a very weak dependence of the ratio between molecular and atomic hydrogen when comparing arm and inter-arm regions in spiral galaxies, concluding that the efficiency of cloud formation is independent of spiral structures \citep[see also][]{Eden2012, Eden2013}. \citet{Moore2012} estimated that $70 \%$ of the increase of the star formation rate (SFR) density is arms is a mere reflection of the higher number density of star forming regions found there, and that only the remaining $30 \%$ result from a more efficient formation \citep[see also][]{Schinnerer2017}.

From this, it follows that spirals are more important at re-organising the ISM and gathering clouds than actually forming them and triggering star formation. Yet, results from simulations emphasize that the gathering of gas in spirals tends to favour the formation of GMCs \citep{Dobbs2011}, while the transition from the compressive, pressurized and deep potential well of spiral to the inter-arm medium leads to an enhancement of the destruction of clouds \citep{Dobbs2006, Roman-Duval2010}. \citet{Koda2009} noted that the giant molecular associations assembled in spirals would then split into smaller clouds when leaving the arm (see also \citealt{Hirota2011} on the mass of star forming clouds downstream the arms, and \citealt{Schinnerer2017} for a cautionary note on not resolving the inner structures of giant molecular associations). \citet{Meidt2013} and \citet{Colombo2014} showed that shear and shocks in the arms would yield a stabilizing effect on the gas structures, preventing star formation.

The velocity pattern of spiral structures also triggers the development of instabilities that give rise to a diversity of features, like beads on a string, feathers and spurs. Not only the spiral structure but also its kinematics set the relative roles of gravity, shear and instabilities. For instance, in spirals with low pitch angles, the steep velocity gradient across the arm due to the differential rotation of the galactic disc could favour Kelvin-Helmholtz instabilities, in the form of spurs (see \fig{paws}, \citealt{Wada2004, Kim2006, Shetty2006, Dobbs2006, Renaud2013b}, but also \citealt{Kim2014, Sormani2015} for different interpretations). Spurs (and feathers when more elongated and populating the inter-arm regions) are offset with respect to the bulk of the spiral and thus outside of the local minimum of the gravitational potential of the spiral. This possibly protect them against dissolution as they do not experience rapid changes of their external pressure and gravity, contrarily to clouds formed in the spiral and then leaving it. Conversely, beads on a string are found along the spiral \citep{Elmegreen1983, Foyle2013}, and could form there in the relative absence of Kelvin-Helmholtz instabilities, i.e. in arms with a higher pitch angle \citep{Renaud2013b}. A given galaxy (observed, \citealt{Schinnerer2013} or simulated, \citealt{Renaud2013b}) can simultaneously host both types of clouds. The effects of this diversity on the formation (and early survival) of clusters remain to be quantified, but it is likely that the variation of external pressure, gravitational energy, tidal field and intrinsic rotation (as induced in spurs) would play a role in shaping the resulting stellar objects.

%%%%%%%%%%%%
\subsection{Cloud-cloud collisions, tips of the bar}
\label{sec:ccc}

Convergent flows in some environments like spiral arms and the tips of bars favour the accumulation of gas structures (e.g. through orbital crowding, see \citealt{Kenney1991}), and thus cloud-cloud collisions. In such events, shocks between the two components at different velocities boosts the local density, in addition to the natural gathering of gas mass. The resulting effect is a clear increase of the Mach number at cloud scale, possibly converting previously sub- or transonic clouds into a supersonic medium \citep{Renaud2015d}. The collision thus triggers star formation \citep{Loren1976, Tan2000, Tasker2009, Inoue2013}, possibly in the form of massive stars \citep{Anathpindika2010, Motte2014, Takahira2014}. \citet{Furukawa2009} and \citet{Fukui2014} further reported that YMC form in the volumes of interaction during cloud-cloud collisions, probably because of the increase of external pressure \citep{Elmegreen2008}. One example of this can be found at the near tip of the Milky Way bar: the molecular complex W43 is thought to be the remnant of a recent ($\approx 20\mh 30 \Myr$) cloud merger and hosts large quantities of gas ($\sim 140 \pc$, $\sim 10^7 \Msun$) leading to an active episode of star formation \citep{NguyenLuong2017}. Due to the complex interactions between gas structures, such giant molecular associations yield a number of sub-structures, and therefore star cluster formation does not proceed in a bulk, monolithic manner, but rather hierarchically by merging sub-clusters. This mode of assembly has been proposed to explain the formation of massive stellar systems, including UCDs in other contexts \citep{Fellhauer2005, Banerjee2015}. As a zone of convergence, tips of the bar(s) would thus be a prime location for the formation of massive stellar systems in isolated galaxies.

In this picture, the gas circulates along the bar on $x_1$ orbits (and is also fuelled inwards along the spiral arms connected at the extremities of the bar). Thus, the crowding and collisions happening at the tips of the bar and leading to cluster formation implies that the newly formed cluster roughly follows a comparable $x_1$ orbit, and thus should be preferentially found on the leading sides of the bar \citep{Renaud2015d}. 

The relatively high degree of symmetry of the bar implies that comparable physical conditions should exist at the other extremity. The presence of a YMC candidate has indeed been reported at the far tip of the Milky Way bar by \citet{Davies2012}, further confirming the role of the kpc-scale (hydro-)dynamics in triggering the formation of massive clusters.

%%%%%%%%%%%%
\subsection{Galactic centre}

The innermost $\sim 100 \pc$ of the Milky Way host a significant fraction of the total molecular gas content of the Galaxy ($\sim 10\%$, see e.g. \citealt{Ferriere2007, Molinari2011}). Classical stability criteria predict that gas at such a high density should form a significant amount of stars. However, observations report a very weak star formation activity despite the large quantity of (theoretically) dense enough molecular gas \citep[e.g.][]{Oka2005, Ginsburg2015}. Identifying young stellar objects in this region of complex extinction requires spectroscopic confirmations, such that the exact SFR remains debated. Yet, in the central molecular zone (i.e. the inner $\sim 300 \pc$ of the Milky Way containing about $10^{7\mh 8} \Msun$ of molecular gas), \citet[see also \citealt{Longmore2013}]{Morris1989} estimated the SFRs of clouds to be about one order of magnitude lower than what expected from the high gas surface density and the classical star formation laws established empirically in other environments \citep[e.g.][]{YusefZadeh2009, Lada2012, Krumholz2012, Kauffmann2017, Barnes2017}. \citet{Kruijssen2014} thus interpreted this as a higher critical density for the onset of star formation in the central molecular zone than in other regions of the galaxy. \citet{Emsellem2015} showed that the physical origin of this peculiarity is set by the galactic rotation curve being steep in this region, inducing strong differential rotation (i.e. shear) on extended structures like clouds. This shear would therefore overcome self-gravity and smooth-out overdensities, preventing gas structures to collapse and form stars, despite their high densities. The tidal field also plays a destructive role in the central molecular zone, but likely of lower amplitude than the shear.

The very centre of the Galaxy offers both the deep potential well and the symmetry to shield clouds and clusters against destruction (with however the notable perturbation from the super massive black hole, \citealt{Zhong2014, ArcaSedda2017}). Star formation could thus happen in the very centre and participate in building the nuclear cluster (see \sect{nc}), but the above-mentioned destructive effects would take over already within offsets of only a few $\sim 10\mh 100 \pc$ from this spot. In this context, it is somewhat surprising to find young massive clusters in the central molecular zone, like the Arches and the Quintuplet. For instance, Arches ($2\e{4}\Msun$, \citealt{Clarkson2012}) hosts very young stars ($\approx 2.5 \Myr$, \citealt{Figer2002, Najarro2004}) which thus indicates an in situ formation, and rules out formation in a more favourable environment, followed by migration. Arches hosts a top-heavy IMF \citep{Espinoza2009, Habibi2013}, which calls for specific, yet unknown, physical conditions for star formation. The origins of these peculiarities and the reason for the very existence of these clusters in such a hostile environment remain unknown, and possibly linked.

In a comparable fashion to the orbital crowding seen at the tips of the bar (i.e. at the apocentres of $x_1$ orbits), it is possible that the intersection of the $x_1$ and $x_2$ orbits (i.e. the elongated orbits along the major and minor axes of the bar, respectively) could favour the accumulation of gas (possibly via cloud-cloud collisions) at the apocentres of the $x_2$ orbit, outside the central molecular zone \citep{Stolte2008}. Clouds could in principle form there, collapse, form massive clusters that would then follow their $x_2$ orbit bringing them back toward the galactic centre. Yet, the measured velocity of Arches seems incompatible with this scenario \citep{Stolte2008}.

The fact that the Milky Way is not currently hosting more than a handful of massive clouds and YMCs is a strong indication that the physical conditions leading to the formation of even more massive clusters ($\gtrsim 10^{4\mh 5} \msun$) must be drastically different than those found in our Galaxy. Clues can be obtained by exploring more extreme, but less resolved, environments.

%%%%%%%%%%%%%%%%%%%%%%%%%%%%%%%%%%%%%%%%%%%%%%%%%%%%%%%%%%%%%%%%%
\section{Young massive clusters in interacting galaxies}
\label{sec:ymc}

The old age of the Milky Way globular clusters has long suggested that massive stellar systems could only form in the early Universe, and that the required physical conditions are not matched anymore. Yet, some galaxies in the local Universe, including the Magellanic Clouds \citep[e.g.][]{Glatt2008}, host young ($\lesssim 0.1\mh 1 \Gyr$) and massive ($\gtrsim 10^{4\mh 5} \Msun$) clusters (YMCs). Interacting galaxies and mergers are particularly good candidates to find such objects, because of their extreme physical conditions being supposedly closer to those at high redshift than those in quiescent discs. Predicted by \citet{Schweizer1987} and \citet{Ashman1992}, starbursting mergers of gas-rich galaxies do trigger the formation of (de facto, young) massive clusters. This has been confirmed in a number of interacting systems \citep[e.g.][]{vandenBergh1971, Arp1985, Lutz1991, Holtzman1992, Whitmore1993, Holtzman1996, Zepf1999, Whitmore2003, Larsen2006, Trancho2012, Roche2015}. Despite many similarities between globular clusters and YMCs (see e.g. \citealt{Brodie2006} and \sect{gc}), it is yet not clear whether YMCs are what globular clusters were $\sim 10 \Gyr$ ago, or whether present-day YMCs observed in $10 \Gyr$ from now would actually resemble present-day globulars (see \sect{ymcglobs}). In any case, YMCs and in particular those in interacting galaxies represent a prime laboratory to explore the physics of massive cluster formation, which is closely related to that of the bursts of star formation.

\citet{Elmegreen2008} proposed that star forming clouds in mergers experience an excess of external pressure, which would keep confined otherwise unbound gas structures \citep[see also][]{Kruijssen2012, Maji2017}. Such a pressure originates from a number of causes, including convergent flows (including cloud-cloud collisions) and tidal compression, as discussed below. 

Note that mergers also play the important role of re-distributing the material of their progenitors. If this can lead to star formation quenching \citep{Martig2009}, it can also eject pre-existing clusters into the halo, i.e. in a low-density, weak tidal environment where the disruption mechanisms are less efficient than in the much denser (and likely more central) regions of their formation. This would then considerably increase the survival rate of clusters that would otherwise get dissolved by shocks and tides \citep[see also Part~\ref{part:evolution}]{Kravtsov2005}.

In interacting galaxies, starbursts are not restricted to the merger phase, i.e. the coalescence of the progenitors, but can take place earlier in the interaction \citep[e.g.][]{Gao2003}. The enhancement of the star formation activity (and thus likely that of YMC formation) is even detected in pairs of galaxies with a large separation \citep[$\sim 10\mh 100 \kpc$,][]{Ellison2008, Scudder2012, Patton2013}. At the largest distances, this probably results from a triggering during a closer passage before the galaxy actually separate. 

Resolving the location(s) of star formation in interacting systems is usually limited to dedicated observational programs focussing on nearby pairs, potentially harbouring a large fraction of the diversity of the physical conditions leading to YMC formation \citep[see][and references therein]{Moreno2015}, but a number of large surveys also address the question in a more systematic way (e.g. S4G, \citealt{Sheth2010}, SAMI, \citealt{Croom2012}, CALIFA, \citealt{Sanchez2014} and MaNGA, \citealt{Bundy2015}).

It happens that the closest ongoing major merger to us encompasses a large variety of these environments: the Antennae galaxies (NGC~4038/39), where the presence of young massive clusters have been reported in several areas of the system \citep{Whitmore1995, Mengel2008, Bastian2009, Herrera2011}. This system is found to be representative for clusters found in larger sample of luminous infrared galaxies \citep{Linden2017}.

%%%%%%%%%%%%
\subsection{Nuclear starburst}
\label{sec:inflows}

The formation activity in interacting galaxies, especially at a late stage of the merger is mostly found in the central region of the remnant \citep{Schweizer1982, Soifer1984}. The reason for this is gas inflows from the strong gravitational torques one galaxy induces on the other \citep{Keel1985, Hernquist1989c, Barnes1991, Barnes1996}. Inside the co-rotation radius (i.e. where the rotation velocity of one galaxy equals the orbital velocity of its companion), the torques on the disc material of a galaxy from its companion are opposed to the rotation motion, which thus induce inflows of material toward the galactic centre \citep{Bournaud2010, Maji2017}. The gas density rapidly increases, leading to the formation of a star cluster (see \sect{nc}). Such nuclear inflows also fuel the central super-massive black hole \citep{Cox2006} and could trigger a violent active galaxy nucleus (AGN) activity, possibly populating the radio-loud class of AGNs \citep{Chiaberge2015}. Then, feedback from the AGN regulates gas flows in the central region and slows down the growth of the black hole itself, the host galaxy and the central cluster \citep{DiMatteo2005, Hopkins2010}.

The nuclear activity being triggered by gravitational torques, it becomes important mainly during close encounters, in prograde discs (i.e. where the disc angular momentum is roughly aligned with the orbital angular momentum of the companion galaxy, see \citealt{Duc2013}). During the interaction, the orbital pericentre distance of the galaxy pair decreases with time, due to dynamical friction, which implies that nuclear bursts are predominantly triggered at the last passage before coalescence. This enhanced activity can however be maintained for $\sim 10\mh 100 \Myr$ after coalescence \citep[e.g.][]{DiMatteo2007}.

Note also that fly-by interactions can trigger the formation of galactic bars \citep{Noguchi1988} which, by creating their own torques, could play an important role in nuclear inflows. Therefore, non-antisymmetric structures formed by a passage could have a comparable effect as a close encounter on star cluster formation in the nuclear region.

%%%%%%%%%%%%
\subsection{Cloud-cloud collisions, shocks}
\label{sec:shock_sf}

The enhancement of star (cluster) formation is not confined to the galactic centres. Off-nuclear activity has been reported, also in the form of YMCs \citep{Wang2004, Hancock2009, Chien2010, Smith2014, Elmegreen2016}, especially before the coalescence stage. \citet{Jog1992} suggested that, during penetrating interactions, the shock between marginally stable clouds from one galaxy and the extended gas reservoir from the other, or even cloud-cloud collisions (recall \sect{environment}) would trigger the necessary instabilities leading to the collapse of the clouds and the formation of star clusters. This is notably visible in the overlap region of the Antennae, where the two galactic discs get imbricated. Number of YMCs have been detected in this region, and their age ($\approx 3\mh 10 \Myr$, e.g. \citealt{Bastian2009}) further confirming the role of the recent galactic collision in triggering their formation \citep[see also][]{Herrera2011, Herrera2012}. In such a region, the shocks increase the pressure on star forming clouds by several orders of magnitude, favouring the formation of massive star clusters \citep{Elmegreen1997, Ashman2001, Maji2017}. However, overlap regions usually are rather obscured by nature, and observational data (particularly kinematics) on the young clusters and the surrounding ISM are difficult to interpret. For instance, it is not clear whether these objects keep signatures of their distinct progenitor clouds (in kinematics and/or chemical content) or not. Shocks and collisions thus play an important in penetrating interactions, which preferentially occurs at the beginning of the final coalescence, and during fly-bys with small pericentre distances\footnote{Note however that, in order to allow the galaxies to separate after a fly-by, dynamical friction must be weak with respect to their relative pre-interaction velocities. Small pericentre passages complying with this requirement thus imply high velocities, which translates into a short interaction phase and a short-lived star formation trigger.}.

%%%%%%%%%%%%
\subsection{Tidal and turbulent compression}
\label{sec:compression}

Finally, enhanced and off-nuclear star cluster formation is also detected in other regions of interacting galaxies, usually before the coalescence. The spatial distribution of such regions is rather complex, but some of them are usually found opposite side of the galaxy with respect to the overlaps. This is the case for instance of the so-called Northern arc of the Antennae \citep[e.g.][]{Bastian2009}, or the so-called ``Feature \emph{i}'' zone of NGC~2207 \citep{Elmegreen2000, Kaufman2012, Mineo2014, Elmegreen2016, Elmegreen2017}. These regions host dense and massive molecular clouds and complexes, high SFRs, and YMCs comparable to those found in overlaps. The absence of inflows (\sect{inflows}) and shocks (\sect{shock_sf}) in such regions calls for another triggering mechanism for star formation.

The symmetry of the starbursting regions with respect to the galactic centre and along the axis connecting the two galaxies originates from the differential aspect of the tidal forces (\citealt{Duc2013}, the same way lunar tides on Earth have the same effect on the side facing the Moon and the opposite one). The combination of the gravitational potentials of the two galaxies (including the contribution of the dark halos, i.e. even without overlap of visible matter\footnote{The absence of stabursting regions between the nuclei and the zones of compressive tides, e.g. between the nucleus of NGC~4038 and the Northern-arc in the Antennae, rules out shocks with diffuse \hi reservoir that would extend the effect of shocks outside the overlap of the dense discs. Only a balance between the (usually) destructive tides from the host galaxy, and the contribution of the other can form the observed star forming regions and their peculiar morphologies.}) induces the formation of cores in the net potential \citep{Renaud2008}. In cores, i.e. the convex regions of the potential, the tidal forces are compressive in all directions \citep{Valluri1993, Dekel2003, Renaud2008}. This situation is found in all interacting systems, with an intensity and duration that depend on the galaxies and their orbits \citep{Renaud2009}. These fully compressive tides then induce an enhancement of the total turbulence of the ISM (as observed, \citealt{Irwin1994, Elmegreen1995, Bournaud2011, Ueda2012}), but also a change of its nature by making it compression-dominated \citep{Renaud2014}.

This forcing drives turbulent motions out of the classical equipartition found (on average) in isolated galaxies \citep[][see \citealt{Hennebelle2012} for a review]{Kritsuk2007, Federrath2010}. The compressive mode (curl-free) of turbulence takes over the classical mixing, solenoidal one (divergence-free), and generates denser and more numerous gas over-densities than usual. Once it has cascaded down to parsec scale, this effect enhances the SFR \citep{Renaud2014}, in particular in the form of massive clusters \citep{Renaud2015, Li2017}. Note that tidal and turbulent compressions also occur in the nuclei, the overlap regions and, to a lower extend, along the tidal tails where TDG-like objects form \citep{Ploeckinger2015}, and thus participate, to various degrees, to the starburst activity in all concerned regions.

Classical stability criteria like the Toomre $Q$ parameter (\citealt{Toomre1964}, used in the context of axisymmetric, mono-component discs, and \citealt{Lin1966}, \citealt{Bertin1988}, \citealt{Elmegreen1995c}, \citealt{Rafikov2001}, and \citealt{Romeo2013} for multi-component discs) and the Jeans formalism (\citealt{Jeans1902}, for propagation of a perturbation in a homogeneous medium) have been adjusted to account for the additional effect of tides \citep{Jog2013,Jog2014}, confirming that compressive tides favour the instabilities of gas structures \citep[see also][in a different context]{Inoue2016}. It is however challenging to propose an observational diagnostic to identify compressive tides and turbulence, such that this theory has not yet been tested observationally. However, no alternative exists to explain extended, off-nuclear, off-overlap starbursts.

%%%%%%%%%%%%%%%%%%%%%%%%%%%%%%%%%%%%%%%%%%%%%%%%%%%%%%%%%%%%%%%%%
\section{In gas-rich, clumpy galaxies}
\label{sec:clumpy}

Along their evolution, at least some disc galaxies experience a phase during which their morphologies are dominated by giant clumps \citep{Cowie1996, Elmegreen2007, Genzel2008, Tacconi2013}. Typically at $z = 1 \mh 4$, once the disc is formed, a galaxy with a large gas fraction ($\gtrsim 30 \%$ of the baryonic mass) can host violent disc instabilities which, because of a high velocity dispersion compared to the disc circular velocity, favour Jeans instabilities in a handful of kpc-scale gas clumps \citep{Agertz2009, Dekel2009b, Bournaud2009, forster2009}. These clumps are typically made of $10^{8\mh 9}\msun$ of gas and stars \citep{Elmegreen2005} and are dense enough to host star formation \citep{Zanella2015}, possibly in the form of massive clusters. Note that the same process also occurs in local dwarf irregulars and blue compact dwarfs which also yield a small ratio of circular velocity to dispersion velocity \citep{Elmegreen2015}. Such galaxies thus host a higher fraction of their star formation in the form of massive, bound clusters.

Observations suggest that such clumps could encompass several sub-clumps (with stellar masses of the order of $10^{6\mh 8} \msun$, \citealt{Dessauges-Zavadsky2017}), although it remains difficult to establish whether such small clumps share a common envelop in a giant clump or not. Details on the cooling, fragmentation and galactic tides set whether these small clumps are gravitationally bound to each other or rather independent structures. In the former case, each clump could form a relatively small star cluster that would contribute to a much larger object by merging with its neighbours, following the hierarchical scenario of \citet{Bonnell2003}. In this process, because all sub-clumps might not be dense enough to form stars (or because all the sub-structures from the same giant clump might not form their stars simultaneously), young clusters could interact with nearby dense gas structures. Such impulsive tidal perturbations could be sufficient to destroy the least massive clusters and damage the others (see \sect{shock}), as proposed by \citet{Elmegreen2010}. This topic is sensitive to the details of the cooling, fragmentation and turbulence, and is still open.

Because of their high gas densities, the giant clumps manage to survive the effects of stellar feedback \citep{Bournaud2015}\footnote{The opposite conclusion has been reached by \citet{Oklopcic2017}. However, their galaxies, due to a too early star formation (as it is often the case in cosmological simulations), yield a gas fraction lower than expected at the redshift examined ($\approx 20\mh 30\%$ instead of $\approx 50\mh 60\%$ of the baryonic mass at $z\sim 2$, see \citealt{Daddi2010, Tacconi2010}). Therefore, the gas overdensities they study are not massive clumps resulting from violent disc instabilities, but rather portions of spiral arms more susceptible to destruction by shear and feedback. The differences between these two types of structures and their survivability depend on their nature and has been proven to be (in this specific case) independent of numerical methods (see Bournaud et al., in preparation, who confirms that the low gas fraction ``clumps'' of \citealt{Oklopcic2017} are indeed short-lived, while giant clumps in higher gas fraction galaxies survive the effects of feedback).} and can, through clump-clump interactions and dynamical friction, merge in the galactic centre and participate in the assembly of the bulge. Therefore, star clusters formed in the massive clumps would not be detected as such at present-day, but instead would be dissolved (by tides) in the bulge.

For such star clusters to survive over a long time-scale, one must invoke a modification of their orbit \citep{Elmegreen2010}. Galaxy interactions provide such an effect and could account for several aspects. First, by scattering the stellar material above and below the original disc, the distribution of star clusters would resemble the flatten spheroid, observed in the Milky Way \citep[e.g.][]{Portegies2010}. Second, by ejecting the clusters off the disc into a low density environment, it would significantly reduce the tidal harassment they experience. However, while long-term, secular tidal effects would be largely decreased (see \sect{secular}), inclined orbits would necessarily lead to disc-crossings and the associated tidal shocks that could severely damage the clusters, and even destroy the least massive ones (see \sect{shock}).

Clumpy galaxies could then play an important role in forming and altering the star cluster populations of disc galaxies, but their exact effects remain to be established.

%%%%%%%%%%%%%%%%%%%%%%%%%%%%%%%%%%%%%%%%%%%%%%%%%%%%%%%%%%%%%%%%%
\section{Globular clusters: formation at high redshift}
\label{sec:gc}

%%%%%%%%%%%%%%%%%%%%%%%%%%%%%%%%%%
\subsection{Uncertainties on the ages, and bi-modalities}

Precise age determination is made difficult by the age-metallicity degeneracy \citep{Worthey1994}, but the bulk of the Milky Way globular clusters is estimated to have formed around $z \approx 3\mh 6$, i.e. $\approx 11.5\mh 12.5 \Gyr$ ago \citep[see e.g.][]{Leaman2013, Forbes2015}. In addition to putting a lower limit on the age of the Universe \citep{Krauss2003} and the age of the host galaxies if they form in situ \citep{Strader2005}, this point could be important to assess the link between globular clusters and the reionization of the Universe, which is actually a twofold question. On the one hand, could globular clusters have played a role in reionization? And on the other hand, how has reionization affected the formation of globulars? On top of the uncertainties on the intensity of the processes involved themselves (see \sect{reionization}), the first topic to address remains the actual concomitance of the formation of globulars (i.e. the presence of massive stars emitting a strong ionisation flux) and the reionization.

Determining the absolute ages of clusters reveals itself to be more complicated than estimating relative differences between cluster populations, yet done with significant uncertainties of $\sim 1\mh 2 \Gyr$ (corresponding to variations in redshift between roughly $z\approx 4$ and 10 at this epoch, \citealt{Strader2005, Forbes2015}). Yet, the features detected in age-metallicity distributions of globular clusters can be used to constrain their formation scenarios, and the number of formation channels involved, as illustrated in the following Sections \citep[e.g.][]{Muratov2010, Renaud2017}.

On top of the ages and extreme stellar densities, the main observational constrain on the formation of globular clusters is the bi-modal colour distribution observed in some (but not all) massive galaxies \citep{Zinn1985, Gebhardt1999,Larsen2001, Peng2006}, which is often translated into a metallicity bi-modality \citep[but see][on the non-linearity of the colour-metallicity relations]{Yoon2006}.

In the Milky Way, the blue clusters (metal poor, with a distribution of [Fe/H] peaking at -1.5, and with no cluster below [Fe/H] $=-2.5$, \citealt{Harris1996}) are preferentially found in the halo. They do not yield any structured kinematics as a population, indicating that they unlikely share a common origin. Conversely, red clusters (metal rich, [Fe/H] peaking at -0.5) are associated with the Galactic disc with which they share the overall rotation motion. The bi-modality aspect of their colour and metallicity distributions (i.e. with a deficit of clusters between two peaks) suggests two distinct modes of formation. \citet{Forbes2015} estimated that the metal-rich clusters are, on average, about $1 \Gyr$ younger than their blue counterparts. Despite uncertainties, the bi-modality is in place before $z \approx 2$, possibly already at $z \approx 3\mh 4$, i.e. $11\mh 12 \Gyr$ ago \citep{Dotter2011, Leaman2013}.

In our neighbour Andromeda, the colour distribution is rather uni-modal, yet spanning a comparable range \citep{Caldwell2016}. Therefore, the formation of globular clusters is strongly dependent on galaxy evolution, such that galaxies with different assembly histories would yields different distributions of globulars. The solution of this puzzle is thus lying in galaxy evolution. Several scenarios have been proposed.

%%%%%%%%%%%%%%
\subsection{Formation in dark matter halos}
\label{sec:formhalo}

In their scenario, \citet{Peebles1968} evoke the formation of globular clusters at very high redshift in low metallicity medium, even before the formation of galaxies themselves. When the Universe has expanded enough such that primordial (metal-free) gas in overdensities reaches $\sim 4000 \U{K}$ and $\sim 10^4 \cc$, the Jeans length and mass setting the lower limit for instabilities are of the order of the typical sizes and masses of globular clusters \citep[see also][]{Fall1985, Bromm2002, Katz2014, Kimm2016, Popa2016, Hirano2017, Penarrubia2017, Kim2018}.

Variants of this idea have proposed to consider the triggering of star formation in otherwise stable halos by ionisation shock fronts \citep{Cen2001}, or by the merger of two or more small gas-rich halos \citep{Trenti2015}. Additionally, \citet{Kimm2016} found that enrichment from the first generation of supernovae could lead to the subsequent formation of other clusters in the same mini halo, and that these structures would eventually merge in the centre, thus leading to a wide spread in metallicity within the resulting globular.

The gathering of gas and its collapse would happen preferentially in potential wells \citep{Diemand2005, Boley2009}, likely containing dark matter (see \citealt{Ricotti2016}, but also \citealt{Naoz2014}). Thus, the absence of dark matter in globular clusters (\sect{dm}, \citealt{Conroy2011}) requires to invoke internal dynamics \citep{Baumgardt2008} and tides to strip at least part of this component \citep{Mashchenko2005}. Determining precisely the kinematics of the outer part of globulars, and inferring the presence of a dark component is thus key to validate this idea.

Such scenarios do not account for the observed bi-modalities, and implies a rather early formation channel with respect to the (however uncertain) age estimates. Therefore, this calls for alternative, or at least complementary, formation scenarios.

%%%%%%%%%%%%%%
\subsection{Formation in wet mergers}

\citet{Schweizer1987} and \citet{Ashman1992} proposed a formation channel in two steps, thus accounting for an age difference between the blue and red populations. The blue clusters would form in the early, metal poor, Universe as before. These clusters (and other stars) would then chemically enrich their still gas-rich host galaxies, that eventually merge. Note that mergers are naturally more frequent at high redshift than in the present-day Universe, because of its density going as $(1+z)^3$. During the merger of gas-rich galaxies, a starburst activity is triggered (recall \sect{ymc}), and by analogy with observed local mergers like the Antennae, massive clusters would thus form from the enriched material, and populate the red group. It is however important to keep in mind that the efficiency of mergers to trigger starbursts (and thus, likely, the formation of globulars) decreases with redshift, as observed \citep{Lofthouse2017} and modelled \citep{Perret2014, Fensch2017}.

With such a scenario, the relative amplitude of the blue and red peaks in the colour distribution, and their potential separation by a deficit of cluster at intermediate colours can easily be regulated with the merger epochs and their cluster formation efficiency. Yet, because there is a priori no particular reason for the merger activity to pause at some epoch, one would expect that most galaxies yield a rather continuous, uni-modal distribution, making the Milky Way an outlier. The efficiency of cluster formation however could be affected by reionization making the merger history a highly inadequate tracer of the cluster formation history.  \citet{McLaughlin1994} noted that the number of globular clusters produced in such events would however be too low compared to the census in massive elliptical galaxies (which are formed through repeated mergers).

\citet{Kravtsov2005} and \citet{Shapiro2010} proposed that globulars form in gas-rich (but not necessary clumpy) disc at very high redshift ($z\gtrsim 3$). Galaxy mergers would then gather the physical conditions to enhance the overall formation, and more specifically the most massive globulars \citep{Li2014, Li2017, Kim2018}. In this case, the variations in metallicity would be due to a spread in formation epochs. In their complementary scenario, \citet{Li2014} invoked late major mergers to reproduce the metallicity spread observed in the massive elliptical galaxies of the Virgo cluster. Such a formation channel however is tailored to violent environments involving recent mergers, and not that of the Milky Way which has not experienced such an event since $z\approx 2$ \citep[e.g.][]{Ruchti2015}.

%%%%%%%%%%%%%%
\subsection{Multiphase collapse}

An alternative scenario is that of \citet{Forbes1997} in which, again, metal-poor clusters form during the collapse of the proto-galaxies. This phase ends early in the formation of galaxy, possibly with the reionization of the Universe that quenches star formation \citep{Beasley2002}. Another phase starts when the formation of the galactic discs make the ISM dense enough to resume the formation of globulars, this time from enriched material and thus making the metal-rich population. The second phase could happen in the early, gas-rich and turbulent clumpy discs (recall \sect{clumpy}). Such a mechanism naturally accounts for the spatial distributions and the kinematics of the two populations, as well as for a clear bi-modality in their metallicity distribution, as observed in the Milky Way, but not in e.g. Andromeda. 

Age estimates from \citet{Forbes2015} indicate that the blue population has a mean age of $12.5 \Gyr$, corresponding to formation at $z\approx 6$, i.e. after (or near the end of) reionization. In addition, the red population yields a mean age of $11.5 \Gyr$, i.e. a formation at $z\approx 3$, which is earlier than the predictions of the onset of disc formation in cosmological simulations ($z\sim 1.5\mh 2.5$, despite difficulties in correctly distributing the angular momentum, likely due to the implementations of a too strong feedback, see discussions in \citealt{Agertz2011} and \citealt{Renaud2017}). If the disc of the Milky Way is indeed not in place at the necessary epoch to form the red clusters, another formation mechanism of this population must be proposed.

%%%%%%%%%%%%%%
\subsection{In situ formation and accretion}

Finally, \citet{Cote1998} and \citet{Tonini2013} proposed that red clusters formed in situ (i.e. in the Milky Way itself), while the blue population has been inherited from other galaxies accreted onto the Milky Way via major and minor mergers (and, to a lower extend, from the capture of clusters during a fly-by), but without necessarily triggering the associated starburst episodes. The actual formation mechanisms are then borrowed from the above-mentioned ideas, i.e. formation of the metal-rich clusters in high-redshift discs \citep{Kravtsov2005, Prieto2008, Shapiro2010}, possibly enhanced during mergers \citep{Li2017}, and formation of the metal-poor cluster in dwarf galaxies \citep{Elmegreen2012}. Note that in situ formation at very high redshift would contribute to the metal-poor population \citep{Brodie2014}.

The accretion of satellite galaxies adds to the complexity of the distribution of globulars. Each accreted galaxy having its own formation history, it brings its own population(s) of globular clusters, which then smooth out potential bi- or multi-modalities of the main galaxy itself (see an example in \citealt{Forbes2010}). The relative importances of in situ and accreted population remain uncertain, due to the difficulty of disentangling the contribution of each satellite, and the large uncertainties on the number (and epoch) of accretion events the main galaxy has encountered \citep[e.g.][]{vandenBergh2000, Mackey2004}.

In such a scenario, there is potentially no age difference between the two populations, but the details depend on the enrichment histories of all galaxy progenitors, and when they form its clusters along these histories, which likely varies with galactic mass. 

The cosmological simulation of the Milky Way of \citet{Renaud2017} confirms this hypothesis, arguing that the details on the bi-modality would mostly depend on the mass of the progenitors of the Milky Way, i.e. their ability to form stars and star clusters, and retain the enriched gas (instead of launching outflows faster than the escape velocity of the galaxy). In that sense, low mass progenitors, including the young Milky Way itself at high redshift, would only contribute to the blue, metal-poor population, while the red clusters would mostly form in situ, but also in galaxies experiencing late major mergers with the Milky Way. Therefore, the epoch of the last major merger would set the bi-modality (or its absence). However, the lack of resolution of present-day cosmological simulations forbids us to conclude on the mass of the clusters formed. (Such simulations form ``stellar particles'' that, while they trace the global SFR relatively well, do not predict whether the formation occurs in massive clusters or not.\footnote{A workaround would be to use a subgrid model calibrated on low-redshift observations to assess the mass of clusters formed, but this would then bias the interpretation toward a universal formation mechanism, thus neglecting the variations of cluster formation with small scale physics ($\lesssim 0.1\mh 1 \pc$) between low and high redshift, metal-rich and metal-poor environments, etc.}) As a result, the relative importance of the red and blue populations derived in these simulations remain hypothetical.

In conclusion, the lack of observational constraints, of understanding of the physics of cluster formation at low redshift and how to extrapolate this to the early Universe leave the field of globular cluster formation full of open questions. Some hints will soon be provided by Gaia on the assembly of the stellar populations of the Milky Way, and by the James Webb Space Telescope on the formation of the clusters themselves at relatively high redshifts.

%%%%%%%%%%%%%%%%%%%%%%%%%%%%%%
\section{Nuclear clusters and their connection with ultra-compact dwarf galaxies}
\label{sec:nc}

About 75\% of spiral and dwarf elliptical galaxies host a nuclear cluster in their centre \citep{Cote2006, Seth2006, Neumayer2012}, which is usually denser ($\sim 1\mh 10\pc$, $10^{4\mh 8} \Msun$, recall \fig{masssize}) than globular clusters \citep{Georgiev2014}, putting them amongst the densest stellar structures known. It was found that the fraction of early-type galaxies observed with a nuclear cluster reaches its maximum for stellar masses of $\sim 10^9 \Msun$, but decreases with mass for smaller galaxies, and is truncated for the most massive hosts \citep{Ferrarese2006, Pfeffer2014}. Most of these points are yet to be explained, but \citet{Antonini2013} proposed that tidal disruption by super massive black holes ($> 10^8 \Msun$) could destroy in-falling clusters and prevent the assembly of nuclear clusters in massive galaxies \citep[see also][]{Arca-Sedda2016}.

A number of scaling relations have been empirically established between the mass of the nuclear clusters and the luminosity of the galaxy, the mass or the velocity dispersion of the galactic bulge \citep{Ferrarese2006, Rossa2006, Wehner2006, Graham2012, Scott2013, Georgiev2016}. Therefore, the formation and evolution of nuclear clusters is connected to that of their host galaxy \citep[e.g.][]{Leigh2015}, such that these objects can be used to explore several aspects of galaxy evolution, especially in the central part, the assembly of the bulge and physics related to super massive black holes. For instance, by measuring the metallicity distribution of stars members of the nuclear cluster of the Milky Way, \citet{Ryde2015} found similarities with that of the bulge stars and concluded that at least a fraction of the nuclear cluster formed in a comparable fashion as the galactic bulge \citep[see also][]{Rich2017}.

%%%%%%%%%%%%%%%%%%%%%%%%%%%%%%%%
\subsection{Two and a half formation scenarios}

In the in situ scenario \citep{Milosavljevic2004}, the nuclear cluster forms in the galactic centre. The deep potential well and the fact that the galactic centre constitutes the destination of most of the large-scale gas transport mechanisms (torques from interacting galaxies, bar, spiral etc.) favour the fuelling of this region with gas from larger radii. When it accumulates, the gas becomes dense enough to form stars as a massive cluster. 

In the migration scenario \citep{Tremaine1975}, star formation occurs elsewhere in the galaxy. Lacking the physical peculiarities of the galactic centre, the cluster formed is usually less dense and less massive than typical nuclear clusters. It then migrates to the centre, because of kpc-scale dynamics (spirals, bar, dynamical friction, see \sect{df}), within a few $\sim 0.1\mh 1 \Gyr$ \citep{Chandrasekhar1943, Mo2010}. Once in the centre, the cluster can further accrete other clusters and grow \citep{Andersen2008, Capuzzo2008, Antonini2012, Antonini2013}. As gas inflows in the previous scenario, the infall of clusters can also lead to a rotating nuclear cluster, for at least some assembly histories \citep{Tsatsi2017}.

Kinematics provide a partial way to disentangle the two scenarios. For instance, counter-rotating populations have been found in massive elliptical galaxies \citep[thus supporting the migration scenario,][]{Seth2010, Lyubenova2013}. In general however, both scenarios allow for complex stellar populations in the nuclear cluster, but the mixture of episodic and more continuous formation histories points toward a combination of both channels \citep{Lyubenova2013}. Thus, it is likely that these two scenarios both participate to the building-up of nuclear clusters \citep{denBrok2014, Cole2016}. Models estimate that 50 to 80\% of the stars in nuclear clusters are formed in situ, while the rest has first migrated from larger galactic radii before being accreted \citep{Hartmann2011, Antonini2015}.

In gas rich galaxies, since the densest clusters could have a sufficiently strong gravitational influence on the nearby ISM to retain a gas reservoir around them, their potential migration to the galactic centre would bring gas with them. Following this idea, \citet{Guillard2016} proposed that the compression of this reservoir at the centre, and the further accretion of other gas would thus lead to a episode of in situ formation. Other clusters, with or without gas can also be accreted onto the nuclear cluster. Such process is highly dependent on the properties of the clusters formed in the galaxy, their masses, their abilities of convoy a gas reservoir with them, and their trajectories toward the galactic centre (Guillard et al., in preparation).

Details on the assembly of nuclear clusters affect their final properties, in term of stellar populations, abundance spreads but also kinematics and morphology. For instance, merging clusters triggers a redistribution of the angular momentum which affects the flattening of the resulting object. (\citealt{Guillard2016} noted that a nuclear cluster formed by the merger of two massive clusters would be significantly flatter than if the merger event did not occur, see also \citealt{Antonini2012, Tsatsi2017}).

%%%%%%%%%%%%%%%%%%%%%%%%%%%%%%%%
\subsection{Ultra compact dwarf galaxies}
\label{sec:ucd}

Being the densest and the most bound objects to their galaxies, nuclear clusters are the most robust structure of their host. This has led \citet{Bassino1994} to propose that a nucleated dwarf galaxy undergoing tidal stripping would lose all its material except the nuclear cluster, which would them form a UCD (see also \citealt{Bekki2001, Drinkwater2003, Pfeffer2013}). In such a case, the IMF of UCDs should be similar to that of NCs (and therefore to the bulk of the globular clusters if the NCs assemble by merging globulars and thus share their IMF). In particular, one would not expect to find top-heavy IMFs in UCDs, contrary to the proposition of \citet{Dabringhausen2009} to explain their high mass-to-light ratios (at a late stage of their evolution, once the massive stars have transformed into dark remnants). Thus, in the formation scenario of UCDs from tidal stripping of galactic nuclei, the UCDs should retain a dark matter halo to account for the observed high mass-to-light ratios \citep[see e.g.][and \sect{dm}]{Hasegan2005, Forbes2014}. However, observations of low-mass X-ray binaries (i.e. binaries made of a neutron star and a low-mass companion), support the excess of dark remnants from the evolution of massive stars, and thus a top-heavy IMF \citep{Dabringhausen2012}. Hence, this suggests a different formation scenario for (at least some) UCDs and NCs (and thus globulars if NCs are mergers of clusters). Another possibility to explain the elevated mass-to-light ratio is to consider the presence of a central massive black hole, accounting for a significant fraction of the total mass of their galactic host ($\sim 10\mh 20\%$), as predicted by \citet{Mieske2013} and supported by the recent detections of super-massive black holes in UCDs \citep{Seth2014, Ahn2017}.

Among the alternatives, \citet{Mieske2002, Mieske2012} suggested that the UCDs constitute the high-mass end of the star cluster distributions, and that they form in comparable environments as young massive clusters (e.g. in galaxy mergers, see \citealt{Renaud2015}). Another option would be that the UCDs result from the merger of several star clusters in large cluster complexes \citep{Kroupa1998, Fellhauer2002, Bruns2011}. It is now likely that, like the nuclear cluster, several formation channels exist for the UCDs, with their relative importance depending on the context \citep{Mieske2006, Brodie2011, Norris2011, Pfeffer2014}.

%%%%%%%%%%%%%%%%%%%%%%%%%%%%%%%%%%%%%%%%%%%%%%%%%%%%%%%%%%%%%%%%%%%%%%%%%%%%%%%%%%%%%%%%%%%%%%%%%%%%%%%%%%%%%%%%%%%%%%%%%%%%%%%%%%%%%%%%%%%%%%%%%%%%%%%%%%%%%%%%%%%%%%%%%%%%%
%%%%%%%%%%%%%%%%%%%%%%%%%%%%%%%%%%%%%%%%%%%%%%%%%%%%%%%%%%%%%%%%%%%%%%%%%%%%%%%%%%%%%%%%%%%%%%%%%%%%%%%%%%%%%%%%%%%%%%%%%%%%%%%%%%%%%%%%%%%%%%%%%%%%%%%%%%%%%%%%%%%%%%%%%%%%%
%%%%%%%%%%%%%%%%%%%%%%%%%%%%%%%%%%%%%%%%%%%%%%%%%%%%%%%%%%%%%%%%%%%%%%%%%%%%%%%%%%%%%%%%%%%%%%%%%%%%%%%%%%%%%%%%%%%%%%%%%%%%%%%%%%%%%%%%%%%%%%%%%%%%%%%%%%%%%%%%%%%%%%%%%%%%%

%%%%%%%%%%%%%%%%%%%%%%%%%%%%%%%%%%%%%%%%%%%%%%%%%%%%%%%%%%%%%%%%%%%%%%%%%%%%%%%%%%%%%%%%%%%%%%%%%%%%%%%%%%%%%%%%%%%%%%%%%%%%%%%%%%%%%%%%%%%%%%%%%%%%%%%%%%%%%%%%%%%%%%%%%%%%%
%%%%%%%%%%%%%%%%%%%%%%%%%%%%%%%%%%%%%%%%%%%%%%%%%%%%%%%%%%%%%%%%%%%%%%%%%%%%%%%%%%%%%%%%%%%%%%%%%%%%%%%%%%%%%%%%%%%%%%%%%%%%%%%%%%%%%%%%%%%%%%%%%%%%%%%%%%%%%%%%%%%%%%%%%%%%%
%%%%%%%%%%%%%%%%%%%%%%%%%%%%%%%%%%%%%%%%%%%%%%%%%%%%%%%%%%%%%%%%%%%%%%%%%%%%%%%%%%%%%%%%%%%%%%%%%%%%%%%%%%%%%%%%%%%%%%%%%%%%%%%%%%%%%%%%%%%%%%%%%%%%%%%%%%%%%%%%%%%%%%%%%%%%%
\part{Evolution}
\label{part:evolution}

When interested in old clusters like globulars, one should not neglect the long-term, post-gaseous phase evolution which shapes the clusters. In order to connect present-day observations to the conditions of cluster formation, it is necessary to understand the physical mechanisms that alter the properties like the mass or the size, and the very survival of clusters.

%%%%%%%%%%%%%%%%%%%%%%%%%%%%%%%%
\section{Collisional systems}
\label{sec:collisional}

Because of the high densities of clusters, their stars interact frequently, which drives exchanges of energy toward equipartition. This leads for instance to mass segregation (i.e. the most massive stars being preferentially in the cluster centre, see e.g. \citealt{White1977, Bonnell1998}, but also \citealt{Trenti2013} and \citealt{Parker2016} for cautionary notes). The time needed to significantly decrease the imprint of a perturbation from energy equipartition of stellar systems is commonly estimated through the relaxation time. \citet{Spitzer1987} provides a simple expression for such a timescale, evaluated at the half-mass radius $\rh$ for a system of $N$ stars of mean mass $m$ (i.e. $\approx 0.5 \Msun$ for old objects) as
\begin{equation}
\trh \approx 0.17 \Myr \left(\frac{\rh}{1 \pc}\right)^{3/2} \left(\frac{1 \Msun}{m}\right)^{1/2} N^{1/2}.
\end{equation}
Applying this expression to dense stellar systems like star clusters ($\rh \sim 1\mh 10 \pc$, $N \sim 10^{3\mh 6}$, $m\approx 0.5\mh 1 \Msun$) shows their half-mass relaxation time is generally shorter than their lifetime. This hence indicates that stellar encounters have likely operated a significant evolution of the internal properties of the clusters since their formation, contrarily to less dense\footnote{Note that the relaxation time does not solely depend on density.} systems like galaxies where the relaxation time is several orders of magnitude longer. This has led \citet{Forbes2011b} to propose to use such a difference to tell apart clusters from galaxies (recall \sect{relax}).

One of the consequences of the collisional nature of clusters is the energy transport from the inner parts to the outskirts (i.e. a negative dynamical heat capacity), that results in the collapse of the central part. This mechanism, known as core-collapse or gravothermal catastrophe \citep{Antonov1962, Lynden1968, Goodman1987}, would eventually lead to an infinite central density if it was not stopped by the injection of energy from binary stars, either primordial, or formed by gravitational capture \citep{Heggie2003}. 

At first order, one can consider the internal physics of the clusters (mainly stellar evolution and stellar encounters) as an engine pumping energy, mostly from the core of the cluster outward \citep{Henon1961}. This translates into the (overall) expansion of the cluster. \citet{Henon1961, Henon1965} proposed a description of energy flows in clusters at a constant rate per relaxation time. According to \citet{Henon1961}, the central source of internal energy driving the expansion does not need to be formally identified, but the hardening of binaries in the cluster's core is a natural candidate \citep{Giersz1994, Baumgardt2002}. Yet, number of works showed that clusters considered in isolation (i.e. when neglecting the galactic context) yield very long dissolution timescales, up to 1000 times their initial relaxation time \citep{Giersz1994, Baumgardt2002}. Therefore, the environment, through the tidal interactions discussed below, is the main actor setting the clusters' lifetimes \citep{Gieles2008b}. Tides contribute to the energy increase of the cluster, but mainly in the outer layers. At some point, stars reach an energy threshold set by the tides and the rest of the cluster, and can (potentially, see \sect{pe}) escape. Consequently, the evolution of star clusters results from the interplay between internal dynamics and external perturbations (of cosmological and galactic nature). It is therefore equally important to understand the internal engine and its evolution along the cluster lifetime, and the role of the environment. The different processes involved and their relative roles vary with time, with the position inside the cluster, and the galactic context \citep[see e.g.][]{Lamers2010}.

Accounting for both the internal and external physics of clusters is the main challenge for theoretical works in this field, because of the multi-scale and multi-physics nature of the problem. Aspects such as stellar evolution, the formation, evolution and destruction of binaries and multiple systems, star-star collisions, the dynamics of the so-called dark remnants (brown dwarfs, neutron stars, stellar mass and intermediate mass black holes, \citealt{Strader2012, Lutzgendorf2013, Mastrobuono2014, Peuten2016}), as well as mass segregation \citep{Webb2017} and multiple populations \citep{Charbonnel2016} all have an non-negligible impact on the internal evolution of clusters, on the interpretation of the observations (e.g. mass to light ratios, \citealt{Strader2011, Kimmig2015, Watkins2015}, velocity dispersions, \citealt{Baumgardt2017} etc.) and deserve dedicated reviews. I recommend exploring the reviews by \citet{Heggie2003} and \citet{Vesperini2010}, and I focus here on the influences of the galactic and cosmological environment in modifying the properties of the clusters, and forming tidal features.

%%%%%%%%%%%%%%%%%%%%%%%%%%%%%%%%%%%%%%%%%%%%%%%%%%%%%%%%%%%%%%%%%%%%%%%%%%%%%%%%%%%%%%%%%%%%%%%%%%%%%%%%%%%%%%%%%%%%%%%%%%%%%%%%
\section{Modelling dense stellar systems}

In this rather technical section, I present the methods and tools that have been developed to study star cluster evolution, and why addressing this gravity-only question is not as simple as it looks at first sight. 

%%%%%%%%%%%%%%%%%%%%%%%%%%%%%%%%
\subsection{A numerical challenge}
\label{sec:nbody}

The challenge of studying the evolution of clusters in their environment is to couple their internal physics to a description of the influence of the environment. The most natural way to do so would be to model jointly a cluster and its host galaxy(ies) and account for all gravitational interactions between all mass elements (stars, gas, dark matter). Taking all interactions into account is commonly done in galaxy and cosmological simulations, where tens of millions resolution elements (particles and/or cells) are now routinely modelled. To speed-up the computation, the gravitational acceleration of distant elements is approximated (see tree codes, particle-mesh, P3M, fast multipole methods and others, e.g. \citealt{Barnes1986}, \citealt{Couchman1995}, \citealt{Dehnen2000b}, among many others), and the potential of each element is smoothed to avoid divergences of the acceleration during close encounters \citep{Aarseth1963, Dehnen2001, Athanassoula1998}.

However, in dense, collisional, stellar systems like clusters, star-star interactions in binaries and multiples have been shown to be an important source of the cluster's internal evolution \citep[e.g.][]{Heggie1975, Lynden1980} by driving an outward flow of energy \citep{Henon1961}. It is thus essential for any method to include their effect, and thus to capture details as short as the orbital period of a binary, and as small as a stellar radius. When put in galactic context, this translates into a wide range of space- and time-scales, making this problem virtually impossible to solve directly with present-day high performance computing resources. To date, the solution adopted consists in using distinct, specialized, algorithms to describe the cluster on the one hand, and its environment on the other, and make then communicate. Even before considering the galactic environment, modelling collisional systems is a challenge by itself, which requires a large amount of algorithmic and engineering tours de force. 

\begin{figure}
\includegraphics[width=\columnwidth]{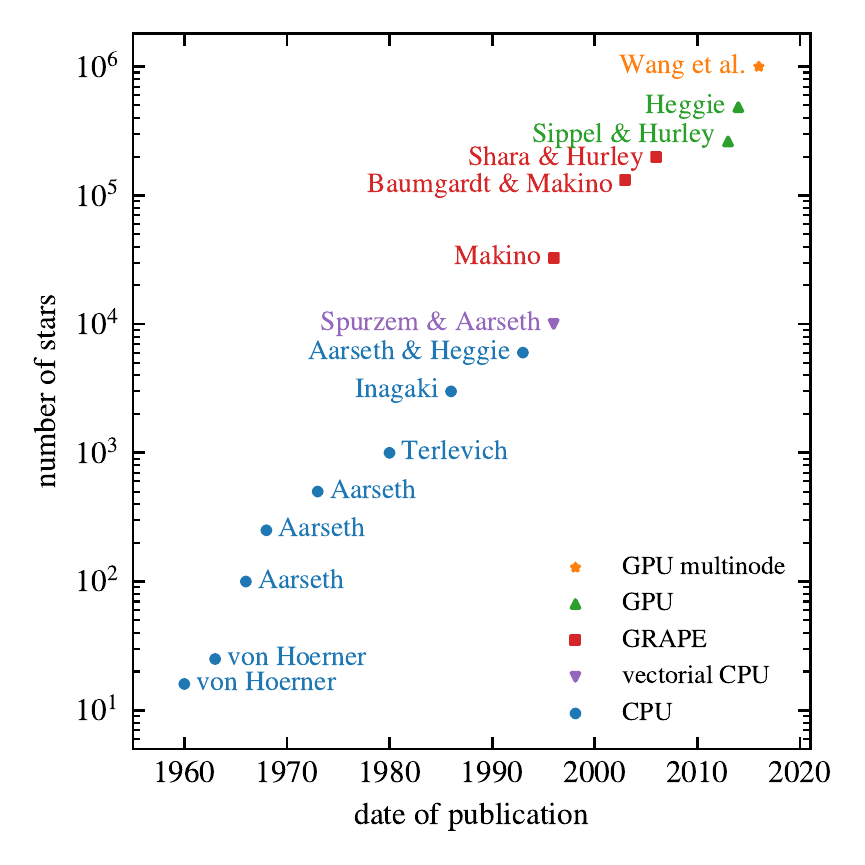}
\caption{Progress of $N$-body simulations of clusters through the years. The original version of this figure was first presented in \citet{Heggie2003}, and it has been updated by Douglas Heggie (private communication). Only works reaching the late evolution stage of clusters are shown \citep[i.e.][]{vonHoerner1960, vonHoerner1963, Aarseth1966, Aarseth1968, Aarseth1973, Terlevich1980, Inagaki1986, Aarseth1993, Spurzem1996, Makino1996, Baumgardt2003, Shara2006, Sippel2013, Heggie2014b, Wang2016}. Rapid progresses have been made by upgrading the technology used, especially with special-purpose hardware (e.g. GRAPE, \citealt{Makino1998}), or existing technology (e.g. GPU, \citealt{Nitadori2012}).}
\label{fig:heggie}
\end{figure}

In the direct $N$-body approach, each of the $N$ stars of a cluster is modelled as a particle which experiences the gravitational acceleration from the $N-1$ others \citep[see][for more complete and detailed presentations]{Spitzer1987, Heggie2003, Binney2008}. This direct method is implemented for instance in \code{NBODY6} \citep{Aarseth2003} and \code{PH4} \citep{McMillan2012}. Short range interactions in the collisional regime are treated accurately through the use of regularization algorithms to resolve the motion of close encounters with no critical modification of the timestep that would otherwise considerably slow down the simulations (see \citealt{Kustaanheimo1965, Heggie1973}, and \citealt{Mikkola2008} for a review). The effect of the environment (i.e. the tidal field) can then be added to the equations of motion of every stars, as described in the next section. The complexity of the $N$-body problem goes as $\mathcal{O}(N^2)$, but this computational cost is significantly reduced by the use of software \citep{Spurzem1999, Wang2016} and/or hardware accelerations \citep{Hut1999, Makino2003, Nitadori2012}. \fig{heggie} shows the progress of the cluster simulations, made possible by always improving computer resources and techniques. Yet, modelling massive clusters with the direct $N$-body method still represents a challenge \citep[see][on solving the million body problem]{Wang2016}, making it impossible to cover a wide parameter space. To circumvent this problem, other methods are also commonly used.

Another approach consists in solving the Boltzmann equation on a distribution function, with the addition of a term describing the close interactions between stars (so-called collisions) which follows a Markov process. For this reason such method is refereed to as Fokker-Planck. The evolution of the distribution function is usually computed numerically in a discrete grid of energy bins \citep{Cohn1979}. However, without knowledge of the position of stars, the treatment of galactic tides is restricted to the idealised cases of clusters on circular orbits.

The Monte-Carlo method is a hybrid of Fokker-Planck and $N$-body: instead of evolving the distribution functions on an energy grid, they are sampled by tracer particles evolving in phase-space. Such method is implemented for example in the \code{MOCCA} code \citep{Giersz2008}. This approach leads to a substantial speed-up compared to $N$-body method, which allows for the study of massive clusters \citep[see e.g. an application in][]{Heggie2014}. However, it is limited to spherically symmetric problems and thus cannot account for the anisotropic nature of tides. Note however, the contribution of the \code{RAGA} code \citep{Vasiliev2014} which allows for arbitrary geometries by implementing relaxation with local diffusion coefficients in velocity, and thus which could include a precise treatment of complex tides.

In addition, \citet{Dehnen2014} proposes a fast multipole method to integrate the motion of individual stars and reach the accuracy of collisional algorithms. By distributing particles in cells, computing cell-cell interactions and then approximating the force on particles using Taylor expansions, this method achieves unprecedented performances of $\mathcal{O}(N^{0.87})$. Yet, the lack of regularisation considerably slows down the code execution when treating dense, collisional systems.

Finally, semi-analytical methods solve a handful of coupled differential equations describing, at first order, the global evolution of the cluster \citep{Ambartsumian1938, Chandrasekhar1942, King1958, Henon1961, Heggie1975, Lee1987, Hut1992, Gieles2011b}. An example of this is the code \code{EMACSS} \citep{Alexander2012, Alexander2014, Gieles2014} which solves for the mass, the half-mass radius and the energy of the cluster. The mathematical complexity of these problems requires a numerical solver but this approach remains doubtlessly the fastest. To estimate the mass-loss of clusters, \code{EMACSS} is calibrated using the results of $N$-body simulations for clusters on circular orbits around galaxies, perfectly suited in the cases of constant tides, or when the tidal field changes adiabatically with respect to the cluster internal evolution (and when tides can be reasonably approximated by a piecewise constant function, see an example in \citealt{Renaud2015c}). However, this method does not retrieve the mass-loss derived from $N$-body methods in non-constant tides. 

%%%%%%%%%%%%%%%%%%%%%%%%%%%%%%%%
\subsection{Accounting for the environment}
\label{sec:nbodyenv}

A very large body of work has been done using the above-mentioned methods to study e.g. the intrinsic evolution of clusters, of their stellar populations and the effect of (potential) intermediate-mass black holes. In their large majority, these studies ignore the effect of the environment on the cluster to focus on internal dynamics. However, the environment plays a paramount role in the questions of e.g., long term evolution, mass-loss, kinematics, especially in the outermost layers of clusters. The simplest and historical approach accounting for this consists in considering clusters on fixed circular or elliptical orbits in an analytical representation of the galactic potential, where the tidal force can be expressed analytically and added to the force every star experiences from the rest of the cluster (e.g. \citealt{Chernoff1990} and \citealt{Gnedin1997} with the Fokker-Planck method, \citealt{Vesperini1997} and \citealt{Portegies1998} with the $N$-body way, and \citealt{Giersz2001} with the Monte Carlo approach). To go further, the galactic potential can also be used to integrate the orbit of the cluster and compute the relevant tidal forces, thus allowing for non-fixed orbits \citep{Renaud2015b}. Such a method provides the galactic acceleration not only at the position of the cluster, but also on its tidal debris (see \sect{streams}). Alternatively to these simulations of clusters, the focus can be shifted to the galaxy, by including a parameterization of star cluster evolution in galaxy simulations \citep{Kruijssen2011, Matsui2012, Brockamp2014}. More detailed prescriptions of the cluster-galaxy connection have been proposed in the last years, opening the field to the study of clusters in cosmological context. 

These methods couple simulations of star clusters with simulations of galaxies. One possibility is to extract the gravitational field along one orbit in the galaxy or cosmological volume and to then pass this information (e.g. through the form of the tidal tensor) to a star cluster simulation (see the \code{NBODY6tt} code, \citealt{Renaud2011}, based on \code{NBODY6}, \citealt{Aarseth2003}). Doing so allows for time-varying, complex galaxies to be considered (e.g. mergers, \citealt{Renaud2013}, satellite galaxies, \citealt{Bianchini2015}, full cosmological context, \citealt{Rieder2013}). However, the orbit in the galaxy must be assumed beforehand, which neglects the non-trivial effect of dynamical friction \citep[see][]{Petts2015}. This limitation disappears when adopting the option of evolving the two simulations simultaneously, by making both codes communicate (see the \code{BRIDGE} code, \citealt{Fujii2007}, and its extension in \code{AMUSE}, \citealt{Pelupessy2013}). However, the different timescales involved on the galactic and cluster sides imply that a numerically efficient coupling is only reached in (sub-)parsec resolution galaxy simulations (i.e. when both timesteps become comparable), of which cost limits these runs to short periods of time. Such approach is thus particularly well suited in simulating the early life of star cluster (in particular when accounting for hydrodynamics), but is less optimum for Gyr-long modelling.

%%%%%%%%%%%%%%%%%%%%%%%%%%%%%%%%%%%%%%%%%%%%%%%%%%%%%%%%%%%%%%%%%%%%%%%%%%%%%%%%%%%%%%%%%%%%%%%%%%%%%%%%%%%%%%%%%%%%%%%%%%%%%%%%
\section{Dynamical friction}
\label{sec:df}

When orbiting the dense regions of galaxies, star clusters experience a loss of their orbital energy caused by dynamical friction. Such a modification of their trajectory does not directly impact their properties like mass and size, but brings clusters in different, usually stronger, tidal fields that alter these properties.

Dynamical friction is the gravitational drag induced by the constituents of a dense medium in which a massive object moves. The object, a cluster or a satellite galaxy for instance, attracts constituents of the medium, like stars, gas or dark matter, toward its position. However, because of its motion, the medium accumulates behind the cluster (i.e. at the position it occupied moments earlier). Such accumulation constitutes an overdensity in the medium of which gravitational force on the cluster drags it against its motion and thus slows it down \citep{Chandrasekhar1943}. This effect is for example responsible for the transfer of orbital energy of interacting galaxies which, under some circumstances, leads to their merger and coalescence \citep{Duc2013}. It can also make clusters spiral in toward galactic centres, where strong tides accelerate their dissolution and help forming galactic nuclei or nuclear clusters \citep[see][and \sect{nc}]{Tremaine1975, Capriotti1996, Lotz2001, Arca2014}. Furthermore, \citet{Fellhauer2007} noted that the stars lost by the cluster could contribute to the medium, thus participating in dynamical friction and accelerating the process.

From this simple explanation, one can deduce dynamical friction is efficient for massive clusters, in dense media, at low velocity, which is depicted in \citet{Chandrasekhar1943} formula. His formalism lacks several physical aspects like the self-gravity of the medium and resonant interactions \citep{Inoue2009}. Yet, it is remarkably accurate at predicting the results of simulations in many configurations \citep[e.g.][]{Tremaine1984, Weinberg1986, Hashimoto2003}. In some other cases however, Chandrasekhar's framework is incomplete. This is for instance the case in uniform density media like the cores of galaxies \citep{Read2006, Goerdt2010, Petts2015}, and when the cluster becomes of comparable mass as the surrounding medium, like in galactic centres \citep{Gualandris2008}.

Modelling precisely dynamical friction can thus requires a self-consistent treatment of the medium, which is often too numerically expensive. The alternative is to use semi-analytic models to mimic the effects of dynamical friction as accurately as possible \citep[see examples in][]{Just2005, Just2011, Arca2015}. In this context, \citet{Petts2016} proposed a complement to Chandrasekhar's formula to reproduce the inefficiency of dynamical friction in the inner parts of galaxies, and in particular in cored dwarfs (the so-called core-stalling phenomenon). This was then applied to clusters orbiting dwarf galaxies to discriminate between cored and cuspy dark matter profiles (\citealt{Contenta2017}, see also \citealt{Hernandez1998, Sanchez2006, Goerdt2006, Cole2012, Amorisco2017}).

It is worth noticing that, in the Chandrasekhar's formalism, the intensity of the drag force varies with the square of the mass of the cluster, meaning that massive clusters are more prone to dynamical friction than their lower mass counterparts. Dynamical friction is thus one of the rare mechanisms preferentially destroying the most massive clusters (by merger with the galactic centre like nuclear clusters or the bulge), typically over a few Gyr timescale.

%%%%%%%%%%%%%%%%%%%%%%%%%%%%%%%%%%%%%%%%%%%%%%%%%%%%%%%%%%%%%%%%%%%%%%%%%%%%%%%%%%%%%%%%%%%%%%%%%%%%%%%%%%%%%%%%%%%%%%%%%%%%%%%%
\section{Adiabatic, secular tides}
\label{sec:secular}
 
Once the gas has been removed from the cluster volume, the only interaction between the cluster and its environment is gravitational. This effect is dual: global and differential. In its global flavour, this interaction sets the trajectory of the cluster as a whole, and does not directly affect its evolution. A notable exception is cluster-cluster interactions and mergers, which mostly occur in galactic centres, and participate in building nuclear clusters (recall \sect{nc}).

The differential effect however directly affects the evolution of the clusters, by inducing an acceleration of which intensity and direction varies across the cluster. Such differential forces are tides, and encompass several aspects, usually sorted according to their timescales. In the rapid, impulsive fashion, the tidal effects are considered as shocks (see \sect{shock}). When the gravitational influence of its environment affects a cluster in a slower, longer manner with respect to its internal evolution, one speaks of adiabatic tides. Details on the evolution of the cluster properties have been intensively studied over the last decades, in particular with $N$-body simulations.

%%%%%%%%%%%%%%%%%%%%%%%%%%%%%%%%
\subsection{Tidal radii and Jacobi surface}
\label{sec:rt}

The tidal radius is a useful quantity to describe the strength of the tidal field on a given object. It is unfortunately rarely explicitly defined in the literature, although it can refer to several, distinct, quantities. It seems that a major source of confusion came from the use of the term ``tidal'' to designate the outermost radius of an observed cluster. The tidal radius appears as a parameter of the \citet{King1966} density profile, as the radius at which the density drops \citep[see also][]{Michie1963, Michie1963b, King1981}. Such radius can exist for clusters in isolation and is, most of the time, independent of tides, as shown observationally by e.g. \citet{Odenkirchen1997}. Even if leaving this confusion aside, the tidal radius remains an ill-defined concept.

In a first definition, one can delineate the volume of influence of a cluster as the sphere centred on the cluster and going through the Lagrange point L$_1$, i.e. the point between the cluster and the galaxy, where the gravitational force from the cluster exactly balances that of the galaxy. The radius of this sphere is called the tidal (or Jacobi) radius. It evolves with the cluster (when it grows, shrinks, moves to volumes of strong or weak tides), and it is thus possible that stars remain well within this radius, such that it does not mark any visible truncation in the stellar distribution.

In this domain, it is convenient and common to adopt the tidal approximation, which consists in assuming that the distance between the cluster and the galaxy is much larger than that between the cluster and its stars. Doing so allows us to linearize (with a first order Taylor expansion) the expressions of the galactic gravitational forces on stars and provides an analytical expressions in several (idealised but illustrative) cases. We will do so in the following, keeping in mind that this formalism is not suited for clusters near steep features in the galactic potential (e.g. the galactic centre), nor for stars far from their cluster.

When considering the cluster and the galaxy as points of respective masses $m$ and $M$ separated by a distance $R$, the tidal radius reads
\begin{equation}
\label{eqn:rt}
r_\textrm{t, point} = R \left(\frac{m}{2M}\right)^{1/3},
\end{equation}
\citep{vonHoerner1957, Spitzer1987}. Variants of this (purely gravitational) expression account for the centrifugal force of clusters on circular orbits, which then brings the equilibrium point toward the cluster:
\begin{equation}
\label{eqn:rtprime}
r'_\textrm{t, point} = R \left(\frac{m}{3M}\right)^{1/3},
\end{equation}
\citep{King1962}, which can also be generalized to circular orbits around any spherically symmetric galactic potential when introducing the orbital angular frequency $\Omega$ and the local galactic gravitational potential $\phi$
\begin{equation}
r'_\textrm{t} = \left(\frac{Gm}{\Omega^2-\frac{\partial^2\phi}{\partial R^2}}\right)^{1/3},
\end{equation}
(see the Appendix~A of \citealt{Renaud2017} for a more detailed explanation of the different expressions and the corresponding assumptions). 

However, the concept of tidal radius is a shortcut often leading to oversimplifications. For instance, the volume of influence of the cluster is not a sphere but rather a flatten ``lemon-like'' shape, defined by the equipotential surface going through L$_1$, called the Jacobi surface (or the Roche surface, see \citealt{Renaud2011} for its equation). The deviation from spherical symmetry induce by tides can have a comparable effect on the shape of a cluster as a degree of intrinsic rotation \citep{Bertin2008, Varri2009}. The Jacobi surface is parameterised by the tidal radius, and its flattening along the other two axes that depends on the exact shape of the local galactic potential. This implies that, for a given tidal radius, different galaxies can lead to different Jacobi surfaces, and thus to different tidal accelerations on cluster members. \citet{Tanikawa2010} illustrated this point by studying the mass-loss rate of clusters in galaxies with a power-law density profile, varying the power-law index, but keeping the tidal radius unchanged (which implies to change the galactocentric distance $R$). They found that shallow galactic profiles (i.e. flattened Jacobi surfaces) induce slow mass-loss. This can be understood when considering the conditions required for a star to leave the cluster, which is itself a rather complicated topic, as discussed e.g. in \citet{Ross1997} and in the next section.

In compressive tides, the mathematical derivation of a tidal radius (using any of the above expression) leads $r_\textrm{t}$ to be the cubic root of a negative quantity, reflecting the fact that there is no point where the force from the galaxy balances that from the cluster since they both point in the same direction. The tidal radius (and the Lagrange points, and the Jacobi surface) cannot be defined. Yet, the acceleration on stars induced by the compressive tides still speeds-up the mass-loss of clusters, with respect to isolated clusters \citep[see also the compressive shocks discussed in \sect{shock}]{Renaud2011}.

Therefore, the tidal radius is an ill-defined quantity, that does not directly relate to the evolution of clusters, but which remains, to some extend, convenient to illustrate \emph{some} tidal effects. A more mathematically rigorous and more general quantity to gauge tides could be the strength of the tidal field along its main axis of action \citep[which is the maximum eigenvalue of the tidal tensor, see the Appendix A of][for details, and the connection to the tidal radius]{Renaud2017}, but this would then neglect the changes of the orientation of the tidal field (i.e. the changes in the eigenbase of the tensor), which necessarily slow down the mass-loss due to a mismatch between the escape funnels (around the Largange point) and the directions of the accelerations the stars experience before getting there.

%%%%%%%%%%%%%%%%%%%%%%%%%%%%%%%%
\subsection{Potential escapers and kinematics in the outer regions}
\label{sec:pe}

\begin{figure}
\includegraphics[width=\columnwidth]{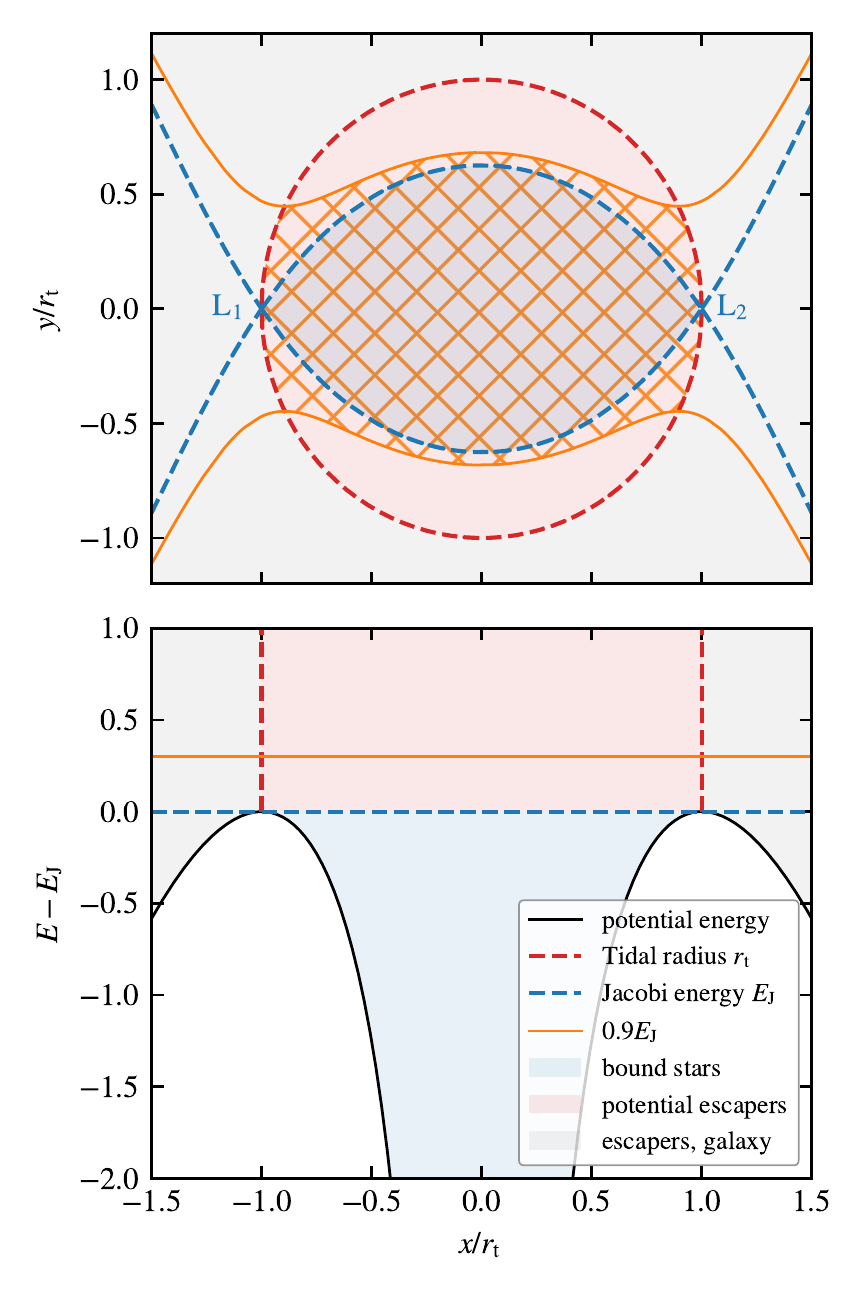}
\caption{Definition of the potential escapers in a geometrical diagram (top) and an energy plane (bottom), illustrated with a point-mass cluster (set at $x=y=0$) experiencing tides from a point-mass galaxy at $x<0$, $y=0$. (An inertial reference frame is considered to compute the energy terms.) The black line represents the potential energy from the cluster and the tides: stars cannot have less energy. Outside the tidal radius ($r_\textrm{J}$, dashed red line) is the galaxy (shaded grey area), including the stars that have escaped the cluster. Inside, one finds bound stars (shaded blue area) with an energy lower than the Jacobi energy $E_\textrm{J}$, and potential escapers (shaded red area) with an energy higher. As an example, the orange line marks 0.9 times the Jacobi energy. A star with this energy inside the tidal radius (i.e. in the hatched area) is a potential escaper.}
\label{fig:pe}
\end{figure}

To escape its cluster, a star needs to cross the gravitational barrier between the potential wells of the cluster and the galaxy. A necessary but not sufficient condition is that the star has a higher (more positive) energy than that of the Jacobi surface (sometimes called critical energy). However, even in this case, a star is likely to remain in the cluster for a significant period of time. \fig{pe} shows diagrams in space and energy of such configuration, with the example of a star having 0.9 times the (negative) energy of the Jacobi surface. Thanks to this excess of energy, the star can travel further than the Jacobi surface. However, to escape, it also needs to actually leave the volume of the cluster, i.e. move through the openings in its own iso-energy surface, near L$_1$ or L$_2$. Only then, the star will definitely leave the cluster (assuming the tides are constant). The size and shape of these openings depend on the energy excess and the local galactic potential \citep[see][for an analytical derivation]{Renaud2011}.

Therefore, even with an energy higher than the Jacobi energy, stars are trapped in the cluster until they reach these openings. Such stars are called potential escapers, as described by \citet[see also \citealt{Henon1970}, \citealt{Fukushige2000}, \citealt{Baumgardt2001}, \citealt{Daniel2017}]{Henon1969}. Depending on the motion of the stars with respect to the (possibly varying) positions of the Lagrange points, and their energies (i.e. the size of the openings), the period spent in this phase can be as long as the cluster lifetime (i.e. possibly longer than a Hubble time), which implies that a significant fraction of the material within the volume of the cluster can be in the potential escaper phase \citep[10 to 35\%, and possibly higher in massive clusters, see e.g.][]{Baumgardt2001, Just2009}.

The presence of potential escapers affects the dynamics and kinematics of clusters, and have been proposed by \citet[see also \citealt{Claydon2017}]{Kuepper2010b} to explain the observed flattening of the velocity dispersion profile in the outermost layers of several globular clusters \citep{Drukier1998, Sollima2009, Scarpa2010, Lane2010, DaCosta2012, Baumgardt2017}. The velocity anisotropies are also affected by this population \citep{Bianchini2017}.

Note however that, the net gravitational acceleration is usually weak in these outer regions and could even, for some clusters, be below the critical acceleration $a_0$ ($\sim 10^{-10} \U{m\,s}^{-2}$) of the modified gravitation framework MOND \citep{Milgrom1983}. In this regime, an extra MONDian acceleration is added by the modification of the law of gravity to that from the baryonic distribution. Such an additional contribution would then account for the observed velocity dispersions \citep{Scarpa2007, Hernandez2013}. An alternative theory is that globulars at large galactic radii could have retained their own dark matter halo \citep{Ibata2013, Penarrubia2017}, remnant of their formation in mini-halos at high redshift \citep[recall \sect{formhalo}]{Peebles1968}. This component would thus have the expected effect in increasing the outer velocity dispersion. In this field, Gaia will soon provide important insights by detecting and estimating kinematic properties of stars at large distances from the centre of their cluster.

%%%%%%%%%%%%%%%%%%%%%%%%%%%%%%%%
\subsection{Stellar streams}
\label{sec:streams}

After leaving the cluster, stars can be used to probe the structure and mass distribution of the galaxy. Because of the shape of the equipotential surfaces around the cluster, most stars leave by flying by the Lagrange points L$_1$ (between the cluster and the galaxy) and L$_2$ (on the opposite side, see \fig{pe}). Those escaping through L$_1$ are thus at smaller galactic radii than the cluster such that, for a given kinetic energy, their orbital motion around the galaxy is faster than that of the cluster. The exact opposite is found for the stars escaping from L$_2$. As a consequence, the tidal debris forms two tails: one leading (connected to the cluster through L$_1$) and one trailing (from L$_2$), the two making a characteristic ``S''-shape. Apart from these small differences, the tails initially have a similar orbit than their progenitor cluster around the galaxy \citep{Johnston1998}, and remain coherent structures for several revolutions \citep{Helmi1999}. Such pairs of tails form in any tidally interacting systems (clusters, galaxies etc.), but their mass, elongation and even detectability depend on several properties \citep[see e.g.][]{Balbinot2018}, like the masses involved, the orbit and the velocity distribution of the stars (e.g. the spin-orbit coupling, see \citealt{Duc2013}). For instance, in a pair of interacting galaxies, the tidal bridges connecting the two members are the leading tails of each galaxy \citep{Toomre1972}.

When originating from a cluster or a dwarf galaxy, these tails are often called streams. Number of them have been detected around galaxies in the Local group, like the Sagittarius, Orphan and Monoceros streams \citep{Ibata2001, Newberg2002, Yanny2003, Grillmair2006, Belokurov2006}, especially in the recent years in all-sky surveys like DES \citep{DES2005}, SDSS \citep{Ahn2012} and VST ATLAS \citep{Shanks2015}. More are expected to be discovered in the near future with data releases from the GAIA mission. These surveys are deep enough to unambiguously detect streams, but characterising their morphology and the potential presence of sub-structures like gaps along them is more disputed (see below). On the one hand, streams intrinsically host variations in their densities, as shown by early numerical experiments by \citet{Combes1999}. \citet{Kuepper2008} explained that the motions of stars along streams lead to orbital crowding at the epicycles (i.e. a slowing down of the stars at a fixed position, see also \citealt{Capuzzo-Dolcetta2005}). This then translates into regularly spaced overdensities in streams originating from stellar systems on a circular or eccentric orbit around their host galaxy \citep{Just2009, Kuepper2010, Lane2012, Mastrobuono2012}. On the other hand, the morphology and structure of streams is also affected by the galactic environment, and is thus used to probe this environment and its evolution.

Note that streams have not been detected around many clusters in the Milky Way, even when experiencing strong enough tides to induce mass-loss. The reason for this could be detection limits: stars in streams originate from the outer layers of the clusters which, because of mass segregation (see \sect{collisional}), mainly encompass low-mass, faint stars \citep[see][in the case of Palomar-5]{Koch2004}. Deeper observations of the outermost regions of clusters, with the necessary corrections for background contamination, could thus reveal faint tidal tails. Furthermore, depending on its age and mass-loss of the cluster (or dwarf galaxy) making it, a stream can be observed after the final dissolution of its progenitor. This is for instance the case of the Ophiuchus \citep{Sesar2015} and Phoenix streams \citep{Balbinot2016}.

Streams can be used to infer some properties of their progenitors. In particular, tidal stripping is less efficient on a dwarf galaxy with a cored dark matter halo than on a cuspy one (\citealt{Penarrubia2010}, see also \citealt{Read2016} for a review on the core/cusp problem and the insights gained when accounting for detailed baryonic physics in cosmological simulations). Thus, the resulting stream (as being directly linked to the mass-loss induced by tides) traces the dark matter distribution of the progenitor \citep{Errani2015}. However, streams are mostly used to map the gravitational potential of the host galaxy, in particular in the Milky Way, by inferring the properties of the dark halo on larger scales than what stellar kinematics and the neutral gas trace \citep{Sackett1999}. 

Encounters with sub-structures alter the morphology and the kinematics of the streams, making them dynamically hotter \citep{Ibata2002, Johnston2002}. Relying on this, streams are used to probe the number and size distribution of dark-matter sub-structures, down to $\sim 10^{6\mh 7} \Msun$ \citep{Erkal2016}, as a complement to gravitational lensing methods \citep[e.g.][]{Mao1998, Vegetti2009}. Several works reported the presence of gaps along streams \citep[e.g.][]{Odenkirchen2001, Odenkirchen2003, Grillmair2006}, which would result from local tidal disruption caused by sub-structures in the galactic potential \citep[e.g.][]{Carlberg2011, Carlberg2012, Sanders2016}. Sub-halos have been proposed as the main suspects since the typical sizes of gaps match the predictions from the cold dark matter framework (\citealt{Yoon2011, Ngan2014, Bovy2017}, but see \citealt{Mastrobuono2012} who excluded this hypothesis for the innermost gaps). The main objective of present-day studies is thus to establish how structures in streams depend on the properties of the galactic halo. In that respect, \citet{Pearson2015} showed that a triaxial halo leads to fanning of the stream at large distance from the progenitor. Furthermore, \citet{Sandford2017} conducted a numerical surveys to highlight the strong sensitivity of the thinness of streams, their symmetry and the regularity of their morphology on the galactic halo shape. Future detections of streams (in particular at distances where the dark matter dominates the mass distribution) could then be matched by numerical models to map the potential of the Galaxy, and reconstruct its assembly.

Yet, deep observations have recently discredited the existence of gaps in the Palomar-5 stream (\citealt{Ibata2016}, see also \citealt{Thomas2016}) which, if generalized to other streams, would have implications on the number of small sub-halos ($\sim 10^{6\mh 9} \Msun$) in the Galaxy. \citet{Garrison-Kimmel2017} proposed that galactic tides can dissolve such halos, and would even deplete the inner Galaxy ($\lesssim 15 \kpc$) from them, when accounting for the tidal effects from the stellar disc (which is ignored when using dark-matter only cosmological simulations).

Theoretically, gaps in streams can also result from interactions with dense baryonic structures (see also \sect{shock}). Using simulations, \citet{Amorisco2016} showed that encounters between a stream and galactic giant molecular clouds would form gaps, comparable to observed ones. Furthermore, \citet{Pearson2017} demonstrated that tidal torques induced by the passage of the Galactic bar could lead to similar results. Therefore, the link between gaps in streams and the dark sub-structures is less direct than previously thought, specially for progenitors of which orbits pass(ed) by the region of the dense disc ($\lesssim 15 \kpc$) and the bar ($\lesssim 3\kpc$).

It has also been found that different dark matter particles, that make different halo shapes, would change the appearance of streams \citep[see examples for several dark matter flavours in][]{Dubinski1991, Yoshida2000, Dave2001}. Streams are thus of prime importance to determine the very nature of dark matter, and its distribution in the outer volumes of galaxies \citep{Koposov2010, Gibbons2014}, and also to test and constrain alternative paradigms. For instance, \citet{Thomas2017, Thomas2017b} modelled the Sagittarius and Palomar 5 streams in the context of modified Newtonian dynamics (MOND, \citealt{Milgrom1983}). They showed that the non-linear nature of MOND, which induces a so-called external field effect \citep{Wu2010}, generates an asymmetric net potential: the potential around L$_1$ is raised with respect to that at L$_2$, by the equivalent of a negative density contribution on top of that of the cluster and the galaxy, which thus as a repulsive effect. Such configuration reproduces the observations of asymmetric streams, with the leading tail being significantly shorter than the trailing one \citep{Odenkirchen2001, Ibata2017}, as well as the high velocity dispersion in the cluster itself (recall \sect{pe}).

Reproducing numerically a real stream remains challenging, due to the size of the parameter space to be explored (orbit, galactic potential, initial conditions of the progenitor) and the non-uniqueness of the model. It also requires a proper description of the (evolving) mass-loss rate of the progenitor which, in a case of collisional systems implies to account for two-body encounters (recall \sect{nbody}). Yet, some level of simplification is usually adopted, like assuming a flux of stars from the Lagrange points of the progenitor centre (i.e. a constant mass-loss rate for the progenitor). This allows simulations to e.g. predict future evolution of observed streams \citep{Dehnen2004}, or estimate local properties of the Milky Way potential \citep{Kuepper2015}. Recent models have started to account for the time-evolution of galactic potential \citep{Carlberg2017}, which introduces a complex, yet necessary, additional degree of freedom when using streams to retrace the assembly history of the Galaxy.

%%%%%%%%%%%%%%%%%%%%%%%%%%%%%%%%
\subsection{Evolution in a fixed, isolated galaxy}

Most of the numerical studies on star cluster evolution taking the galactic environment into account assumed a fixed, constant galactic potential in which the cluster evolves, usually on a circular or an elliptical orbit. While these strong assumptions were first justified by technical and methodological limitations, they remain valid with present-day resources when considering that most of the growth of the Milky Way (in mass, size and in term of the formation of structures like the discs) occurred before redshift unity \citep[see e.g.][]{Behroozi2013}, i.e. in the first $5 \Gyr$ of the Universe. Adopting an idealised description of the Galaxy has allowed for a better understanding of cluster evolution, on which recent studies in more complex and realistic frameworks rely.

Following the analytical study of \citet{Henon1961}, \citet{Hayli1970} presented the first numerical models of star clusters in an external potential mimicking a galaxy. Both established that introducing a galactic potential creates the saddle points and the general features shown in \fig{pe}, which leads to an accelerated mass-loss of the cluster with respect to isolation, as described above. Since then, studies on this topic have aimed at deriving how this mass-loss depends on the galaxy, the cluster, and its orbit \citep[e.g.][]{Lamers2010, Madrid2017}. The commonly adopted approach is to split the problem into individual mechanism, then to considers several combinations of them representative of specific phases of galaxy evolution, and finally to consider the full cosmological context accounting for the formation and evolution of the host galaxy(ies).

%%%%%%%%%%%%%%
\subsubsection{Constant tides: circular orbits}

In the absence of tides, the internal dynamics of a cluster drives its expansion \citep[recall \sect{collisional}]{Henon1965}. However, when considering the galactic environment, the tides slow down this expansion. By introducing a self-similar model of clusters which conserves their average density, \citet{Henon1961} showed that the mass-loss (accelerated by the tides) must be associated with the contraction of the cluster. In other others words, escaping stars remove (negative) energy from the cluster which must therefore re-adjust. \citet{Gieles2011b} pushed forward the formalism of H\'enon and showed that clusters in constant tides experience expansion for about half of their life, followed by contraction. Such a behaviour has been noted in a very large number of simulations. In such models, the evolution of quantities is normalised to characteristic times like the half-mass relaxation time and the dissolution time, which thus remain to be estimated.

Early works showed that the dissolution time of clusters on circular orbits is rather insensitive to their initial structure (e.g. its concentration), but is mostly dependent on their initial mass and the strength of the tides, characterised by the galactic-centric distance and the orbital velocity \citep{Chernoff1990, Takahashi2000, Vesperini1997, Aarseth1998}. \citet{Fukushige2000} then combined simulations and analytical derivations to establish an expression for the dissolution time of clusters on circular orbits around point-mass galaxies, as a function of the excess of energy of stars above the critical energy. They highlighted the complication introduced by potential escapers that delay the mass-loss in a yet not fully understood manner (recall \sect{pe}). Furthermore, the approach of \citet{Tanikawa2010} of setting the tidal radius and varying the slope of the galactic potential (which implies playing with the galactocentric distance) showed the importance of the three-dimensional aspect of tides. This further underlines that, even for clusters on circular orbits, the evolution is not well represented by the sole tidal radius (recall \sect{rt}). Indeed stars experience a three-dimensional acceleration from the galaxy, which is not necessary aligned with the cluster-galaxy axis, i.e. not always toward the Lagrange points. So far, the proposed analytical derivations of a dissolution timescale still fail at reproducing the simulations results \citep{Fukushige2000, Tanikawa2010, Renaud2011, Claydon2017}. Recently, \citet{Daniel2017} used simulations to build a dynamical model for potential escapers (based on a \citealt{Woolley1954} model). Such a model opens new perspectives in understanding the mass-loss rate of clusters and the kinematics in their outer regions.

Furthermore, the escape condition depends on the energy of the star and is thus different for all of them. \citet{Keenan1975} illustrated this by showing that stars on prograde orbits (i.e. with an orbital angular momentum vector roughly aligned with that of the cluster around the galaxy) escape more easily than stars on retrograde orbits \citep[see also][]{Fukushige2000, Read2006, Ernst2007}. This can be understood when considering the centrifugal effect on these stars, which provides an additional positive (respectively negative) contribution to their energy with respect to that of the cluster in the case of prograde (respectively retrograde) orbits\footnote{We note that the same situation if found in interacting disc galaxies: prograde galaxies yield long tidal tails, while retrograde ones only undergo mild disruptions \citep{Duc2013}.}. Over a long evolution, this phenomenon results in the depletion of the star on prograde orbits, and thus induces a net retrograde spin of clusters \citep{Tiongco2016b}. This is particularity noticed in the outer layers, more sensitive to tides, and less to possible disruption by two-body encounters than in the central regions \citep{Baumgardt2003}. The spin-orbit coupling of the cluster is thus an important parameter when studying its evolution and mass-loss. Simulations are however often initialized in the so-called phase-lock (also known as tidal locking) setup, such that the Jacobi surface and the Lagrange points are fixed in the (spinning) reference frame of the cluster \citep[e.g.][]{Heggie2003}. Despite this, the asymmetry of the tidal effects induces differential rotation in clusters as observed \citep[e.g.][]{vandeVen2006} and modelled \citep{Boily2001, Theis2002, Vesperini2014}. This is strongly enhanced around core-collapse when the Coriolis effect alters the inward (in the contraction phase, respectively outward in the re-expansion phase) radial motions of stars by making them prograde (respectively retrograde). \citet{Vesperini2014} found that the layers close to $1\mh 2$ times the half mass radius rotate significantly faster than others. Therefore, the potential initial rotation of clusters inherited from their formation phase is disturbed and possibly erased by the tidal field.

%%%%%%%%%%%%%% 
\subsubsection{Time-varying tides: the example of eccentric orbits}
\label{sec:eccentric}

The non-spherically symmetric potential of galaxies implies that most clusters are on non-circular orbits \citep{Dinescu1999, Casetti2013}. Therefore, this requires studying their evolution in time-varying tides and understanding how it differs from the constant tidal field of circular orbits. The first, simplest step in this direction is to keep the galaxy spherically symmetric, but to set the clusters on eccentric orbits.

The complexity of the problem of time-varying tides lies on the delay between the tidal acceleration undergone by stars and their actual escape from the cluster. Let's consider a cluster on an eccentric orbit, approaching its pericentre. Because of tidal acceleration and two-body relaxation, the cluster expands and the energy of its stars in the outer regions increases (on average). In the same time, the Jacobi surface shrinks. Close to pericentre, some of the stars yield an energy above that of the critical one, and become potential escapers (recall \sect{pe}). Some of them escape rapidly (i.e. within a crossing time), while others do not reach the openings in their Jacobi surface and are trapped in the cluster for a longer period of time.

The tidal configuration is symmetric on both the approaching and receding side of the pericentre such that, if one neglects the mass-loss the cluster experiences around pericentre, both the tidal radius and the Jacobi surface are identical before and after the pericentre. Yet, when moving away from pericentre the cluster has already lost the stars susceptible to escape: the high energy population has already been almost completely depleted. The only remaining stars to be stripped are the potential escapers trapped since before the pericentre, and those which have gained energy since then and thus recently became potential escapers. As a consequence, despite symmetric tides, the mass-loss is highly non-symmetric with respect to the pericentre passage. Furthermore, when receding from the galaxy, the cluster yields a growing Jacobi surface and can potentially re-capture stars previously lost, if they fall back in the cluster potential well \citep[e.g.][]{Webb2013}. (It is however likely that such stars will be among the firsts to escape again, due to their high energy, unless they get trapped in the potential escaper phase, or lose energy through two-body encounters.) One can conveniently picture this process by considering that potential escapers constitute a population buffer, filled from the bound material (by relaxation and tidal acceleration), and emptied through the openings around the Lagrange points of which size (and position, see below) varies along the cluster orbit. Hence, mass-loss only occurs when the emptying mechanism is active (i.e. strong tides), \emph{and} when the buffer actually contains stars to be ejected. Because these two conditions do not share the same timescale, it is likely that they are not always simultaneously met in time-varying tidal fields.

The evolution of the mass of a cluster on an eccentric orbit exhibits the classical ``stair-case'' functional form, with steep mass-loss just before pericentre, and slower evolution elsewhere \citep[e.g.][]{Baumgardt2003}. Depending on how cluster membership is defined (e.g. within the tidal radius or below some energy threshold), the mass of a cluster can slightly increase after the pericentre, when stars that recently escaped are re-captured by the cluster as its Jacobi volume increases.

An additional complexity arises when considering the motion of the Lagrange points in the cluster reference frame. On eccentric orbits, this reference frame is non-inertial, and accelerated with respect to an inertial one (e.g. that of the galaxy). Therefore, the Euler effect must be taken into account, on top of the centrifugal and Coriolis effects already present in the case of circular orbits. By altering the positions where the force from the cluster balances that from the environment, it induces a motion of the Lagrange points, as oscillations around the galaxy-cluster axis \citep[see e.g.][]{Renaud2011}. Therefore, stars that are tidally accelerated toward a Lagrange point at time $t$ could well face a higher potential barrier when arriving at this position at $t+\dd t$, as the Lagrange point has moved away. This increases the population of potential escapers and lower the mass-loss rate, for a given tidal radius.

Although inferring the instantaneous mass-loss of a cluster on eccentric orbits is analytically involved\footnote{Attempts are currently being made, following the perturbation approach from the circular case (Bar-Or et al, in preparation).}, empirical relations have been established to express the total lifetime of clusters, using numerical experiments. \citet{Baumgardt2003} found that the lifetime of a cluster on an eccentric orbit is the same as that of the cluster on a circular orbit passing through the pericentre, but corrected by a linear factor of the eccentricity \citep[see also][]{Webb2014b}. \citet{Cai2016} then found that it exists a circular orbit on which the evolution of the cluster (in terms of mass and half-mass radius) is the same than on an eccentric orbit when averaged over the long term (i.e. for longer than the orbital period). They found the corresponding orbital radius by trial and error, and noted it is neither the semi-major axis of the eccentric orbit, nor the average orbital radius \citep[see also][]{Brosche1999, Webb2014b}. \citet{Cai2016} concluded that the evolution of a cluster on an eccentric orbit can be approximated with that of a cluster on a circular orbit with the same dissolution time, when averaging over the entire cluster lifetime. This conclusion can however not be extended to the cases of non-periodic tides.

%%%%%%%%%%%%%%
\subsubsection{In non-axisymmetric, barred potentials}
\label{sec:bar}

Moving away from spherically symmetric galactic potentials, several works explored the effect of the galactic bar on the evolution of clusters. (The role of the disc and spiral arms is discussed in \sect{shock}.) The mass-loss (or even dissolution) of clusters in these volumes participates in assembling the galactic bulge. To understand how galaxies like the Milky Way form, it is therefore important to pin down where the various stellar populations observed originate from, whether they formed in situ or not, when they reached the bulge, and what is the role of the bar in such process.

The first order effect of a non-axisymmetric potential is to modify the orbits of clusters. \citet{Pichardo2004} showed the presence of a bar induces dispersion in the orbital energy and angular momentum of clusters, that could explain the observed distribution of prograde and retrograde orbits \citep[see also][]{Moreno2014}. On top of such aspects, a bar contributes to the tidal field experienced by cluster and alters their evolution. \citet{Berentzen2012} found that the dissolution time of a cluster in a barred potential is mainly set by the average tidal forcing along the orbit. They showed the bar induces periodic expansion and contraction of the tidal debris, making characteristic morphologies of under- and over-densities that correlates with the orbital period (see also \sect{streams}). Going further (mainly by including collisional aspects in their simulations), \citet{Rossi2015} highlighted the importance of the orbital type in the evolution of clusters within the zone of influence of the bar (i.e. $\lesssim 4 \kpc$, as they found little effect at larger radii). They found that clusters on non-chaotic orbits (i.e. co-rotating, anti-rotating, or with a time-varying rotation with respect to the bar) only experience mild perturbations when passing by the bar, and not a dramatic change in their mass-loss timescale. However, clusters on chaotic orbits undergo strong disruptions at their pericentre passages. This quickly depletes this family of orbits such that \citet{Rossi2015} suggested that such clusters should be rare.

\citet{Martinez2017} showed that tidal compression generated by local cores in the overlapping potentials of the bar and spiral arms (i.e. mainly at the tips of the bar) play a protective role on clusters \citep[see also][for considerations on cluster formation in these areas]{Renaud2015}. They found that clusters in such regimes could survive twice longer than if outside of these structures, at a given galactic radius.

Yet, these studies neglected the formation, evolution, possible destruction and re-formation of the bar itself (see e.g. \citealt{Cole2002}), and the evolution of its strength, which would have required a detailed prescription of the galactic evolution (in particular through galaxy mergers and the accretion of gas, see \citealt{Kraljic2012}).

%%%%%%%%%%%%%%%%%%%%%%%%%%%%%%%%
\section{Tidal shocks}
\label{sec:shock}

As opposed to adiabatic tides, short-lived perturbations (with respect to the internal crossing time of clusters) fall into the shock regime. Encounters with nearby dense stellar and/or gaseous structures generate such tidal shocks. This happens when clusters cross a galactic disc, fly by a spiral arm, or interact with dense gas clouds. During such events, the impulsive perturbation rapidly and strongly accelerates the stars of clusters, shrinks their Jacobi surface and can strip the least bound layers. This can severely damage or even destroy clusters, especially the fragile ones, at the low-mass end of their mass function.

%%%%%%%%%%%%%%
\subsection{Disc crossing}

Some star clusters being observed at high galactic latitudes, their orbits must be inclined with respect to the plane of the disc. This naturally implies that clusters fly through the disc and experience then a short but intense tidal acceleration. \citet[see also \citealt{Spitzer1987}]{Ostriker1972} derived the vertical acceleration induced by a disc (modelled as an infinite plane) on a cluster moving perpendicularly toward it. They noted that the vertical tidal forces experienced by stars of such a cluster point toward the cluster centre. Hence, one speaks of compressive tidal shocks, which increase the accelerations of stars (in the cluster reference frame), and thus constitute a source of (dynamical) heating for the cluster \citep[see also][in the context of destroying dark matter sub-structures]{Donghia2010}. 

In this formalism, one classically neglects the high order moments in the energy input by the shock, and focus on the first order $\ave{\Delta E}$ (i.e. energy shift). However, the quadratic moment $\ave{(\Delta E)^2}$ (i.e. energy diffusion) can be as important as the first order term in the outskirts of the clusters, and even more at smaller radii \citep{Aguilar1988, Kundic1995}. In fact, stars with an energy close the critical energy could gain energy through the energy dispersion term, that could make them unbound \citep[see also][]{Madrid2014}. The second order term could thus govern cluster mass-loss (see also a detailed mathematical derivation in \citealt{Kundic1995}, and an early discussion in \citealt{Spitzer1973}). On top of accelerating the dissolution of clusters \citep{Gnedin1997, Vesperini1997, Gnedin1999c}, the contribution of shocks to the energy of stars can also speed-up its internal evolution, for instance toward core-collapse \citep{Chernoff1986, Gnedin1999b}.

Furthermore, because the energy gain from tidal shocks depends on the position in the galaxy and the distance to the cluster centre (and thus the mass of the individual stars when the cluster is mass-segregated), disc shocking could explain the observed relation between the position of clusters and the slope of their (present-day) stellar mass function \citep[see also \citealt{Capaccioli1993, Djorgovski1993}]{Stiavelli1991}. Disc shocking tends to flatten the stellar mass-function of cluster, by depleting preferentially the low-mass end \citep{Vesperini1997}.

By varying the orbital inclination, \citet{Martinez2017} showed, on the one hand, that clusters with a low maximum orbital latitude (i.e. staying near the disc) cross the disc at a lower speed than those on a more inclined orbit. This results in a lower dynamical heating, a weaker tidal shock, and thus a longer cluster lifetime \citep[see also][]{Webb2014}. On the other hand, clusters with a high orbital inclination cross the disc less frequently, which also results in a longer lifetime. By combining these two effects, \citet{Martinez2017} found that clusters with a maximum altitude of $600 \pc$ above the disc plane yield the shortest lifetime. At large galactic distances (i.e. typically further than the dense stellar disc), \citep{Webb2014}. Orbital inclination, and thus the origin of clusters and the process that put them at high latitude are thus key for their long-term survival.

%%%%%%%%%%%%%%
\subsection{Passages of spiral arms}
\label{sec:spirals}

On top of playing a role on their formation (recall \sect{spiral}), spiral arms can also alter the orbits of clusters. For instance, \citet{Martinez2016} showed that the passage of a spiral arm could lift the clusters to high galactic altitude and contribute to establishing their observed thick distribution above (and below) the disc plane. Furthermore, spirals can have a strong tidal effect on clusters.

\citet{Gieles2007} showed that the heating induced by the slow passage of a spiral arm (with respect to the cluster crossing time, i.e. in the adiabatic regime), is mostly damped. Thus, the broad (stellar) component of a spiral arm has little net effect on clusters crossing it. However, encounters with the thinner dense gas component of spirals lie in the impulsive regime, which could potentially severely damage the clusters. Yet, \citet{Gieles2007} noted that most of the energy gain is converted to high velocity, but only for a few stars in the outer layers of the cluster, and thus does not induce an important mass-loss. Furthermore, they showed that the effect of subsequent passages of an arm depends on the period between successive encounters, which is thus a function of the galactocentric radius (with respect to co-rotation). All together, spiral arms do not constitute a major source of perturbation of massive clusters.

%%%%%%%%%%%%%%
\subsection{High speed encounters with clouds}

In addition to compressive shocks due to disc crossing and the passage of spiral arms, clusters can also experience shocks when encountering local, dense sources of gravitation like giant molecular clouds. Such encounters are expected to be a major player in altering the clusters' properties (or even dissolving them) at high redshift, before the galactic disc is fully formed, and when the baryonic content of galaxies is dominated by gas, more turbulent than in the Local Universe \citep{Genzel2006}, and possibly in the form of massive, dense clumps (at $z \sim 2\mh 3$, see e.g. \citealt{Tacconi2010, Zanella2015}).

\citet{Spitzer1958} considered interactions between a cluster and several clouds along its orbit (in the impulsive regime, but also with an extension to slow encounters), and found analytically that the tidal heating induced by such shocks leads to disruption on a timescale that scales with the cluster density. This was later confirmed with simulations by \citet{Gnedin1999} and \citet{Gieles2006c}, who also found that the dissolution time due to GMC encounters is several times shorter than that from typical adiabatic tides. \citet{Wielen1985} showed however that low-mass gas structures have a limited impact on star clusters, and that only passages of giant molecular clouds significantly affect them. 

The dense gas clouds being in the Galactic disc, cloud-clusters encounters have been proposed as a driver of the disruption of low-mass clusters there, and thus to explain the dearth of old open clusters in the Milky Way disc \citep{Wielen1985, Terlevich1987}.

\citet{Elmegreen2010b} proposed that, because of the hierarchical structure of the ISM (particularly at high redshift), the tidal field felt by a young cluster is dominated by local gaseous sub-structures in the cloud complex where it forms. This situation endures until the cluster drifts away from the complex, i.e. for about $100 \Myr$, or until the gas structure is dissolved. During this time, shocks with dense gas structures participate in shaping the clusters, and can help reproducing the observed mass-age distributions. \citet{Elmegreen2010} pushed this argument further to propose that this mechanism would drive the evolution of an initial cluster mass function of the Schechter type (recall \sect{cmf}), into the observed present-day log-normal shape \citep[see also][]{Kruijssen2015}. 

\begin{figure}
\includegraphics[width=\columnwidth]{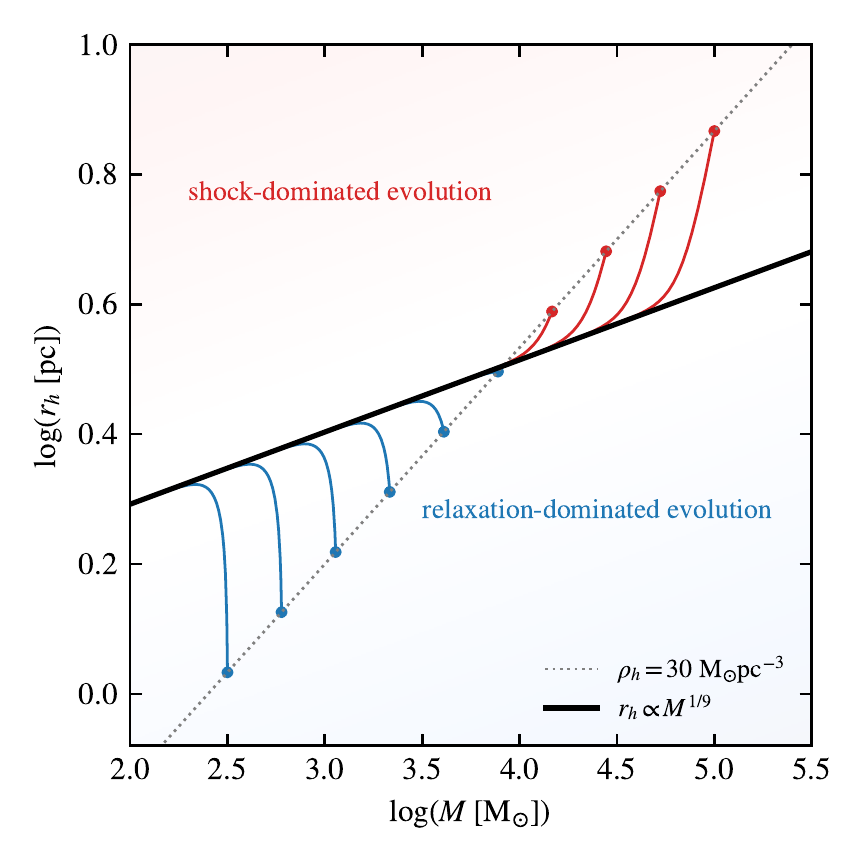}
\caption{Adaptation of the figure 3 of \citet{Gieles2016}, showing evolutionary tracks in the mass-size diagram of clusters subject to repeated shocks with giant molecular clouds in a Milky Way-like ISM. The formalism and all numerical values are identical to that of \citet{Gieles2016}. The solid black line marks the equilibrium relation between the half-mass radius and the mass: $r_h \propto M^{1/9}$. The dots represents clusters of different masses but same half-mass density. Clusters initially denser than the equilibrium relation lie in the relaxation-dominated regime. Internal dynamics first make them expand, toward the equilibrium relation. Conversely, clusters initially less dense than the equilibrium are more sensitive to tidal harassment, which truncates them and thus reduces their size, toward the equilibrium relation.}
\label{fig:gieles}
\end{figure}

\citet{Gieles2016} noted that repeated shocks do not yield a cumulative effect: tidal shocks are self-limited \citep[see also][]{Gnedin1999b}. While the first shock does potentially significantly affect the clusters properties by stripping the outermost stars as discussed above, it takes the cluster a long time to replenish these layers (through two-body relaxation and adiabatic tidal acceleration). In the mean time, it is little sensitive to repeated shocks, and its evolution is thus mainly driven by secular processes. As a consequence, clusters that are not fully destroyed by a shock are likely to survive subsequent shocks of similar (and lower) amplitude, over a period of a few relaxation times, until they expand again. The result of combining shocks and relaxation is an equilibrium in the mass-size relation, as shown in \fig{gieles}. \citet{Gieles2016} showed that clusters denser than this equilibrium lie in a relaxation-dominated regime that makes them expand, while those less dense are more sensitive to tides that reduce their mass and their size, until they reach the equilibrium. The weak dependence of the size on the mass could explain the observations of a typical radius of $\approx 3 \pc$ for the low-mass and young clusters (i.e. before the secular tides dominate their evolution, \citealt{Larsen2004}).

\citet{Gnedin1997} modified the formalism of \citet{Spitzer1958} of point-mass clouds to consider extended objects to study tidal shocks induced by the galactic bulge \citep[see also][]{Aguilar1985, Weinberg1994}. They concluded that, when integrating internal evolution, secular tides and shocks, the dissolution time of clusters is such that most clusters have already dissolved, and that their remnants account for a significant fraction of the old stellar components of the Galaxy (i.e. the bulge and the halo).

In all the above-mentioned studies, the importance of tidal shocks has been established in the context of the present-day Milky Way. However, changes in the properties of the ISM (in particular the distribution and density of molecular clouds), the formation of the disc, the bulge and the spiral arms must be taken into account when inferring the evolution of clusters at high redshift. Unfortunately, there are little observational constraints on this, and the exact role of tidal shocks in shaping the cluster population is yet to be established.

%%%%%%%%%%%%%%%%%%%%%%%%%%%%%%%%%%%%%%%%%%%%%%%%%%%%%%%%%%%%%%%%
\section{Evolution in cosmological context}

When studying the evolution of old clusters ($\gtrsim 6\mh 8 \Gyr$), it becomes crucial to account for the evolution of their context. Studying the dynamics of clusters for several Gyr in a static galactic potential, as representative of a present-day galaxy it can be, misses the important aspects of galaxy evolution and the variations of the tidal field it induces. The formation and evolution of galactic substructures (bars, spirals), the secular growth of the galaxy itself, interactions and mergers with neighbour galaxies and satellites must be considered. To date, these aspects have been mostly ignored, but numerical methods allowing to couple galaxy and clusters simulations have recently opened new possibilities (recall \sect{nbodyenv}). 

%%%%%%%%%%%%%%
\subsection{Adiabatic growth}
\label{sec:adiabatic}

The first step when accounting the evolution of the environment of star clusters is to consider the adiabatic, secular evolution of the galaxy itself. This encompasses its growth form the accretion of dark matter and gas, the conversion of this gas into stars, the (possible) formation of the discs, the bulge, the spiral arms, the bar, the stellar and the gaseous halo etc. Numerical and observational studies on galaxy formation provide constraints on these aspects, but their exact causes, when these events occurs, and at which speed remain poorly known \citep[see a review in][]{Freeman2002}. Furthermore, they vary from galaxy to galaxy and thus, modelling them to study the populations of clusters requires exploring a large parameter space.

\citet{Renaud2015c} considered the adiabatic growth of the dark matter halo (in mass and in size) across cosmic time (neglecting the baryonic components) to infer how such a time-varying tidal field affects star clusters. They found that, because most of the growth of the galaxy is done at high redshift ($z \gtrsim 1\mh 2$), the details of the growth have close to none influence on clusters and their tidal debris, and that only the final state of the halo sets their properties (mass, size, morphology) at $z=0$. Since the Milky Way has not experienced any major merger over the last $6 \mh 9 \Gyr$ \citep{Deason2013, Ruchti2014}, these conclusions can be applied to clusters over this period in this galaxy, but the evolution of the baryonic structures must be accounted for in the case of clusters orbiting in the innermost regions. In particular, the formation of the bar and the evolution of its strength could alter the evolution of clusters (recall \sect{bar}). The same could be concluded about the formation of (transients, see e.g. \citealt{Roskar2012}) spiral arms, but \citet{Gieles2007} showed the inefficiency of spirals in altering clusters (\sect{spirals}). Furthermore, the disc(s) are supposedly formed at $z \approx 2$ \citep{Wyse2001}, i.e. close to the end of the period during which the growth is dominated by major mergers, i.e. in a violent, non-adiabatic phase \citep{Renaud2017}. The evolution of the importance of disc crossings thus remains difficult to address without a clear understanding of the processes forming the stellar and gas discs \citep{Mo1998, delafuentemarcos2009}.

Furthermore, the mechanism of radial migration within the disc (induced by the bar and the spirals, \citealt{Minchev2010, Aumer2017}) is expected to change the tidal field of the Galaxy, but also to modify the orbits of the disc clusters \citep[e.g.][]{Grand2012, Martinez2017} and possibly speed up or slow down their dissolution \citep{Fujii2012}, depending on the origin and destination of their migration.

%%%%%%%%%%%%%%
\subsection{Accretion of satellite galaxies}

On top of the adiabatic growth, galaxy also accrete material in impulsive ways, by merging with their neighbours. Let's consider a star cluster in a dwarf galaxy, which approaches a bigger galaxy and becomes one of its satellites. Dynamical friction induced by the dark matter halo, the hot gas corona and any dense component of the main galaxy makes the dwarf lose orbital energy and spiral in (\sect{df}). In the process, the dwarf is tidally stripped by the main and the cluster eventually escapes the dwarf (or the dwarf dissolves completely). The cluster then experiences a transition from (mainly) the tidal field of the dwarf to (mainly) that of the main galaxy. Before the accretion, the velocities of the stars of the cluster are set according to the potential of the dwarf. Equilibrium is not necessarily reached, but for simplicity, we consider here that the cluster is virialised. The transition to a different tidal field implies that the velocity of the stars are likely not adequate to the new potential to maintain virialisation, and the cluster needs to adjust to its new environment (see \citealt{Hills1980, Boily2003, Boily2003b, Baumgardt2007} for a comparable formalism but in a different context).

This situation is particularly interesting when the cluster initially lies in the core of the dwarf, i.e. in compressive tides. Such a tidal configuration increases the energy of the stars while limiting the mass-loss of the cluster (with respect to classical extensive tides, see \citealt{Renaud2011}). When switching to the extensive tidal field of the main galaxy, stars likely yield an excess of kinetic energy with respect to the new gravitational potential, i.e. the cluster is super-virialised and it expands (\citealt{Bianchini2015, Webb2017b}, see also \citealt{Miholics2014, Miholics2016} for a comparable evolution but without compressive tides). Such an expansion could possibly explain the properties of the extended clusters (also known as faint fuzzy clusters, recall \fig{masssize}).

This expansion is much faster than one a cluster which would have always been in the tidal field of the main could experience, and likely makes the cluster filling its Jacobi volume (or, in other terms, be tidally-limited). Therefore, the cluster from the dwarf could be more extended than one from the main galaxy if the latter is not (yet) tidally limited \citep{Webb2017}, i.e. if it is still in the expansion phase of its evolution \citep{Baumgardt2010, Gieles2011b}. However, if the cluster from the main were already in the tidally-limited phase, the one from the dwarf would have a comparable size, set by the Jacobi surface \citep{Bianchini2015, Miholics2014, Miholics2016}. Another mechanism, likely linked to the formation of these clusters (e.g. primordial binaries and/or gas expulsion), should then be invoked to explain observational data \citep[see e.g.][]{Zonoozi2011, Zonoozi2014, Leigh2013, Leigh2015}. In any case, the transition from compressive to extensive tides induces an important mass-loss, and possibly the dissolution of the cluster.

%%%%%%%%%%%%%%
\subsection{Interacting galaxies and major mergers}
\label{sec:evolmerger}

Interactions and mergers between galaxies of comparable masses induce a number of transformations of the properties of the progenitors and of their cluster populations. One effect is to efficiently form clusters, in particular massive ones, during the possible starburst episode(s) such encounters trigger (recall \sect{ymc}).

The role galaxy encounters play on the evolution of existing clusters is dual. On the one hand, galactic tides, inflows, outflows, shocks and mergers modify the mass distribution of the galaxies \citep[see a review in][]{Duc2013}, and thus change their tidal fields. In particular, many (if not all) interactions trigger a significant enhancement of compressive tides, over large volumes, thus changing the very nature of the tidal field \citep{Renaud2008, Renaud2009}. Such compression is however short-lived ($30 \mh 100 \Myr$, \citealt{Renaud2009}) and is thus not expected to significantly alter the evolution of pre-existing star clusters (but see \sect{ymc} on the enhancement of their formation). On the other hand, galactic interactions significantly alter the orbits of their constituents, which indirectly modify the tidal fields in which they evolve. The evolution of clusters in interacting galaxies is thus non-trivial, and highly depends on their position within the system.

\citet{Renaud2013} conducted a survey of star cluster simulations, imposing different tidal fields derived from a simulation of a merger. They found that the collisions themselves have little effects on the existing clusters, because the tides are too short-lived and too weak to affect dense clusters. However, the modifications of the clusters' orbits significantly alter their long-term evolution. For instance, clusters ejected into the galactic tidal debris experience much weaker tides than before the interaction, which increases their lifetime. On the contrary, clusters dragged toward the galactic central regions lose mass faster after the interaction than before. Because the tidal debris represent a small fraction of the total mass of a merger (and originate from low density environments in the progenitors), the latter case is much more frequent than the former.

A full description of the tides, especially the shocks induced by small, dense gas structures requires pc-resolution (e.g. on the softening of the gravitational potential), which was not the case in \citet{Renaud2013}, and remains rare in modern galaxy simulations. As a consequence, the tidal shocks were not properly accounted for. It is however likely that the collision of two galaxies increases the number of shocks, including disc crossing (due to their relative inclination), and the relative velocity of clusters with respect to dense material, which increases the energy input \citep[recall the formalism of][see also \sect{shock}]{Ostriker1972}. The re-distribution of orbits during interaction events has been proposed by \citet{Kravtsov2005} and \citet{Elmegreen2010} as an efficient way to send clusters to weak tidal environments (e.g. at high altitudes above the disc remnant), and ease the disruptions they undergo. This would preserve a fraction of the low-mass clusters, while the most massive ones are supposedly dense enough to survive most of the tidal harassment.

%%%%%%%%%%%%%%
\subsection{Full cosmological context}

To establish the relative importance of all the above-mentioned mechanisms on the evolution of star clusters, one must study the global cosmological context of their evolution, i.e. the full formation and evolution of the galaxy(ies) in which they evolve. Such a holistic approach requires to resolve pc-size perturbations (e.g. dense clouds) over cosmological volumes, which is still out-of-reach of large-volume simulations, but can be obtained in zoom-in setups. Present-day works thus focus on either a large-volume with a statistical diversity of galaxies, or on a narrower space but at higher resolution \citep[see e.g.][in the context of globular cluster formation]{Kravtsov2005, Li2017, Kim2018}.

\citet{Rieder2013} derived the evolution of tidal fields along several orbits in dark matter only simulations describing two different assembly histories (i.e. different merger trees) of Milky Way size halos. They then followed the evolution of clusters on these selected orbits and found rather smooth, continuous mass-losses, with no abrupt changes (except those from pericentre passages on eccentric orbits, recall \sect{eccentric}), but still accelerated by major merger events. Since the mass of the host galaxy directly influences cluster mass-loss, a Milky Way assembled by many minor mergers inherits clusters that have experienced (on average) less severe mass-loss than one assembled by a few major mergers. Therefore, despite forming comparable halos at $z=0$, \citet{Rieder2013} found that their two assembly histories have different imprints on clusters. 

\citet{Renaud2017} included the baryonic component and a prescription for star formation in their simulation of the assembly of the Milky Way and also tracked the evolution of the tidal field. Without modelling the clusters themselves, they found that clusters accreted by the Milky Way (i.e. formed in another galaxy) experience (on average) weaker tides than those formed in situ, but that the overall behaviour these evolutions remain similar. They are marked by phases in galaxy evolution like the early assembly by major mergers, the formation of the disc, and the late adiabatic growth. On average, all types of clusters experience tides significantly weaker in the past, even those formed in the central parts of dwarf satellites. The strength of the tidal field average over the cluster population increases the most when the galactic disc forms, that the average galactocentric radius of cluster becomes smaller (due to more clusters forming in the inner regions of a massive galaxy). But again, the lack of numerical resolution (on the softening length of the gravitational force) forbids to infer the effects of small-scale structures and thus the tidal shocks.

The most recent simulations can reach resolutions of a few parsecs, meaning that the gravitational effects of external molecular clouds can be (barely) reproduced and that (some) tidal shocks can be accounted for (Li and Gnedin, in preparation). Yet, it remains difficult to disentangle signal from numerical artefacts, especially because of the use of smoothing kernel modifying the local gravitational field, and thus setting a minimum scale to derive gravitation-related quantities.

In conclusion, the secular tidal histories of clusters are rather well understood now, despite some uncertainties on e.g. disc formation. However, the early evolution of clusters and the dissolution of the least resistant ones is yet to be quantified. This suffers from the lack of observational constraints on galaxy formation and their ISM at high redshift, and on the formation of the clusters themselves. Although the physical processes described above are rather well understood individually, their non-linear combination and its exact effects on clusters is still to be established.

%%%%%%%%%%%%%%%%%%%%%%%%%%%%%%%%%%%%%%%%%%%%%%%%%%%%%%%%%%%%%%%%%%%%%%%%%%%%%%%%%%%%%%%%%%%%%%%%%%%%%%%%%%%%%%%%%%%%%%%%%%%%%%%%%%%%%%%%%%%%%%%%%%%%%%%%%%%%%%%%%%%%%%%%%%%%%
%%%%%%%%%%%%%%%%%%%%%%%%%%%%%%%%%%%%%%%%%%%%%%%%%%%%%%%%%%%%%%%%%%%%%%%%%%%%%%%%%%%%%%%%%%%%%%%%%%%%%%%%%%%%%%%%%%%%%%%%%%%%%%%%%%%%%%%%%%%%%%%%%%%%%%%%%%%%%%%%%%%%%%%%%%%%%
%%%%%%%%%%%%%%%%%%%%%%%%%%%%%%%%%%%%%%%%%%%%%%%%%%%%%%%%%%%%%%%%%%%%%%%%%%%%%%%%%%%%%%%%%%%%%%%%%%%%%%%%%%%%%%%%%%%%%%%%%%%%%%%%%%%%%%%%%%%%%%%%%%%%%%%%%%%%%%%%%%%%%%%%%%%%%
\part{Open questions}

Most of the still open questions in this field originate from our lack of understanding of the relative importance of formation and evolution processes, in particular for old objects like globulars. This nature versus nurture puzzle hinders our attempts to connect well-observed present-day properties and objects with their equivalent in the early Universe. I identify below, in no particular order, a few open questions for which the community has started to provide answers, with various degrees of completeness.

%%%%%%%%%%%%%%%%%%%%%%%%%%%%%%%%%%%%%%%%%%%%%%%%%%%%%%%%%%%%%%%%%
\section{Are YMCs young globular clusters?}
\label{sec:ymcglobs}

The similar masses between YMCs formed in the present-day Universe (e.g. in starbusting galaxy mergers, recall \sect{ymc}) and the globular clusters formed in the early Universe (\sect{gc}) suggest that the buildings of two populations share a number of physical properties and mechanisms. Yet, the question ``Are YMCs young globular clusters?" can be interpreted in different ways. 

First, one could rephrase it as ``will the present-day YMCs observed in $\sim 12 \Gyr$ from now resemble the present-day globulars today?''. Assuming that YMCs do resemble young globulars (which is a strong assumption discussed below), the answer only depends on the evolution of the two populations, and is therefore clearly negative. Indeed, the early Universe hosts physical conditions significantly different than the present-day and the future Universe, in term of density, galaxy interaction and merger rate, gas content, etc. Therefore, the tidal effects (impulsive and secular) the YMCs will experience in the next few Gyr are necessarily different than those experienced by the globular clusters in the past few Gyr. First, the consumption of gas for star formation will progressively decrease the importance of tidal shocks (\sect{shock}). Second, the expansion of the Universe will continue to slow down the merger rate and thus the increased mass-loss of the majority of clusters (\sect{evolmerger}). Finally, the fact that galaxies are and will be more massive than before implies that their (average) tidal field will be stronger than it has ever been \citep[see an illustration in][their figure 10]{Renaud2017}. The combination of these factors make unlikely that the environment-induced mass-loss will proceed the same way in the next $12 \Gyr$ than it has in the last $12 \Gyr$. This statement must however be balanced by the fact that the most massive clusters are only weakly affected by tides, and thus the differences between the two populations (YMCs and globulars) after $\sim 12 \Gyr$ of evolution could be moderate.

Another way to interpret the question above-mentioned could be to take the opposite perspective, i.e. ``were the present-day globulars $12 \Gyr$ ago resembling the present-day YMCs today?''. The interest of doing so is seeking whether one can observed the relatively nearby YMCs to better understand the formation of globular clusters, which remains out of reach observationally (but see the recent use of gravitational lensing to probe young globulars at redshifts up to 6, \citealt{Vanzella2017, Vanzella2017b}). In this case, our lack of understanding of the process of star cluster formation, in particular for the most massive objects makes this question more complicated than the previous one. 

Some of the differences between the formation sites of massive clusters at high redshift and in the Local Universe are the chemical composition (including metallicity and dust content) and possibly the turbulence. \citet{Clark2008} suggested that dust cooling dominates over H$_2$ at extremely low metallicities ($\lesssim 10^{-5} \U{Z_{\odot}}$) and high densities ($\gtrsim 10^{12} \cc$, see also \citealt{Dopcke2013}). Note however that this in not the case in lower densities \citep[see a discussion in][]{Glover2014}. Therefore, it seems critical to distinguish the formation of the cloud (for which cooling at low densities matters) from the formation of the cluster itself (at much higher densities). Different cooling could result in different fragmentation and hierarchical structure of the star forming regions, through e.g. the formation of sub-clusters that would eventually merge to build a massive system. The associated physics likely relates to the observed minimum metallicity detected in Milky Way globulars ([Fe/H] = -2.5, \citealt{Harris1996}), of which origin remains to be understood.

Furthermore, turbulence is much stronger at high redshift, due to the larger amount of gas and the repeated pumping by kpc-scale phenomena (interactions, mergers and gas accretion, \citealt{Wise2007, forster2009, Swinbank2011}). However, the absence of measurement of turbulence of the dense gas at very high redshift ($z \gtrsim 3$) makes it difficult to assess whether the formation sites of the first globular clusters are strongly more turbulent than that of YMCs in galaxy mergers for instance. With such variations of metallicity and turbulence, one could question the universality of the stellar IMF, which is of prime importance for many aspects of astrophysics, and in particular for the internal dynamics of clusters (segregation, binarity, stellar mass-loss), the formation and evolution of binary stars, the feedback processes (in term of intensity and timing, due to possible delays for supernovae, see \citealt{Walcher2016, Zapartas2017}) and thus the subsequent star formation, gas removal and early shock-driven evolution of the clusters. Analytical models suggest that a compression-dominated turbulence (like that measured in mergers, \citealt{Renaud2014b}, and expected to be stronger and longer-lived in the early Universe, \citealt{Renaud2017}) would lead to a bottom-heavy IMF \citep{Chabrier2014}. Yet, how such a turbulence cascades and decays from its galactic injection scale down to that of the pre-stellar cores is still unknown (due to the enormous range of scales that must be captured self-consistently), and it remains possible that some mechanism(s) like early feedback or magnetic fields could erase the imprint of the large scales in the sub-parsec regime. Despite uncertainties, the observed variations in massive elliptical galaxies and dwarfs tend to show that the environment does alter the IMF \citep[see e.g.][]{Cappellari2012, Geha2013, Dib2017}.

Remains the early evolution of these clusters. The self-limited aspect of tidal shocks and the strong dependence on the properties of the ISM \citep[see also \sect{shock}]{Gieles2016} suggests yet another evolution in the dense, gas-rich, clumpy ISM in which globular formed, compared to the significantly smoother and more diffuse ISM of the local Universe, even in the formation sites of YMCs. Uncertainties on the density of gas structures and their velocity dispersions leave the problem without a clear conclusion.

The importance of tides on shaping the cluster mass-function has been demonstrated qualitatively and quantitatively \citep[e.g.][]{Elmegreen2010b, Rossi2015b, Gieles2016}, but as mentioned before, the physical conditions of cluster formation at high redshift are not fully known, and it is possible that the initial mass function of present-day globulars differs from that of present-day YMCs. 

All the points mentioned above seems to indicate different formation and evolution schemes for the globular clusters and the YMCs, but many uncertainties and the lack of quantitative description still forbids to conclude. Probing the early stages of cluster evolution at high redshift with e.g. the forthcoming James Webb Space Telescope will certainly provide insights on this question. 

%%%%%%%%%%%%%%%%%%%%%%%%%%%%%%%%%%%%%%%%%%%%%%%%%%%%%%%%%%%%%%%%%
\section{Zone of avoidance}

Mass - size diagrams of stellar systems show what \fig{masssize} already hints in the luminosity - size plane: no stellar systems (galaxies and star clusters) is found below an oblique line setting a maximum mass as a function of projected size. This corresponds to a maximum surface density of $\sim 10^5 \Msun\pc^{-2}$, which defines the boundary of the so-called zone of avoidance\footnote{not to be confused with the region behind the Milky Way centre, where observations suffer from strong extinction \citep{Kraan2000}.} \citep[see e.g.][]{Norris2014}, irrespective of the type of object considered (globular clusters, nuclear clusters, ultra-compact dwarfs, compact elliptical galaxies, elliptical galaxies). The existence of the common boundary in objects ranging from central ellipticals to compact globulars tends to suggest a common physical process setting such an upper surface density limit. \citet{Hopkins2010b} underlined that the limit is detected in surface density and not volume density (because its spans a wide diversity of objects), which already provides a constrain on the physical process(es) responsible for it.

Naturally, the first idea coming to mind invokes regulation at formation. Using a timescale argument, \citet{Hopkins2010b} noted that feedback from supernovae go off too late to set the maximum surface density. However, wind and radiation from massive stars are active early enough to account for the observed zone of avoidance in young objects. The radiative pressure from young massive stars on a dusty medium (i.e. optically thick) would then be sufficient to balance collapse and could possibly set the maximum surface density. As noted by the authors, this explanation suffers from several shortcuts: first, dense systems experience relaxation that erase within a few relaxation time any structure, in particular in the densest regions. Therefore, if set at formation, the maximum surface density will be erased for old objects. Furthermore, the rate of erasement would depend on the collisional nature of the system, and would then be different in dense globular cluster and in extended galaxies, making a common zone of avoidance rather unlikely. Another problem of this theory is that objects formed in low-metallicity ISM would be surrounded by an optically thinner medium, making radiative pressure less efficient.

Furthermore, the formation mechanisms, epochs, environments and timescales of these objects differ, making the existence of such a common limit somewhat surprising. For instance, the formation of a globular cluster takes place in a small region of space and in a relatively short period of time. On the contrary, that of massive elliptical galaxies results from the successive assembly of galaxies with different sizes and masses, formed in different regions and at different epochs, thus in contrast with the rapid, in-situ formation mode of globular clusters. As a result, one would not expect these two processes of in-situ formation and assembly would generate a common zone of avoidance \citep{Norris2014}. The same line of arguments holds for nuclear clusters, which also (at least partly) assemble by merging pre-existing structures (see \sect{nc}).

To date, the physical origin(s) of the zone of avoidance and, most importantly, the reason making it relevant for a broad range of objects with different formation mechanisms and epochs, remains unidentified, which further demonstrates our lack of understanding of cluster and galaxy formation.

%%%%%%%%%%%%%%%%%%%%%%%%%%%%%%%%%%%%%%%%%%%%%%%%%%%%%%%%%%%%%%%%%
\section{Maximum cluster mass and star formation rate}
\label{sec:mc}

The observational evidence that starbursting mergers host the formation of massive star clusters (recall \sect{ymc}) tends to suggest a relation between the galactic SFR and the mass of the most massive young cluster \citep{Kravtsov2005, Whitmore2010}. \citet{Larsen2002} conducted an analysis of cluster luminosity functions confirming this idea, and in particular in galaxies with lower SFRs than starbursts \citep[see also][]{Larsen2000, Billett2002, Larsen2004, Cantiello2009}. \citet{Johnson2017} also reported a comparable relation, more precisely between the surface density of SFR ($\Sigma_\textrm{SFR}$) and the characteristic truncation mass of the Schechter mass function ($M_c$, see \eqn{schechter}), i.e. a proxy for the mass of the most massive cluster (recall \sect{cmf})\footnote{Recall however that the high-mass end of the cluster mass function suffers from low number statistics \citep[see e.g.][]{Chandar2017}.}. They considered a sample of four galaxies: M~31, M~83, M~51 and the Antennae, i.e. a green-valley galaxy with a low SFR, a grand-design star forming spiral, a minor merger and a starbursting major merger, and found a power-law of index close to unity connecting the two quantities (see their figure 6). The diversity of galaxies considered then suggests that the physical process(es) setting the mass of the most massive cluster depends on the global star formation activity, but not on the physical conditions setting star formation itself. For instance, the existence of two regimes of star formation (discs and mergers, see e.g. \citealt{Daddi2010b} and \citealt{Genzel2010}) demonstrates that a given SFR can have different physical origins, which thus questions the origin of the relation observed.

Simulations and analytical models show that these two regimes originates from a different conversion from the local 3D gas volume density ($\lesssim 100 \pc$, more or less directly linked with the SFR), to the large scale, galactic surface density of gas ($\gtrsim 100 \pc$, see \citealt{Renaud2012} and \citealt{Renaud2014b}). For a given surface density of gas, a starbursting merger galaxy yield a different turbulence, which compresses its gas more efficiently\footnote{Efficiency should not be confused with rate!} into dense clumps (i.e. potential sites for cluster formation), which in turn creates an excess of star formation, with respect to an isolated galaxy. The distributions of gas densities (i.e. the very organisation of the ISM) is thus not directly linked to the large scale surface density of gas.

Let's now change our perspective and consider for instance a gas-rich disc galaxy at $z \sim 2$ yielding a SFR of $\sim 50 \Msunyr$, with $\Sigma_\textrm{SFR} \approx 0.3 \Msunyr \kpc^{-2}$. Such a galaxy is comparable to those in the BzK sample of \citet{Tacconi2010}. Now, let's consider a local starbursting merger, with the same SFR and $\Sigma_\textrm{SFR}$, which could be considered as a LIRG \citep{Kennicutt1998}. Because the high redshift galaxy lies in the normal star formation mode while the merger is in starburst mode, the efficiencies of their star formation (i.e. the inverse of their gas reservoir depletion time) are not the same: for the same $\Sigma_\textrm{SFR}$, the LIRG can have a surface density of gas (typically) one order of magnitude lower than that of the BzK \citep{Daddi2010b}. (See also \citealt{Bournaud2015} showing different physical triggers leading to comparable SFRs and $\Sigma_\textrm{SFR}$'s in clumpy discs and in mergers.)

Furthermore, in the merger simulation of \citet{Renaud2015}, the two consecutive galaxy interactions\footnote{The first passage is rather distant and rapid, while the second (because of dynamical friction experienced during the first) is more penetrating and marks the onset of final coalescence.} yield comparable SFRs and $\Sigma_\textrm{SFR}$'s, but very different cluster formation rates (see their figure 11). The reason for this is again the differences in the physical processes setting the organisation of the ISM and triggering the starburst (tidal and turbulent compression, shocks and nuclear inflows, recall \sect{ymc}).

Although these points do not inform us directly about the most massive cluster per se, it would be rather surprising that masses of the clusters formed in the two sets of physical conditions (i.e. the two galactic types, with two different ISM structures) would be the same, as the relation observed by \citet{Johnson2017} suggests. Solving this paradox requires a better understanding of the shortcuts followed in the reasoning above. In particular, does the most massive cluster form in the most massive gas clump? Do, for some reason, the changes in the structure of the ISM only affect the bulk of the star forming regions and not the most massive ones? The hierarchy of the ISM and the potential fragmentation of clouds into sub-structures should be taken into account, which requires a scale-dependent description of instabilities accounting for external effects (tides, shear etc., see e.g. \citealt{Jog2013, Jog2014}). 

A convenient but unsatisfactory, solution lies in the observational errors reported by \citet{Johnson2017}, which allows for different scaling relations between isolated galaxies and mergers. A larger galaxy sample is needed to conclude on this point. If the relation is confirmed, it would then be interesting to understand how it would behave at high redshift, at the formation epoch of globulars, in line with questions raised in \sect{ymcglobs}, but observational limitations makes this task out of our current and short term reach. 

%%%%%%%%%%%%%%%%%%%%%%%%%%%%%%%%%%%%%%%%%%%%%%%%%%%%%%%%%%%%%%%%%
\section{The halo mass - cluster mass relation}

Estimating the total mass of a galaxy (i.e. including the dark matter component) is key in many dynamical studies, but represents an observational challenge. Usual techniques are based on rotation curves \citep[e.g.][]{Sofue2001}, the detection of X-ray gas in giant elliptical galaxies \citep[e.g.][]{OSullivan2004}, the kinematics of clusters \citep[e.g.][]{Romanowsky2009} and stellar streams \citep[e.g.][]{Kuepper2015}, and weak and strong lensing of background objects \citep{Hoekstra2005, Ferreras2005, Mandelbaum2006}. However, each of these approaches is often restricted to a particular galaxy type, which introduces different biases. 

\begin{figure}
\includegraphics[width=\columnwidth]{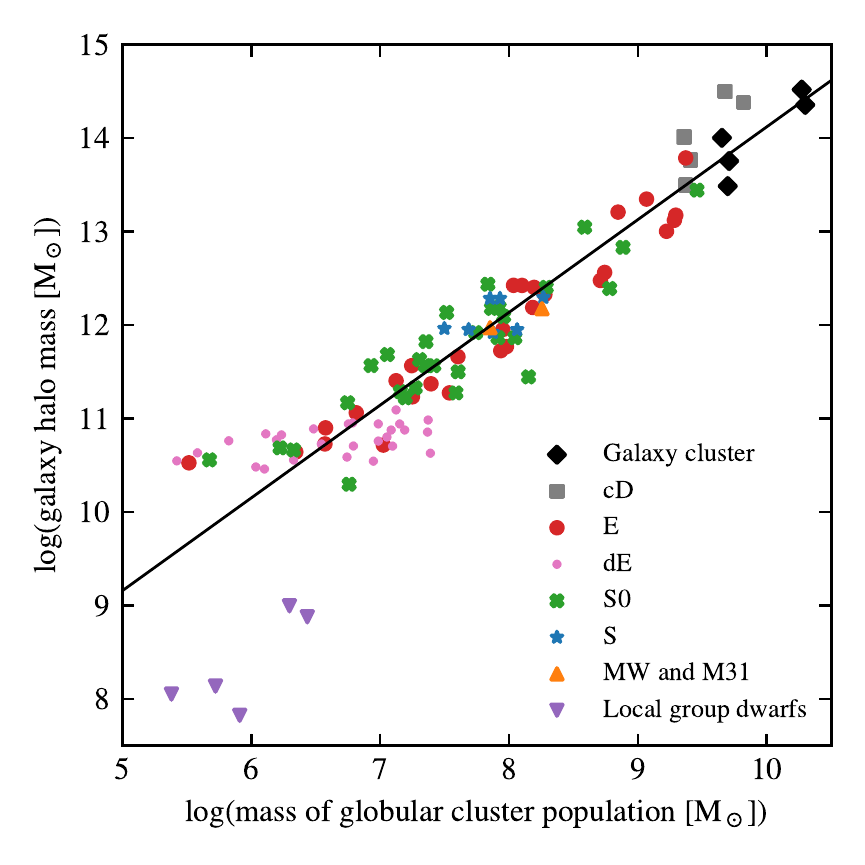}
\caption{Correlation between the mass of the globular cluster population and the halo mass, over a range of galaxy types: galaxy clusters, central dominants in galaxy clusters (cD), ellipticals (E), dwarf ellipticals (dE), lenticulars (S0), spirals (S), the Milky Way (MW) and Andromeda (M31), and a few local dwarf galaxies. The line represents a power-law of slope unity. Data points from \citet{Spitler2009}.}
\label{fig:spitler}
\end{figure}

\citet{Spitler2009} proposed a method to derive the total mass of the galaxy from its globular cluster population. They showed empirically that the mass of the population scales linearly with the total mass of the galaxy (including baryons and dark matter), almost independently of the galactic mass or morphological type: $\approx 0.007\%$ of the total galactic mass is found in the form of globular clusters (see \fig{spitler} and also \citealt{Blakeslee1997}, \citealt{Harris2017}). However, a deviation from this relation is observed at the low mass end (galactic mass $\lesssim 10^9 \Msun$), i.e. in the regime of dwarf ellipticals. The reason for this is likely due to small number statistics, as these galaxies counts only a handful of globular clusters\footnote{The authors argued that the discrepancy noted for the local dwarfs originates from uncertainties in the conversion of the measured velocity dispersions to total mass.}. The relation is otherwise remarkably tight for more massive galaxies, indicating a strong link between the global properties of the galaxies and the formation and evolution of its globular clusters.

Note that no such relation exists between the mass of the globular cluster population and the stellar mass of the galaxy. (It however might have been the case at the epoch of formation, if the cluster specific frequency was universal, before universality got erased by evolution of clusters in a diversity of galactic environments, see \citealt{Zaritsky2016, Elmegreen2017}.) This derives from the fact that the stellar mass does not correlates with the total mass across galactic types, but remains intriguing as one would expect a tighter relation between baryonic components than between clusters and dark matter. The physical reason remains to be understood.

In this context, the enhanced formation of massive clusters in galaxy mergers (see \sect{ymc}) raises a problem. Let's consider the merger of two equal-mass gas-rich galaxies. It is reasonable to neglect the galactic mass-loss due to tidal ejection (as most of the tidal debris will eventually fall back onto the remnant), and thus to consider that galactic mass is doubled during the merger process. However, the physical conditions encountered during the interaction lead to the formation of massive clusters, which add up to the existing population of clusters from the two progenitors which thus more than doubles. As a result, the cluster mass-halo mass relation should break, in particular in central galaxies of clusters which have likely experienced (at least) one merger event. This paradox can be solved in two ways.

The first relies on cluster evolution: by reducing the mass of cluster system by the same amount that what is created, the galaxy would stay on the relation. The increased mass of the merger remnant relative to that of the progenitors does strengthen the tidal field experience by the clusters and accelerate their dissolution. This enhanced mass-loss of clusters in mergers has been noted numerically, due to more frequent tidal shocks, and from the secular, long term tidal field (recall \sect{evolmerger}). This effect has a more dramatic impact for low-density and/or low-mass clusters which are more sensitive to tidal harassment than their massive counterparts. Thus, a large number of small clusters should be destroyed to balance the formation of YMCs during the interaction events.

The second option to solve the paradox is based on cluster formation. The above-mentioned problem only holds if one considers that the young massive clusters are the analogues, or the progenitors, of globular clusters. If one lifts this hypothesis (see \sect{ymcglobs}), then the ageing massive clusters could not resemble present-day globulars, and thus would not be accounted for in the mass of the globular cluster population. A merger would then have a different impact on the cluster mass, such that the cluster mass - halo mass relation could still hold.

Exploring these two options requires a precise quantitative modelling of both the formation and destruction mechanisms of clusters. Yet, even once the effects of galaxy evolution on this relation studied and understood, it will remain to explain the very origin of the relation between a baryonic component and the mass of the dark matter halo. The solution is likely hidden in galaxy formation, including cooling processes.

On top of tracing the mass of the galaxies, globular clusters are also used to map the dark halos at large radii, i.e. where the stellar component is too faint \citep[e.g.][]{Pota2013, Kartha2016}. Along those lines, \citet{Hudson2017} and \citet{Forbes2017} noted a correlation between the size of the cluster population and the virial radius of the galaxy. There is an ongoing debate on the nature of the scaling but, in any case, one can ask the same questions as for the mass relation, on the origin of this relation and why it apparently remains insensitive to galaxy evolution.

%%%%%%%%%%%%%%%%%%%%%%%%%%%%%%%%%%%%%%%%%%%%%%%%%%%%%%%%%%%%%%%%%
\section{The role of globular clusters in reionization}
\label{sec:reionization}

Reionization of the Universe is a major step of galaxy evolution as it alters the star formation activity and can even quench it in the most fragile, low-mass galaxies \citep[e.g.][]{Brown2014}. It has been invoked for instance to solve the so-called missing satellite problem \citep{Moore1999, Bullock2010}, i.e. the over-production of halos populated with stars in cosmological simulations, with respect to the observation of nearby dwarfs.

Despite its importance, the lack of strong observational constraints on reionization hinders our understanding of the physics driving it. Simulations also cannot provide a definite answer, due to the huge range of scales to cover. It is indeed crucial that ionising stars and the medium in which the flux propagates and is absorbed are captured together. At present-day resolutions, one has to rely on sub-grid recipes to describe the star formation process, the emission of the ionising flux (and the other feedback mechanisms), and its ability to propagate in the ISM. This escape fraction is thus largely linked to the porosity of the ISM which can only be described by capturing its turbulence.

In particular, the sources of ionising photons have not been clearly identified. Naturally, one first considers powerful sources like active galactic nuclei. However, such objects are rare before $z\sim 3$ \citep{Ricci2017}, and despite being powerful, they might not be sufficient to produce (alone) the flux needed for re-ionising the Universe \citep{Hassan2018}. The next candidates are massive star forming galaxies \citep{Barkana2001, Loeb2001}, more numerous at high redshift. For instance, \citet{Robertson2015} showed that early star formation (including population III stars) could play an important role. Such scenarios suffer from uncertainties on the escape fraction, which is only a few percent in massive systems \citep{Grazian2016, Vasei2016}, but significantly the higher in low mass galaxies \citep{Vanzella2016, Bian2017}. Therefore, yielding an higher escape fraction and being more numerous, dwarf galaxies could also be a strong driver. \citet{Weisz2017} showed that most of the stars in low mass galaxies are formed early, and could thus participate in the re-ionising flux. The limitation here is the ability of the lowest-mass systems to actually form stars efficiently, which supposedly sets a mass threshold to participate in the reionization, and is yet to be established.

Following this idea, it appears that young globular clusters formed as early as $z \gtrsim 6$ \citep{Katz2013} could also provide a sufficient ultra-violet flux to be major actors in the reionization process. \citet{Boylan2018} estimated the evolution of the cluster luminosity function by adopting a star cluster formation history and assuming that the shape of the luminosity function remains the same for a given population of clusters. This assumption is motivated by our lack of understanding of the dissolution processes affecting clusters (and their dependence on mass and luminosity, as discussed before). Yet, even with these conservative arguments and others (like the universality of the IMF), \citet{Boylan2018} found that young globular clusters would emit a sufficient ultra-violet flux to reionize the Universe. By better understanding the condition of globular cluster formation (including its epoch), of their multiple stellar populations, with more precise constrains on their dissolution, and finally with a description of the ISM in which their ionising flux propagates, one could quantify their role and date reionization. Such prediction should then be confronted to observational data, in particular in the deep fields of the Hubble Space telescope (HST), and the forthcoming James Web Space Telescope.

As often, it is quite possible that the net ionising flux results from the combination of several types of sources. In this picture, it is for instance not clear whether dwarf galaxies are the main drivers of the reionization, its victim, or both. While HST observations provides constraints on the ultra-violet flux at the end of the reionization era \citep{Finkelstein2015, Bouwens2015}, the escape fraction remains largely uncertain \citep[see a discussion in][]{Leitet2013}. This situation will improve with JWST that will push further the boundaries in term of redshift and luminosity, but the faintest galaxies will still remain out of reach.

%%%%%%%%%%%%%%%%%%%%%%%%%%%%%%%%%%%%%%%%%%%%%%%%%%%%%%%%%%%%%%%%%
\section{And many more ...}

Only a few open questions are highlighted above, but many more deserve deep investigations. In no particular order, one could mention:
\begin{itemize}
\item What sets the initial binary fraction, and is it the same at high redshift?
\item How do extended (faint-fuzzy) clusters form?
\item How to explain the high specific frequency of clusters in dwarf galaxies?
\item How do multiple stellar populations form?
\item Do intermediate mass black holes exist in clusters, and how do they relate to stellar-mass and super massive black holes?
\item and many more.
\end{itemize}
These open questions, as well as all the topics addressed in this review show that the previously distinct fields of star clusters and galaxies, but also stellar evolution and cosmology are increasingly getting closer to each others, and might well be on the verge of merging in a few years. The huge and diverse sample of multi-scale and multi-physics topics will keep many astrophysics busy and excited for a long time.

%%%%%%%%%%%%%%%%%%%%%%%%%%%%%%%%%%%%%%%%%%%%%%%%%%%%%%%%%%%%%%%%%%%%%%%%%%%%%%%%%%%%%%%%%%%%%%%%%%%%%%%%%%%%%%%%%%%%%%%%%%%%%%%%%%
\section*{Acknowledgements}

It is a pleasure to thank colleagues and friends for many discussions on the topics of this review over the years, and for pointing me to useful works and datasets. I am particularly grateful to Oscar Agertz, Mark Gieles, Fr\'ed\'eric Bournaud, Christian Boily, Eduardo Balbinot, Michelle Collins, Francoise Combes, Filippo Contenta, Maxime Delorme, Pierre-Alain Duc, Bruce Elmegreem, Eric Emsellem, Jeremy Fensch, Duncan Forbes, Morgan Fouesneau, Iskren Georgiev, Nicolas Guillard, Oleg Gnedin, Douglas Heggie, Michael Hilker, Nathan Leigh, Nicolas Martin, Alessandra Mastrobuono-Battisti, Ramon Rey-Raposo, Aaron Romanowsky, Anna Sippel, Enrico Vesperini and Karina Voggel. I also thank the members of the astrophysics groups at the Universities of Lund (Sweden) and Surrey (UK) for numerous interesting, stimulating and motivating discussions. I am indebted to the editor, Cathie Clarke, for her kind invitation to write this review. 
I acknowledge support from the European Research Council through grant ERC-StG-335936, and the Knut and Alice Wallenberg Foundation. The preparation of this contribution made intensive use of the NASA's Astrophysics Data System, as illustrated below.

\section*{References}
\bibliographystyle{elsarticle-harv} 
% \bibliography{biblio}

\end{document}